\documentclass[letterpaper,11pt,xcolor=dvipsnames]{article}
\pdfoutput=1
\usepackage{float}
\usepackage{hyperref}
\usepackage{cite}

\usepackage[left=2cm,right=2cm,top=2cm,bottom=3cm]{geometry}
\usepackage{amssymb}
\usepackage{amsmath}
\usepackage{amsfonts}
\usepackage{graphicx}
\usepackage{color}
\usepackage[dvipsnames]{xcolor}
\usepackage{xspace}
\usepackage{latexsym} 
\usepackage{mathrsfs}
\usepackage{xcolor}
\usepackage{cancel}
\usepackage{anysize}
\usepackage{wrapfig}
\usepackage{slashed}
\usepackage{lipsum}
\usepackage{multirow}
\usepackage{ctable}
\usepackage[utf8]{inputenc}
\usepackage{extarrows}

\usepackage{ltablex}
\usepackage{gensymb}
\usepackage{pifont}
\usepackage{setspace}
\usepackage{booktabs}
\usepackage{bbm}
\usepackage{pifont}
\usepackage{marginnote}

\definecolor{nicered}{rgb}{0.6,0,0}
\definecolor{nicegreen}{rgb}{0.1,0.5,0.1}
\definecolor{niceblue}{rgb}{0,0.4,0.8}
\definecolor{newgreen}{rgb}{0,0.667,0}
\hypersetup{
	colorlinks=true,      
	linkcolor=blue,      
	citecolor=blue,       
	filecolor=magenta,     
	urlcolor=blue
}

\allowdisplaybreaks

\newcommand{\Op}{\mathcal{O}}

\newcommand{\M}{\mathcal{M}}

\newcommand{\lr}[1]{\left(#1\right)}
\newcommand{\Lag}{\mathcal{L}}
\newcommand{\nnbar}{n\text{\--}\bar{n}}
\newcommand{\eq}{\text{eq}}

\begin{document}
	\allowdisplaybreaks
	\setcounter{footnote}{0}
	\vspace*{-1.5cm}
	\begin{flushright}
	TUM-HEP-1337/21 \\

	\vspace*{2mm}
	%\today
	\end{flushright}

	\begin{center}
	\vspace*{1mm}
	%%%%%%%%%%%%%%%%%%%%%
	\vspace{1cm}

\vspace*{.5cm}
{\Large\bf Probing baryogenesis with neutron-antineutron oscillations}

\vspace*{0.8cm}

{\bf K\aa re Fridell\footnote{email: {\tt kare.fridell@tum.de}}, Julia Harz\footnote{email: {\tt julia.harz@tum.de}}, and Chandan Hati\footnote{email: {\tt c.hati@tum.de}}}

\vspace*{.5cm}
 Physik Department T70, Technische Universit\"at M\"unchen,\\
James-Franck-Stra{\ss}e 1, D-85748 Garching, Germany

\end{center}

\vspace*{2mm}
%%%%%%%%%%%%%%%%%%%%%%%%%%%%%%%%%%%%%%%%%%%%%%%%%%%%%%%%%%%%%%%%%%%%%%%%%%%%%%%%%%%%%%%%%
\begin{abstract}
\noindent
In the near future, the Deep Underground Neutrino Experiment and the European Spallation Source aim to reach unprecedented sensitivity in the search for neutron-antineutron ($n\text{\--}\bar{n}$)~oscillations, whose observation would directly imply $|\Delta B| = 2$ violation and hence might hint towards a close link to the mechanism behind the observed baryon asymmetry of the Universe. In this work, we explore the consequences of such a discovery for baryogenesis first within a model-independent effective field theory approach. We then refine our analysis by including a source of CP violation and different hierarchies between the scales of new physics using a simplified model. We analyse the implication for baryogenesis in different scenarios and confront our results with complementary experimental constraints from dinucleon decay, LHC, and meson oscillations. We find that for a small mass hierarchy between the new degrees of freedom, an observable rate for $n\text{\--}\bar{n}$~oscillation would imply that the washout processes are too strong to generate any sizeable baryon asymmetry, even if the CP~violation is maximal. On the other hand, for a large hierarchy between the new degrees of freedom, our analysis shows that successful baryogenesis can occur over a large part of the parameter space, opening the window to be probed by current and future colliders and upcoming $n\text{\--}\bar{n}$ oscillation searches.
\end{abstract}
%%%%%%%%%%%%%%%%%%%%%%%%%%%%%%%%%%%%%%%%%%%%%%%%%%%%%%%%%%%%%%%%%%%%%%%%%%%%%%%%%%%%%%%%%
\setcounter{footnote}{0}

\section{Introduction}\label{sec:into}
Baryon number ($B$) is an accidental global symmetry in the Standard Model (SM), which explains the empirically severe experimental limits from the non-observation of the proton decay. Baryon number conservation is violated in the SM only at finite temperature through non-perturbative instanton effects. However, $B$ is expected to be violated in many well motivated ultraviolet (UV) extensions, e.g. grand unified theories (GUTs) naturally violate $B$, given the fact that quarks and (anti)leptons
are often placed in the same representation(s) of the GUT gauge group. Finally, one of the big questions of particle physics is the observed baryon asymmetry in the Universe: the overabundance of baryons over antibaryons, quantified by the measured baryon-to-photon number density ratio \cite{Aghanim:2018eyx}
\begin{equation}
 \eta_{B}^{\mathrm{obs}}= (6.20 \pm 0.15) \times 10^{-10}\,.
\end{equation}
A theoretical explanation of the dynamical generation of such a baryon asymmetry requires the three Sakharov conditions \cite{Sakharov:1967dj} -- (i) $B-L$ violation (where $L$ is the lepton number), (ii) C and CP violation and (iii) a departure from thermal equilibrium -- to be fulfilled, where the first condition can be induced via $B$ violation. 

Proton decay modes, e.g. $p\rightarrow e^+ \pi^0$, mediated via dimension-six operators, attributed to $|\Delta B|=1$ and $|\Delta (B-L)|=0$ can directly probe very high scales, $\mathcal{O}(10^{16} \text{GeV})$. The severity of the experimental limits on the non-observation of the proton decay might lead to the na\"ive expectation that dimension-nine $|\Delta B|=2$ operators mediating neutron-antineutron ($\nnbar$) oscillations must be even more suppressed when compared with single-nucleon decay modes. However, this is true only when a single heavy new physics (NP) mass scale is involved. In fact, the presence of more than one new scale beyond the SM might suppress single-nucleon decay, while mediating $\nnbar$ oscillations at a level comparable to the current experimental limits. Intriguingly, $|\Delta B|=2$ observables such as $\nnbar$ oscillations or dinucleon decays can be intimately connected to the baryon asymmetry of the Universe as these processes violate $B-L$ by two units ($|\Delta (B-L)|=2$). Therefore, $|\Delta B|=2$ processes like $\nnbar$ oscillations can be used to verify and probe baryogenesis mechanisms through $B$ (and $B-L$) violation directly\footnote{Note that in some early baryogenesis realisations in GUT theories \cite{Weinberg:1979bt,Fry:1980bd,Yoshimura:1978ex,Ellis:1978xg} containing baryon number violation, $B-L$ is conserved while $B+L$ is violated. In such scenarios any asymmetry generated below the unification scale gets washed out by electroweak sphaleron interactions \cite{Kuzmin:1985mm} and therefore a successful baryogenesis mechanism cannot be realised. Some alternatives include electroweak baryogenesis (which does not work within the SM alone, but can potentially work in some SM extensions, see e.g. Ref. \cite{Morrissey:2012db}), leptogenesis~\cite{Fukugita:1986hr} connected to the seesaw mechanism~\cite{Minkowski:1977sc,GellMann:1980vs,Yanagida:1979as,Mohapatra:1979ia,Schechter:1980gr,
Mohapatra:1980yp,Schechter:1981cv} of neutrino masses, as well as high-scale baryogenesis. In this work we will primarily be interested in the latter scenario.}. 

During the recent years, on the one hand, lattice-QCD calculations have been improved tremendously in computing the QCD matrix elements that connect amplitudes of $B$-violating interactions to $\nnbar$ oscillations~\cite{Rinaldi:2019thf}, on the other hand, the current experimental sensitivities and future prospects for the observation of $\nnbar$ oscillations have improved significantly. Currently, the most stringent constraint from the bound $\nnbar$~oscillation is due to the Super-Kamiokande experiment~\cite{Abe:2020ywm}, which provides a limit on the $\nnbar$~oscillation lifetime $\tau_{\nnbar}^{\text{SK}}\ge 4.7 \times 10^8$~s. The current best limit from the free $\nnbar$~oscillation is due to the ILL experiment~\cite{BaldoCeolin:1994jz} $\tau_{\nnbar}^{\text{ILL}}\ge 0.86 \times 10^8$~s. Experimental sensitivities from both, free and bound $\nnbar$~oscillation times, are expected to be improved significantly in future experiments. The DUNE experiment \cite{Acciarri:2015uup} is expected to achieve a sensitivity of $\tau_{\nnbar}^{\text{DUNE}}\ge 7 \times 10^8$~s using $^{40}\text{Ar}$ nuclei, while NNBAR \cite{Addazi:2020nlz} will exploit the Large Beam Port of the ESS facility to search for free $\nnbar$ oscillations and is expected to achieve an impressive sensitivity of $\tau_{\nnbar}^{\text{NNBAR}}\ge 3 \times 10^9$~s. Given the current stringent limits and expected future improvements in the experimental sensitivities for $\nnbar$ oscillations, a detailed study of the phenomenological implications of such searches for baryogenesis mechanisms is timely and of high theoretical importance, since $\nnbar$ oscillations are among very few observables which provide an opportunity to directly probe baryogenesis mechanisms and to distinguish underlying NP scenarios in synergy with direct searches at the high-energy frontier. 

In this work, we explore the phenomenological possibility of probing baryogenesis using $\nnbar$ oscillations and other complementary observables at the high-energy and high-intensity frontiers. We commence with an effective field theory (EFT) framework for $\nnbar$ oscillations and study the impact of the current and future experimental limits of the $\nnbar$~oscillation lifetime on the viability of realising a successful baryogenesis mechanism. Taking into account the latest lattice-QCD computations of the QCD matrix elements and relevant renormalisation group (RG) running effects, we first present a general framework for estimating the washout processes to derive model-independent limits on the viable scale for baryogenesis. In order to accommodate the possibility of different hierarchies of NP within the effective
operator and an additional new source of CP violation, we explore then a simplified set-up to perform a comprehensive phenomenological study of the viability of baryogenesis above the electroweak scale in the context of an observable $\nnbar$~oscillation lifetime. 
In particular, we focus on the $B$ (and $B-L$) violating trilinear scalar coupling topology for $\nnbar$~oscillation originally proposed in Ref.~\cite{Mohapatra:1980qe}. In order to make our analysis as general as possible, we consider a minimal simplified extension of the SM with diquark scalar fields coupling to SM quark fields and a $B$ (and $B-L$) violating trilinear scalar coupling involving diquark scalar fields. Interestingly, a split scenario featuring some diquarks at TeV scale and some around GUT scale leads to the interesting possibility of realising a high-scale baryogenesis mechanism that can be readily embedded in many well motivated UV realisations. Moreover, it can be probed using the synergy between $\nnbar$ oscillations and direct searches at the colliders, where the scalar diquarks are subject to extensive searches.  The most stringent current constraints on the mass of the diquarks are already at the level of a few TeV for order unity couplings and are expected to be improved significantly in future searches. 
However, phenomenologically, diquark masses can generally also lie within the range from a few TeV to the GUT scale\footnote{Another alternative baryogenesis mechanism that can occur in a comparable set-up is post-sphaleron baryogenesis, see e.g. \cite{Babu:2006xc,Babu:2008rq,Babu:2013yca}. We leave a detailed analysis of this realisation for future work~\cite{Fridell:2021aaa}.}. While there are a few instances of relevant studies for baryogenesis for some comparable scenarios in the literature~\cite{Babu:2012vc,Baldes:2011mh,Herrmann:2014fha}, in this work we present for the first time a detailed and consistent prescription for the derivation of applicable Boltzmann equations and study the different relevant cases of phenomenological interest. We include in a comprehensive manner all experimental and theoretical constraints relevant for constraining the parameter space comprising $\nnbar$ oscillations, meson oscillations, dinucleon decay, LHC constraints, and limits from a colour preserving vacuum. For instance, in contrast to $\nnbar$ oscillations, the dinucleon decay is particularly relevant for TeV-scale masses of diquarks due to the stringent constraints on the couplings of diquarks to first generation quarks from meson oscillations. 

Our findings suggest that the complementarity between $\nnbar$ oscillations and LHC searches for diquarks can probe the baryogenesis mechanism extensively, ruling out the possibility of successful baryogenesis in some scenarios. The current best limit from the Super-Kamiokande experiment on the $\nnbar$~oscillation lifetime together with the latest CMS limits exclude a large part of the viable parameter space for successful high-scale baryogenesis where one of the diquarks features a GUT scale mass while another one lies in the collider-accessible TeV scale range. However, we demonstrate that in case the future searches for $\nnbar$ oscillations observe a signal, high-scale baryogenesis still remains a viable option to generate the correct observed baryon asymmetry of the Universe. On the other hand, for a scenario with all scalar diquark masses being in a similar mass range ($\lesssim 10^5$ TeV), the corresponding washout processes prove to be too strong to create a sizeable baryon asymmetry. Therefore, if the relevant scalar diquark fields are to be discovered by the LHC or future collider searches to lie within a few tens of TeV, then the possibility of a viable baryogenesis scenario featuring no additional particles at a high scale is completely ruled out. Baryogenesis scenarios with all diquarks having masses $\gtrsim 10^5$~TeV can still work successfully, however, remain unfortunately largely inaccessible by current and future experiments.   

The paper is organised as follows:  We introduce in Sec.~\ref{sec:EFT}, an effective field theory (EFT) framework for the operators mediating $\nnbar$ oscillations, which we then use to set limits on the corresponding Wilson coefficients and consequently on NP mass scales and couplings in the subsequent sections. In Sec.~\ref{sec:boltzmann}, we first present a model-independent analysis of the washout of a pre-existing baryon asymmetry due to the effective operators mediating $\nnbar$ oscillations without introducing any new sources of CP violation. To explore the possibility of different hierarchies of NP within the effective operators and the impact of a new source of CP violation, we present two possible topologies for realising $\nnbar$ oscillations and study one of them using a simplified model set-up to perform a comprehensive phenomenological analysis. To conclude this section, we present the Boltzmann equation framework that we used to study the evolution of the baryon asymmetry including a detailed discussion of the CP violating decays and all relevant washout processes. In Sec.~\ref{sec:constraints}, we discuss all phenomenologically relevant constraints on such a set-up including the limits from LHC and future collider searches, neutral meson oscillations, dinucleon decay, colour preserving vacua and comment on some other observables. In Sec.~\ref{sec:results}, we combine all experimental constraints with the predictions for the final baryon asymmetry for two distinct scenarios (classified by the hierarchies of the NP scales involved) to present the parameter space for successful baryogenesis and discuss the implications for current and future experiments. Finally, in Sec.~\ref{sec:conclusion} we summarise and make concluding remarks.

\section{Neutron-antineutron oscillations in effective field theory}\label{sec:EFT}
 Neutron-antineutron ($\nnbar$) oscillations violate baryon number by two units ($|\Delta B|=2$), and therefore must be induced by some NP beyond the SM in case of an observation. Given any new physics model (e.g. the simplified model that we will consider in Sec~\ref{sec:boltzmann}), it is convenient to match the new physics operators mediating $\nnbar$ oscillations to the effective operators involving only light fields at the scale where the heavy NP is integrated out. The effect of the heavy new physics then can be encoded in the Wilson coefficients of the effective operators, while the operators can be rundown to the scale of $\nnbar$ oscillation and be identified with hadronic matrix elements, which are available from the lattice QCD computations. Therefore, we proceed to discuss below the relevant EFT formalisms which are of particular interest and provide an independent set of operators (and their relation to other commonly used operator bases in the literature) and hadronic matrix elements, which we then subsequently use for our simplified model.
 
 At the QCD scale, the effective Lagrangian for $\nnbar$ oscillations (after integrating out the heavy degrees of freedom) consists out of six SM quark fields with associated Wilson coefficients. These operators correspond to scattering processes violating baryon number at the temperature of interest for baryogenesis (up to the effects of RGE running of the Wilson coefficients). Therefore, the relevant effective operators (and the associated Wilson coefficients) are directly correlated with the washout processes which provide the possibility of probing the effectiveness of a given baryogenesis mechanism using the current and expected future experimental limits on the $\nnbar$~oscillation lifetime. Since the $\nnbar$~oscillation operators will be important for our study, we briefly summarise the formalism for the effective operator bases and the relevant RG running effects that we will later use for the analysis to obtain the corresponding $\nnbar$~oscillation rates. In subsection~\ref{subsec:2.1}, we first survey one of the commonly used $SU(3)_c\times U(1)_{\text{\normalfont EM}}$ invariant EFT bases, and comment on possible connections of this basis with a SMEFT formalism. In subsection~\ref{subsec:2.2}, we introduce the operator basis that we use in the rest of this work. This basis is also $SU(3)_c\times U(1)_{\text{\normalfont EM}}$ invariant and additionally obeys a chiral $SU(2)_L\otimes SU(2)_R$ symmetry, which is commonly used in the literature for lattice QCD computations of the relevant hadronic matrix elements. We also provide the relations between the operator basis in subsection~\ref{subsec:2.1} and the one in subsection~\ref{subsec:2.2}, and provide a list of independent hadronic matrix elements necessary to compute the $\nnbar$ oscillation rate for a given NP model. Finally, in subsection~\ref{subsec:2.3}, we provide the prescription to compute the $\nnbar$ oscillation rate including the RG running effects for the hadronic matrix elements.
 
  %%%%%%%%%%%%%%%%%%
  %%%%%%%%%%%%%%%%%%
  \subsection{$SU(3)_c\times U(1)_{\text{\normalfont EM}}$ invariant basis}\label{subsec:2.1}
  Note that an effective Lagrangian for $\nnbar$ oscillations relevant at scales below the electroweak symmetry breaking must ensure that the relevant six-quark operators preserve $SU(3)_c\times U(1)_{\text{EM}}$. A complete basis of such six-quark operators of the form $uudddd$, relevant for $\nnbar$ oscillations, can be constructed as follows~\cite{Chang:1980ey,Kuo:1980ew,Rao:1982gt,Rao:1983sd,Caswell:1982qs}:
\begin{equation}
\label{eqn:nnbar_obasis}
  \begin{split}
  \Op^{1}_{\chi_1 \chi_2 \chi_3}
    &= (u_i^T CP_{\chi_1} u_j) (d_k^T CP_{\chi_2} d_l) (d_m^T CP_{\chi_3} d_n)
        T_{\{ij\}\{kl\}\{mn\}}^\text{SSS}\,, \\
  \Op^{2}_{\chi_1 \chi_2 \chi_3}
    &= (u_i^T CP_{\chi_1} d_j) (u_k^T CP_{\chi_2} d_l) (d_m^T CP_{\chi_3} d_n)
        T_{\{ij\}\{kl\}\{mn\}}^\text{SSS}\,, \\
  \Op^{3}_{\chi_1 \chi_2 \chi_3}
    &= (u_i^T CP_{\chi_1} d_j) (u_k^T CP_{\chi_2} d_l) (d_m^T CP_{\chi_3} d_n)
        T_{[ij][kl]\{mn\}}^\text{AAS}\, .
  \end{split}
\end{equation}
Hereby $\chi = \{L, R\}$ indicates the chirality with $P_{L,R} = \frac{1}{2}(1 \mp \gamma_5)$ being the chiral projection operators. 
The contraction of the spinor indices are implicitly assumed in the parentheses, $C$ is the charge-conjugation operator and the quark colour tensors are defined as
\begin{equation}
  \begin{split}
	T^{SSS}_{\{ij\}\{kl\}\{mn\}} &= \varepsilon_{ikm}\varepsilon_{jln} + \varepsilon_{jkm}\varepsilon_{iln} + \varepsilon_{ilm}\varepsilon_{jkn} + \varepsilon_{jlm} \varepsilon_{ikn},\\
	T^{AAS}_{[ij][kl]\{mn\}} &= \varepsilon_{ijm}\varepsilon_{kln} + \varepsilon_{ijn}\varepsilon_{klm},
  \end{split}
\end{equation}
where $\{\;\}$ denotes index symmetrisation and $[\;\;]$ denotes index antisymmetrisation. Note that the operators involving the diquark invariants of the vector form $(q^T C P_\chi \gamma_\mu q)$ or tensor form $(q C P_\chi \sigma_{\mu\nu} q)$ are not independent and can be expressed as linear combinations of the operators given in Eq.~\eqref{eqn:nnbar_obasis} by performing Fierz transformations on a relevant subset of four fermions. Each of the operators in Eq.~\eqref{eqn:nnbar_obasis} leads to eight distinct operators when all possible combinations of chiralities are considered, leading to 24 operators in total. However, imposing the relations due to antisymmetrisation
\begin{equation}
\Op^{1}_{\chi_1 L R} = \Op^{1}_{\chi_1 R L}\, , \quad
  \Op^{2,3}_{L R \chi_3} = \Op^{2,3}_{R L \chi_3} ,
\end{equation}
each of the operators in Eq.~\eqref{eqn:nnbar_obasis} leads to only six distinct operators making the total number of operators 18. Out of these 18 possible operators, four can further be eliminated due to the relation
\begin{equation}
\Op^2_{\chi\chi\chi^\prime} - \Op^1_{\chi\chi\chi^\prime} = 3\Op^3_{\chi\chi\chi^\prime}\,,
\end{equation}
where $\chi,\chi'\in \left[L,R\right]$. If the NP fields mediating the $\nnbar$ oscillations are much heavier than the electroweak symmetry breaking scale, then the relevant effective six-quark operators in the effective Lagrangian (valid above the electroweak symmetry breaking scale), after integrating out the heavy NP degrees of freedom, must be SM gauge group invariant.

In passing we note that, since at the energy scale of $\nnbar$ oscillations the whole unbroken SM gauge group need not to be respected a $SU(3)_c\times U(1)_{\text{\normalfont EM}}$ invariant basis is the more appropriate choice of EFT. In case the requirement of invariance under $SU(2)_L \times U(1)_{Y}$ is imposed in addition to the $SU(3)_c\times U(1)_{\text{\normalfont EM}}$ (e.g. in the case of a SMEFT~\cite{Buchmuller:1985jz,Grzadkowski:2010es,Jenkins:2013zja,Jenkins:2013wua,Alonso:2013hga} formulation), then a subset of only four independent operators survive~\cite{Rao:1983sd,Caswell:1982qs}, e.g.
\begin{equation}
\label{eqn:nnbar_SMEFT_basis}
  \begin{split}
  \Op^{\text{SM}}_{1}
    &= \Op^{1}_{R R R}\,, \\
  \Op^{\text{SM}}_{2}
    &= \Op^{2}_{R R R}\,, \\
  \Op^{\text{SM}}_{3}
    &= 2 \Op^{3}_{L R R}=(q_i^{T\alpha} C P_L q_j^\beta) (u_k^T C P_{R} d_l) (d_m^T CP_{R} d_n)
        \varepsilon_{\alpha\beta} T_{\{ij\}\{kl\}\{mn\}}^\text{SSS}\,,\\
  \Op^{\text{SM}}_{4}
          &= 4 \Op^{3}_{L L R}=(q_i^{T\alpha} CP_{L} q_j^\beta) (q_k^{T\gamma} CP_{L} q_l^{\delta}) (d_m^T CP_{R} d_n)
          \varepsilon_{\alpha\beta}\varepsilon_{\gamma\delta}    T_{[ij][kl]\{mn\}}^\text{AAS}\, ,
  \end{split}
\end{equation}
where $q$ denotes $SU(2)_L$ quark doublet and the Greek indices $\alpha, \beta, \gamma, \delta=1,2$. 
  %%%%%%%%%%%%%%%%%%
  %%%%%%%%%%%%%%%%%%
\subsection{$SU(3)_c\times U(1)_{\text{\normalfont EM}}$ invariant basis with chiral $SU(2)_L\otimes SU(2)_R$ symmetry}\label{subsec:2.2}
Most of the recent robust calculations for the hadronic matrix elements are performed using lattice-QCD simulations including nonzero quark masses and matched to massless chiral perturbation theory in which the chiral symmetry $SU(2)_L\otimes SU(2)_R$ is approximately preserved~\cite{Buchoff:2015qwa}. Since the relevant latest hadronic matrix elements are readily available in an $SU(3)_C\times U(1)_{\text{EM}}$ invariant chiral basis with a $SU(2)_L\otimes SU(2)_R$ symmetry~\cite{Rinaldi:2019thf}, we find it convenient to work with it for the numerical evaluation of the $\nnbar$~oscillation rates. In this framework the $\nnbar$ oscillations can be described by the effective Lagrangian
\begin{equation}{\label{eq:nnbarLag}}
    \Lag^{\bar{n}\text{\--}n}_{\text{eff}} = \sum_{i=1,2,3,5} \big(C_i(\mu)\Op_i(\mu) +C_i^{P}(\mu) \Op_i^P(\mu)\big)+\text{h.c.},
 \end{equation}
where $C_i$ are the Wilson coefficients corresponding to the set of effective operators $\Op_i$ defined as~\cite{Rinaldi:2019thf}:
\begin{equation}
\begin{aligned}
\label{eq:operatorlist}
    &\Op_{1} =-4\Op^{3}_{RRR}= (\psi CP_R i\tau^2 \psi)(\psi CP_R i\tau^2 \psi)(\psi CP_R i\tau^2 \tau^+ \psi)T^{AAS},\\
    &\Op_{2}=-4\Op^{3}_{LRR} = (\psi CP_L i\tau^2 \psi)(\psi CP_R i\tau^2 \psi)(\psi CP_R i\tau^2 \tau^+ \psi)T^{AAS},\\
    &\Op_{3}=-4\Op^{3}_{LLR} = (\psi CP_L i\tau^2 \psi)(\psi CP_L i\tau^2 \psi)(\psi CP_R i\tau^2 \tau^+ \psi)T^{AAS},\\
    &\Op_{4}=-\frac{4}{5}\Op^{1}_{RRR}-\frac{16}{5}\Op^{2}_{RRR}\\
    &\phantom{\Op_{4}} =\left[ (\psi CP_R i\tau^2 \tau^3 \psi)(\psi CP_R i\tau^2 \tau^3 \psi)-\frac{1}{5}(\psi CP_R i\tau^2 \tau^a \psi)(\psi CP_R i\tau^2 \tau^a \psi)\right] (\psi CP_R i\tau^2 \tau^+ \psi)T^{SSS},\\
    &\Op_{5}=\Op^{1}_{RLL} = (\psi CP_R i\tau^2\tau^- \psi)(\psi CP_L i\tau^2\tau^+ \psi)(\psi CP_L i\tau^2 \tau^+ \psi)T^{SSS},\\
    &\Op_{6}=-4\Op^{2}_{RLL} = (\psi CP_R i\tau^2\tau^3 \psi)(\psi CP_L i\tau^2\tau^3 \psi)(\psi CP_L i\tau^2 \tau^+ \psi)T^{SSS},\\
    &\Op_{7}=-\frac{4}{3}\Op^{1}_{LLR}-\frac{8}{3}\Op^{2}_{LLR}\\
    &\phantom{\Op_{7}} =\left[ (\psi CP_L i\tau^2 \tau^3 \psi)(\psi CP_L i\tau^2 \tau^3 \psi)-\frac{1}{3}(\psi CP_L i\tau^2 \tau^a \psi)(\psi CP_L i\tau^2 \tau^a \psi)\right] (\psi CP_R i\tau^2 \tau^+ \psi)T^{SSS},\\
\end{aligned}
\end{equation}
which are related to the remaining seven independent operators $\Op_i^P$ by a parity transformation, accounting for total 14 independent operators. Here $\psi$ corresponds to the isospin doublet $\psi=(u,d)^T$, $C$ corresponds to the charge conjugation operator, $\tau^a$ denote the Pauli matrices for $i=1,2,3$ and $\tau^{\pm}=\tfrac{1}{2}(\tau^1\pm i \tau^2)$. We have dropped the colour subscripts of the fields and the colour tensors $T^{AAS(SSS)}$ for brevity. In Eq.~\eqref{eq:operatorlist}, the first equalities provides the relation between the new basis and the $SU(3)_c\times U(1)_{\text{\normalfont EM}}$ invariant basis defined in Eq.~\eqref{eqn:nnbar_obasis}.

As an useful remark, we note that many of the NP models (e.g. the simplified model considered in this work in Sec.~\ref{sec:boltzmann}), introduce two additional operators $\tilde{\Op}_1$ and $\tilde{\Op}_3$, given by~\cite{Buchoff:2015qwa}
\begin{equation}
\begin{aligned}
\label{eq:oplisttilde}
    &\tilde{\Op}_{1} =-4/3( \Op^{2}_{RRR}- \Op^{1}_{RRR})= (\psi CP_R i\tau^2\tau^a \psi)(\psi CP_R i\tau^2 \tau^a \psi)(\psi CP_R i\tau^2 \tau^+ \psi)T^{SSS},\\   
    &\tilde{\Op}_{3}=-4/3( \Op^{2}_{LLR}- \Op^{1}_{LLR}) = (\psi CP_L i\tau^2 \tau^a \psi)(\psi CP_L i\tau^2\tau^a \psi)(\psi CP_R i\tau^2 \tau^+ \psi)T^{SSS}.\\
\end{aligned}
\end{equation}
However, these operators are not independent with respect to the complete basis of 14 operators included in Eq.~\eqref{eq:operatorlist}. In fact, in dimension $D=4$, the operators $\tilde{\Op}_1$ and $\tilde{\Op}_3$ are equal to ${\Op}_1$ and ${\Op}_3$, respectively, by Fierz relations. However, such Fierz relations are broken by dimensional regularisation, therefore, in addition to the operators in Eq.~\eqref{eq:operatorlist} one must also include ${\Op}_{1}-\tilde{\Op}_{1}$ and ${\Op}_{3}-\tilde{\Op}_{3}$ as evanescent operators (vanishing for $D=4$) for a complete treatment of the EFT at an arbitrary $D$. Alternatively, one can choose to include $\tilde{\Op}_1$ and $\tilde{\Op}_3$ explicitly as a part of the physical basis of EFT operators. 

To compute the $\nnbar$ oscillation rate, we will be interested in the hadronic matrix elements associated with the operators defined in Eq.~\eqref{eq:operatorlist}. To this end, we note that the isospin symmetry approximation further reduces the number of the relevant $\nnbar$ matrix elements, making the matrix elements associated with three of the operators in Eq.~\eqref{eq:operatorlist} redundant~\cite{Rinaldi:2019thf}, as follows. The hadronic matrix element for $\Op_{4}$ vanishes
\begin{equation}{\label{eq:vanO4}}
    \langle \bar{n}| \Op_4| n \rangle= 0\, ,
 \end{equation}
in the approximate limit $m_u=m_d$ (even after including the isospin breaking effects this matrix element is suppressed by powers of $(m_u-m_d)/\Lambda_{\text{QCD}}$). On the other hand, in the presence of isospin symmetry the hadronic matrix elements for $\Op_{6}$ and $\Op_{7}$ are related to that of $\Op_{5}$ by 
\begin{equation}
    \langle \bar{n}| \Op_5| n \rangle= \langle \bar{n}| \Op_6| n \rangle=-\frac{3}{2} \langle \bar{n}| \Op_7| n \rangle\, .
 \end{equation}

Therefore, for all practical purposes we work with in total four independent hadronic matrix elements to\footnote{Note that the operators $\Op_{1}$, $\Op_{2}$, and $\Op_{3}$ are $SU(2)_L \times U(1)_Y$ singlets, while $\Op_{5}$ (and the related $\Op_{6,7}$) is an $SU(3)_C\times U(1)_{\text{EM}}$ singlet, but is not invariant under $SU(2)_L$ and therefore can arise from a higher dimensional electroweak symmetry invariant operator from the SMEFT~\cite{Buchmuller:1985jz,Grzadkowski:2010es,Jenkins:2013zja,Jenkins:2013wua,Alonso:2013hga} point of view.} Here we will not try to construct a complete SMEFT invariant basis but we will rather assume that the SMEFT basis is matched to the $SU(3)_C\times U(1)_{\text{EM}}$ invariant EFT introducing the relevant electroweak vacuum expectation values (VEVs). $\Op_{1}$, $\Op_{2}$, $\Op_{3}$, $\Op_{5}$. Given an explicit model, we first compute the relevant Wilson coefficients corresponding to a given operator in Eq.~\eqref{eq:operatorlist} at the scale where the heavy NP is integrated out and then identify the operator with one of the four independent hadronic matrix elements, which are subject to running effects between the scale where the lattice-QCD nucleon matrix elements are available in the $\overline{\text{MS}}$ scheme and the heavy NP scale where the effective Lagrangian is defined, as discussed in the following subsection.

  %%%%%%%%%%%%%%%%%%
  %%%%%%%%%%%%%%%%%%
\subsection{$\nnbar$ oscillation lifetime and RG running effects}\label{subsec:2.3}
The transition rate for the $\nnbar$ oscillations, $\tau_{\nnbar}^{-1}$, is related to the Lagrangian in Eq.~\eqref{eq:nnbarLag} via
\begin{equation}
\tau_{\nnbar}^{-1} = \langle\bar{n}|\Lag^{\bar{n}\text{\--}n}_{\text{eff}}|n\rangle = \big|\sum_{i=1,2,3,5} (C_i(\mu)\M_i(\mu) +C_i^{P}(\mu) \M_i^P(\mu))\big|.
\end{equation}
Here, $\M_i(\mu)$ is the transition matrix element for operator $\Op_i(\mu)$, with $\M_i(\mu)\equiv \langle \bar{n}| \Op_i| n \rangle$. These can be determined using lattice-QCD techniques as described in~\cite{Rinaldi:2019thf}, which also provides the relevant numerical values for $\M_i(\mu)$ in the $\overline{\text{MS}}$ scheme at $\mu = $ 2 GeV. Since we are interested in the case where the NP scale is much higher as compared to the $\nnbar$ scale, i.e. where the relevant Wilson coefficients are defined at some heavy NP scale, the running of the operators between the different scales needs to be considered.

\begin{figure}[ht!]
	\centering
	\includegraphics[width=0.7\textwidth]{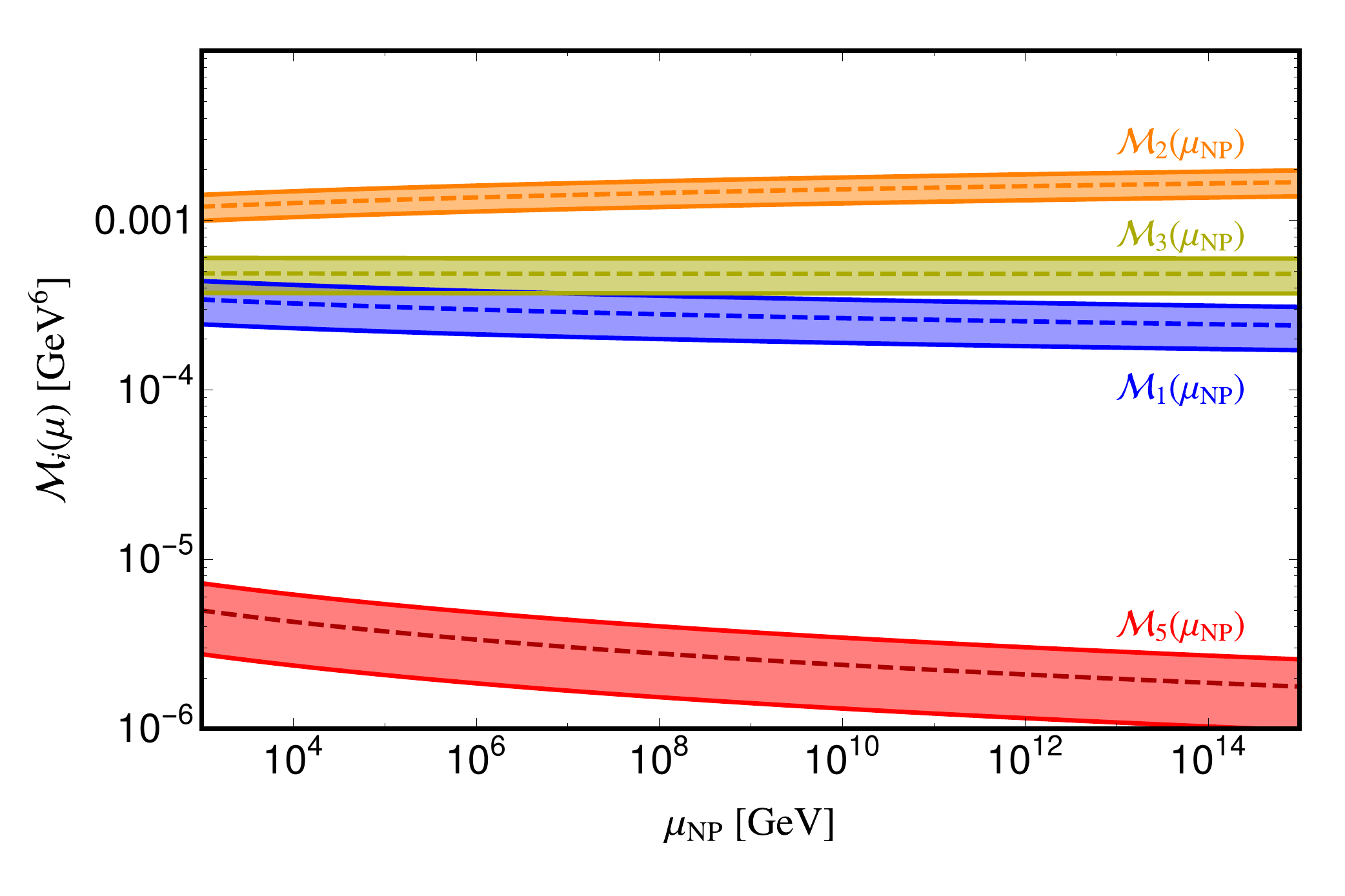}
	\caption{Running of the relevant transition matrix elements with respect to the NP scale $\mu_{\text{NP}}$. The dashed line represents the central value, and the width shows the errors added in quadrature as tabulated in Tab.~\ref{tab:gammar0}.}
	\label{fig:RG}
\end{figure}

Hence, to evolve the transition nuclear matrix elements from the relevant lattice-QCD scale to some heavy NP scale, it is necessary to perform the RG running of the EFT operators, for which we follow the prescription of~\cite{Buchoff:2015qwa}, as detailed below. The running of the operators from the lattice scale $\mu_0$ to a higher scale $\mu_{NP}$, to first order in the strong coupling constant $\alpha_S$, is described by the following equation~\cite{Buchoff:2015qwa}:
\begin{equation}
\label{eq:operatorrunning}
    \Op_i(\lambda) = U'_i(\mu_{NP},\mu_0) \Op_i(\mu_0)\equiv U_i^{N_f=6}(\mu_{NP},m_t) U_i^{N_f=5}(m_t,m_b) U_i^{N_f=4}(m_b,\mu_0)\Op_i(\mu_0),
\end{equation}
for $m_c < \mu_0 < m_b$, where
\begin{equation}
\label{eq:runningfactor}
    U_i^{N_f}(q_1,q_2) = \lr{\frac{\alpha_S(q_2)}{\alpha_S(q_1)}}^{-\gamma_i^{0}/8\pi\beta_0}\left[1-\delta_{q_2,\mu_0}r_i^{(0)}\frac{\alpha_S(\mu_0)}{4\pi}+\lr{\frac{\beta_1\gamma_i^{(0)}4\pi}{2\beta_0^2}-\frac{\gamma_i^{(1)}}{2\beta_0}}\frac{\alpha_S(q_2)-\alpha_S(q_1)}{16\pi^2}\right],
\end{equation}
up to $\mathcal{O}(\alpha_s^2)$. Here, $q_1$ and $q_2$ are two mass scales, with $q_2 < q_1$, and $N_f$ is the number of quark flavours with masses above $q_2$. Furthermore, the one- and two-loop $\overline{\text{MS}}$ anomalous dimensions $\gamma_i^{(0)}$ and $\gamma_i^{(1)}$, and the one-loop Landau gauge Regularisation-Independent-Momentum (RI-MOM) matching factor $r_i^{(0)}$, are summarised in Table~\ref{tab:gammar0}.
\begin{table}[h!]
    \begin{center}
    \begin{tabular}{|c|c|c|c|c|c|}
    \hline
       $\Op$  &  $\gamma_i^{(0)}$ & $\gamma_i^{(1)}$ & $r_i^{(0)}$ & $\M(\mu=2\text{ GeV})$ [GeV$^6$]
       \\
       \hline
       $\Op_1$  &  $4$  & $335/3-34N_f/9$ & $101/30+8/15\ln 2$ & $-46(13)(2)\times10^{-5}$
       \\
       $\Op_2$  &  $-4$  & $91/3-26N_f/9$ & $-31/6+88/15\ln 2$ & $95(15)(7)\times10^{-5}$
        \\
       $\Op_3$  &  $0$  & $64-10N_f/3$ & $-9/10+16/5\ln 2$ & $-50(10)(6)\times10^{-5}$
       \\
       $\Op_5$  &  $24$  & $238-14N_f$ & $49/10-24/5\ln 2$ & $-1.06(45)(15)\times10^{-5}$\\
       \hline
    $\vphantom{\tilde{\tilde{\Op}}_1}\tilde{\Op}_1$  &  $4$  & $797/3-118 N_f/9$ & $-109/30+8/15\ln 2$ & -
       \\
	 $\tilde{\Op}_3$  &  $0$  & $218-38 N_f/3$ & $-79/10+16/5\ln 2$ & -
	 \\
		\hline
    \end{tabular}
    \caption{The one- and two-loop running factors $\gamma_i^{(0)}$ and $\gamma_i^{(1)}$, and the one-loop matching factor $r_i^{(0)}$, for each operator $\Op_i$ taken from~\cite{Rinaldi:2019thf} and~\cite{Buchoff:2015qwa}. }
    \label{tab:gammar0}
    \end{center}
\end{table}
The scale-dependent strong coupling constant $\alpha_S$ is given, at 4-loop order, by~\cite{Bethke:2009jm}
\begin{equation}
\label{eq:alphastrong}
\begin{aligned}
    \alpha_S(q) &= \frac{1}{\beta_0L}-\frac{1}{\beta_0^3L^2}\beta_1\ln L + \frac{1}{\beta_0^3L^3}\left[\frac{\beta_1^2}{\beta_0^2}(\ln^2L-\ln L - 1)+\frac{\beta_2}{\beta_0}\right]\\
    &+ \frac{1}{\beta_0^4L^4}\left[\frac{\beta_1^3}{\beta_0^3}\lr{-\ln^3L + \frac{5}{2}\ln^2L+2\ln L - \frac{1}{2}} - 3\frac{\beta_1\beta_2}{\beta_0^2}\ln L - \frac{\beta_3}{2\beta_0}\right],
    \end{aligned}
\end{equation}
where
\begin{equation}
L = \ln\lr{\frac{q^2e^{1/(\beta_0\alpha_S(q_{\alpha}))}}{q_{\alpha}^2}},
\end{equation}
with $q_{\alpha}$ corresponding to the scale (with $N_f$ quarks on-shell) at which $\alpha_S$ is known. The relevant $\beta$-functions in Eqs.~\eqref{eq:runningfactor} and \eqref{eq:alphastrong} are given by
\begin{equation}
    \begin{aligned}
        \beta_0 &= \frac{33-2N_f}{12\pi},\\
        \beta_1 &= \frac{153-19N_f}{24\pi^2},\\
        \beta_2 &= \frac{77139-15099N_f+325N_f^2}{3456\pi^3},\\
        \beta_3 &\approx \frac{29243-6946.3N_f+405.089N_f^2+1.49931N_f^3}{256\pi^4}.
    \end{aligned}
\end{equation}
Using Eq.~\eqref{eq:operatorrunning}, we obtain the matrix elements for each operator at a scale $\mu_{\text{NP}}$, by running them from the scale $\mu_0 = 2$ GeV to the corresponding scale of NP $\mu_{\text{NP}}$. For a ready reference, we show the the relevant running for the matrix elements as a function of the NP scale in Fig.~\ref{fig:RG}. 

Neglecting the next-to-next-to-leading-order perturbative renormalisation effects the $\nnbar$ transition rate can be expressed in terms of the nuclear matrix element at the scale $\mu = $ 2 GeV as~\cite{Rinaldi:2019thf}
\begin{eqnarray}
\label{eq:taunn}
\tau_{\nnbar}^{-1}
&=&  1.52\times 10^{18} \; \left|\sum_{i=1,2,3,5} \frac{\M_i(\mu)}{{\text{(GeV)}}^6} \left[\left(\frac{C_i(\mu)}{{\text{(TeV)}}^{-5}}\right) - \left(\frac{C_i^{P}(\mu)}{{\text{(TeV)}}^{-5}} \right)\right]\right|_{\mu=\mu_{\text{NP}}} \times 10^{-9} \text{ s}^{-1} \\
&=&  1.52\times 10^{18} \;
\left|\sum_{i=1,2,3,5}  U'_i(\mu,2\,\text{GeV}) \frac{\M_i(2\, \text{GeV})}{{\text{(GeV)}}^6} \left[\left(\frac{C_i(\mu)}{{\text{(TeV)}}^{-5}}\right) - \left(\frac{C_i^{P}(\mu)}{{\text{(TeV)}}^{-5}} \right)\right]\right|_{\mu=\mu_{\text{NP}}}\hspace{-1cm}\times 10^{-9} \text{ s}^{-1}\,,\nonumber
\end{eqnarray}
where $\M_i(2\, \text{GeV})$ is given in Tab.~\ref{tab:gammar0} and $U'_i(\mu,2\,\text{GeV})$ in Eq.~\ref{eq:operatorrunning}. The relevant Wilson coefficients are obtained by computing the NP diagrams in a BSM scenario and integrating out the heavy NP degrees of freedom to match $\Op_i(\mu)$ and consequently $\M_i(\mu)\equiv \langle \bar{n}|\Op_i(\mu)| n\rangle$. Notice that Eq.~\eqref{eq:taunn} is expressed to make all the relative fractions involved dimensionless. It is then straightforward to notice that for $C_i^{(P)}\sim \Lambda^{-5}$, with $\Lambda$ being the relevant NP scale, and taking $\M_i(\mu)\sim \Op(10^ {-4})$ we obtain a $\nnbar$~oscillation rate within the reach of current and future generation of $\nnbar$~oscillation experiments ($\tau_{\nnbar}\gtrsim \Op{(\text{few})} \times 10^8$ s) for $\Lambda\lesssim \Op(100)$ TeV. We further notice from Fig.~\ref{fig:RG} that the the matrix elements run rather moderately as a function of the NP scale keeping the order of magnitude for the matrix elements the same across the relevant mass scales. Therefore, to demonstrate the model-independent implications of an observable rate of $\nnbar$ oscillations for baryogenesis in the following section, we will take $\mathcal{O}_1$ as a working example with the corresponding matrix element $\M_1$ shown in Fig.~\ref{fig:RG}. As a numerical example, considering the operator $\mathcal{O}_1$, we take the Wilson coefficient to be $C_1 (\mu)\sim \Lambda_1^{-5}$, expressed in terms of a NP scale $\Lambda_1$, such that for the most stringent current experimental limit from the Super-Kamiokande experiment~\cite{Abe:2020ywm} $\tau_{\nnbar}^{\text{SK}}\ge 4.7 \times 10^8$ s, we obtain $\Lambda_1\ge 7.04 \times 10^5 \text{ GeV}$. For the NNBAR experiment with an expected improved future sensitivity by an order of magnitude $\tau_{\nnbar}^{\text{NNBAR}}\ge 3 \times 10^9$ s we find the limit $\Lambda_1\ge 1.02 \times 10^6 \text{ GeV}$.

\section{Implications for baryogenesis}\label{sec:boltzmann}
In the following, we want to study first the model-independent consequences of a possible observation of $\nnbar$ oscillations on baryogenesis models. In the second part, we then study in particular two baryogenesis scenarios of our simplified model set-up that can lead to successful baryogenesis and confront the resulting parameter space with current and future constraints.

In order to generate a baryon asymmetry the three Sakharov conditions (i) $B$ violation, (ii) CP violation and (iii) departure from thermal equilibrium have to be fulfilled. While the first condition is in principle fulfilled within the SM via the so-called sphaleron processes, the latter two are not sufficiently realised: The known CP violation in the SM is not enough and as the Higgs is too heavy in order to establish a first order phase transition, either new physics has to alter the potential or another mechanism is required such as out-of-equilibrium decays of heavy particles. 

We will study in the following the consequences of baryon number violating operators relevant for $\nnbar$ oscillations on baryogenesis. Hereby, it is important to keep in mind that above the electroweak phase transition, electroweak sphalerons, violating $B+L$, are highly active such that only 
\begin{align}
 \dot{\eta}^{\text{sph}}_{\Delta(B-L)} = \dot{\eta}^{\text{sph}}_{\Delta B} - \dot{\eta}^{\text{sph}}_{\Delta L} = 0 \, ,
\end{align}
is conserved. This means that when we study baryon number violating interactions, it is convenient to consider the change in $B-L$ 
\begin{align}
 \dot{\eta}_{\Delta(B-L)} =  \dot{\eta}^{\text{sph}}_{\Delta(B-L)} +  \dot{\eta}^{\text{new}}_{\Delta(B-L)} =  \dot{\eta}^{\text{new}}_{\Delta B}\,,
\end{align}
where $\eta_X\equiv n_X/n_\gamma^{\text{eq}}$ describes the number density of quantity $X$ normalised to the photon density $n_\gamma$, and where $\dot{\eta}^{\text{new}}_{\Delta B}$ tracks the yield of the new baryon-number-violating interactions. Below around $T \sim 10^{12}~\mathrm{GeV}$, when the electroweak sphalerons reach chemical equilibrium, one can easily use the known relation~\cite{Harvey:1990qw}
\begin{align}
{\eta}_{\Delta(B-L)} =  \frac{79}{28}{\eta}_{\Delta B}\,,
\end{align}
in order to solve for the final baryon asymmetry with the collision term $\dot{\eta}^{\text{new}}_{\Delta B}$ arising from new baryon number violating interactions
\begin{align}
\frac{79}{28}\dot{\eta}_{\Delta B} =  \dot{\eta}^{\text{new}}_{\Delta B}(\eta_{\Delta B})\,.
\end{align}
%
%%%%%%%%%%%%%%%%%%%%%%%%%%%%%%%%%%%%%%%%%%%%%%%%%%%%%%%%%%%%%%%%%%%%%%%%%%%%%%%%%%%%%%%%%%%%%%%%%%%%%%%%%%%%%%%%%%%%%%%%%%%%%%%%%%%%%%%%%%%%%%%%%%%%%%%%%%%%%%%%%%%%%%%%%%%%%%%%%%%%%%%%%%%%%%%%%%%%
\subsection{Model-independent implications of $\nnbar$~oscillation on baryogenesis models}{\label{subs:MIWO}}
In Sec.~\ref{sec:EFT}, we have introduced the $|\Delta B| = 2$ effective operators that are relevant for $\nnbar$ oscillations. When the NP mediators are much heavier than the external quarks in the effective operators, then at any temperature below their mass scale, the operators relevant for $\nnbar$ oscillations correspond to potential washout processes for any baryon asymmetry generated at a comparably higher scale. This mechanism can be either due to some CP-violating decay of any of the unstable mediators or due to a completely disconnected mechanism. Hence, an observed rate of $\nnbar$ oscillations at experiments directly indicate washout effects in a model-independent way. In order to estimate their corresponding washout effect on baryogenesis scenarios, we will use the generalised Boltzmann equation formalism as described in Ref.~\cite{Deppisch:2017ecm}. Hereby, we consistently account for the running of relevant Wilson coefficients of the $\nnbar$ operators between the scale of $\nnbar$ oscillations and the scale at which the washout effects are of relevance.

The generic form of a Boltzmann equation for a particle species $X$ is given by (see, e.g. Refs.~\cite{Griest:1990kh, Edsjo:1997bg, Giudice:2003jh})
\begin{align}
\label{eq:BoltzmannY}
	z H n_\gamma \frac{d\eta_X}{dz}
	= -\sum_{a,i,j,\cdots} [Xa \cdots\leftrightarrow ij\cdots],
\end{align}
where 
\begin{align}\label{eq:bol_sd}
	[Xa\cdots\leftrightarrow ij\cdots]
	&= \frac{n_X n_a \cdots}{n_X^{\rm eq}n_a^{\rm eq}\cdots}
	\gamma^\text{eq}(Xa \cdots \to ij\cdots) - \frac{n_i n_j \cdots}{n_i^{\rm eq}n_j^\text{eq}\cdots}
    \gamma^\text{eq}\left(ij \cdots \to Xa \cdots \right)\,,
\end{align}
where $\gamma^{\mathrm{eq}}$ is the scattering density in thermal equilibrium. The Hubble rate $H$ is given by
\begin{equation} \label{eq:hubble}
H(T) = \frac{1.66\sqrt{g_*}}{m_{\text{Pl}}}T^2\, ,
\end{equation}
with the effective number of degrees of freedom $g_* \sim 107$ in the SM and the photon density
\begin{equation}\label{eq:pd}
n_{\gamma}^{\text{eq}} = 2\frac{\zeta(3)}{\pi^2}T^3\,.
\end{equation}
We can use this formalism to describe the evolution of baryon number over time. With the baryon number density per comoving photon defined as 
\begin{eqnarray}
  \label{eq:def_B}
 \eta_{\Delta B} &=&\sum_{u,d} \frac{1}{3} [(\eta_{u_L}-\eta_{\bar{u}_L})+(\eta_{d_L}-\eta_{\bar{d}_L})+(\eta_{\bar{u^c}}-\eta_{u^c})+(\eta_{\bar{d^c}}-\eta_{d^c})]
\end{eqnarray}
where the sum over $(u,d)$ indicates the number of generations in thermal equilibrium. Note that we have used the left-handed fields along with the CP conjugates of the right handed fields, which are left handed antiparticles, $(\psi^c)_L=(\psi_R)^c$. In the 2-component Weyl spinor notation, the 4-component Dirac spinors are then given e.g. as $u=(u_L, \bar{u^c})^T$. We summarise the notation for the SM fields in Tab.~\ref{tab:SMfields}.
\begin{table}
	\begin{center}
	\begin{tabular}{l | c | c | c }
		\specialrule{.2em}{.3em}{.0em}
		\text{Field} & $SU(3)_C$ & $SU(2)_L$ & $U(1)_Y$ \\[1mm]
		\hline
		$Q\equiv (u,d)_L^T$ & $3$ & $2$ & $\frac{1}{6}$  \\[1mm]
		$(u^c)_L$ & $\overline{3}$ & $1$ & $-\frac{2}{3}$   \\[1mm]
		$(d^c)_L$ & $\overline{3}$ & $1$ & $\frac{1}{3}$  \\[1mm]
		$L\equiv (\nu,e)_L^T$ & $1$ & $2$ & $-\frac{1}{2}$  \\[1mm]
		$(e^c)_L$ & $1$ & $1$ & $1$   \\[1mm]
		$H\equiv(h^+,h^0)^T$ & $1$ & $2$ & $\frac{1}{2}$  \\[1mm]
		\specialrule{.2em}{.0em}{.3em}
	\end{tabular}
	\caption{Notation for the representations of the SM fermion and Higgs fields.}
	\label{tab:SMfields}
	\end{center}
\end{table}
In thermal equilibrium, we can relate the number densities with their corresponding chemical potentials
\begin{align}
 n_q-n_{\bar{q}}=\frac{g_q \mu_q T^2}{6}\,,
\end{align}
with $g_q=3$ being the number of degrees of freedom of the quarks. The chemical potential of a particle species $a$ is related to the chemical potential of its antiparticle via $\mu_a = -\mu_{\bar a}$.  When the SM Yukawa interactions and the sphalerons are in equilibrium, all relevant chemical potentials can be expressed in terms of a single chemical potential $\mu_{u_L}$~\cite{Harvey:1990qw},
\begin{eqnarray}
\label{eq:mu_rel_1}
	\mu_H = \frac{4}{21}  \mu = -\frac{12}{7}   \mu_{u_L}\, , \quad
	\mu_{\bar{u^c}} = \frac{5}{63}   \mu = -\frac{5}{7}   \mu_{u_L}\,
   \, , \quad
   	\mu_{\bar{d^c}} = - \frac{19}{63}  \mu= \frac{19}{7}   \mu_{u_L} \, ,
\end{eqnarray}
with $\mu=\sum_{e,\mu,\tau} \mu_{e_L}$.

Given the relations among the chemical potentials, we can express the baryon number density in terms of the chemical potential of a single species (which we choose to be $u_L$)
\begin{eqnarray}
  \label{eq:def_B_chem}
  \eta_{\Delta B} =\sum_{u,d}\frac{g_q T^2}{6 n_\gamma}(\mu_{u_L}+\mu_{d_L}+\mu_{\bar{u^c}} +\mu_{\bar{d^c}}) =\frac{\pi^2}{\zeta(3)}\frac{\mu_{u_L}}{T}\,.
\end{eqnarray}

In the following, we want to consider the $\nnbar$~oscillation operator $\Op_1$, which corresponds to $\Op_1$ in Eq.~\eqref{eq:operatorlist}.  We want to address the question, what the observation of $\nnbar$ oscillations would imply for the washout of a baryon asymmetry that might have been created at a higher scale. We can write down the Boltzmann equation for the baryon-to-photon density by differentiating Eq.~\eqref{eq:def_B}
      \begin{align}
      \label{eq:deltaBdef}
     z H n_\gamma   \frac{d \eta_{\Delta B}}{d\, z}=\frac{1}{3} z H n_\gamma \left[\frac{d\, (\eta_{\bar{d^c}}-\eta_{{d^c}}) }{d\, z}+ \frac{d\, (\eta_{\bar{u^c}}-\eta_{{u^c}}) }{d\, z}+ \frac{d\, (\eta_{u_L}-\eta_{{\bar{u}_L}}) }{d\, z}+ \frac{d\, (\eta_{d_L}-\eta_{{\bar{d}_L}}) }{d\, z}\right]\,.
      \end{align}
	In the presence of the operator $\Op_1$  the evolution of the abundance for $\bar{u^c}$, $\bar{d^c}$, and their antiparticles can be obtained by writing the corresponding Boltzmann equation for a given species abundance; e.g.
      \begin{align}
        z H n_\gamma \frac{d\, \eta_{\bar{d^c}} }{d\, z}
        &= - \left[ \bar{u^c}\bar{d^c}\bar{d^c}\leftrightarrow u^c d^c d^c \right]
           + \left(\text{other possible permutations}\right) \nonumber\\
        &= - \left(\frac{n_{ \bar{u^c}}n_{\bar{d^c}}^2}{n^\text{eq}_{\bar{u^c}}
           (n^\text{eq}_{\bar{d^c}})^2}
           - \frac{n_{ {u^c}}n_{{d^c}}^2}{n^\text{eq}_{{u^c}}
             (n^\text{eq}_{{d^c}})^2} \right)
             \gamma^\text{eq} (\bar{u^c}\bar{d^c}\bar{d^c}\rightarrow u^c d^c d^c )
             + \cdots \nonumber\\
        &= - \frac{66 \, \mu_{u_L}}{7\, T}
        \gamma^\text{eq} (\bar{u^c}\bar{d^c}\bar{d^c}\rightarrow u^c d^c d^c )  + \cdots \nonumber\\
      &= - \frac{66\, \zeta(3)}{7 \pi^2}
      \eta_{\Delta B}
      \gamma^\text{eq} (\bar{u^c}\bar{d^c}\bar{d^c}\rightarrow u^c d^c d^c )  + \cdots\, ,
      \end{align}
      where we have assumed three generations of fermions and a universal chemical potential among the three quark generations. The ellipsis denote other possible permutations of $3\leftrightarrow 3$ and
      $2\leftrightarrow 4$ processes. In order to arrive at the last line, we used the relation derived in Eq.~\eqref{eq:def_B_chem}. Similarly, one can write down the evolution for  $\bar{u^c}$, $u^c$ and $d^c$. On the other hand, in the absence of any $B-L$ violating interactions involving $u_L$ and $d_L$ we can make the simplifying approximation $\frac{d}{dz}(\eta_{u_L}-\eta_{{\bar{u}_L}})\simeq 0$ and $\frac{d}{dz}(\eta_{d_L}-\eta_{{\bar{d}_L}})\simeq 0$. The evolution of baryon number density per comoving photon $\eta_{\Delta B}$ is then given by
\begin{align}
	z H n_\gamma \frac{d\, \eta_{\Delta B} }{d\, z}
	&= - \frac{4}{3} \left[ \bar{u^c}\bar{d^c}\bar{d^c}\leftrightarrow u^c d^c d^c \right]
	   +  \cdots \nonumber\\
&= - \frac{88\, \zeta(3)}{7 \pi^2}
\eta_{\Delta B}
\gamma^\text{eq} (\bar{u^c}\bar{d^c}\bar{d^c}\rightarrow u^c d^c d^c )  + \cdots \, .
\end{align}
\begin{figure}[htb!]
	\begin{center}
		\includegraphics[width = 0.49 \textwidth]{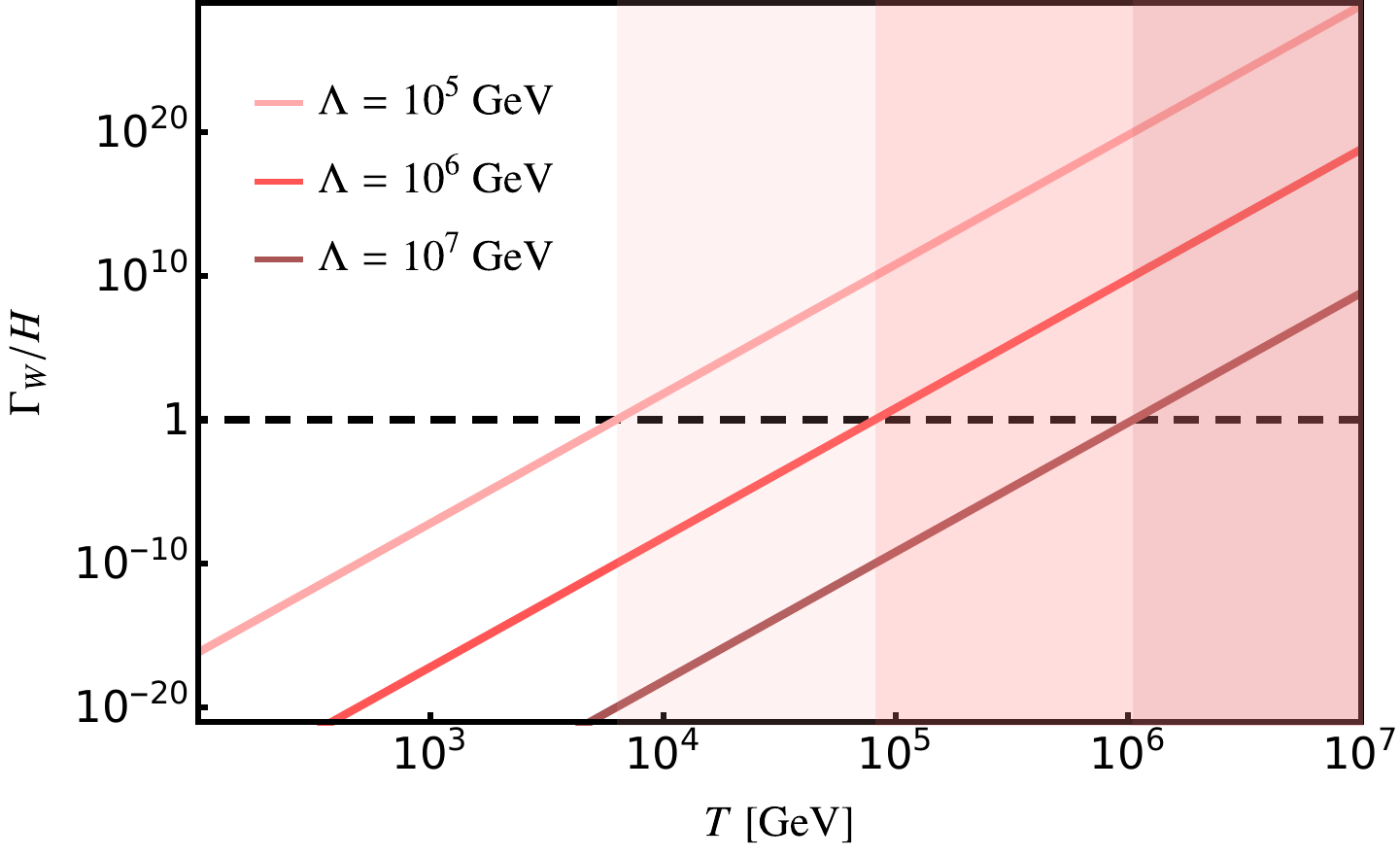}
		\includegraphics[width = 0.49 \textwidth]{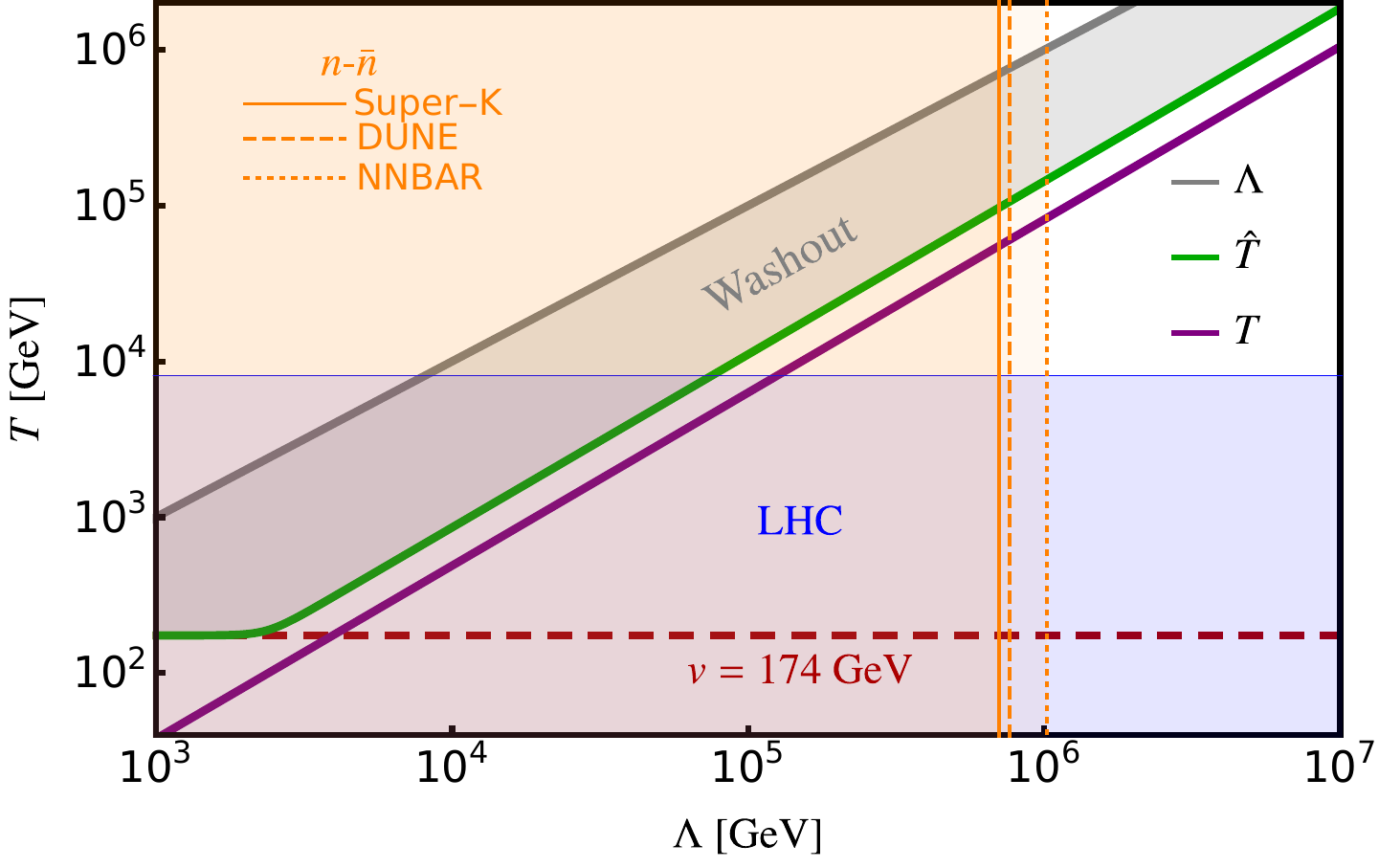}
		\caption{{\textit Left:} Ratio between the width of the interaction induced by operator $\mathcal{O}_1$ and the Hubble rate $H$, as a function of temperature $T$. When this fraction is greater than 1, as indicated by shaded areas, the interaction is assumed to provide a strong washout of baryon asymmetry. {\textit Right:} Temperature at which the fraction $\Gamma_W/H$ falls below 1 (purple), and a more accurate limit on the out-of-equilibrium temperature coming from Eq.~\eqref{eq:washout2} (green). In blue and orange shaded areas, the experimental reach of the LHC and different $\nnbar$ experiments are shown, respectively.
		}
		\label{fig:MIWO}
	\end{center}
\end{figure}

Following the prescription of~\cite{Deppisch:2017ecm,Deppisch:2015yqa,Deppisch:2020oyx} for the computation of thermal rate $\gamma^{\text{eq}}$, the total washout effect from the operator $\Op_1$ can be expressed as
\begin{align}{\label{eq:O3_be}}
	z H n_\gamma \frac{d\, \eta_{\Delta B} }{d\, z}
         & = -\frac{5}{3} \frac{ 2200 \zeta(3) \, T^{14} }{ 7 \pi^{11} \Lambda^{10}} \eta_{\Delta B}\equiv - c \frac{T^{14}}{\Lambda^{10}} \eta_{\Delta B}\, ,
\end{align}
where we have only included the $3\leftrightarrow 3$ scatterings since the $2\leftrightarrow 4$ scatterings are phase space suppressed. Therefore, the washout process corresponding to the $\nnbar$ operator $\Op_1$ with an interaction rate $\Gamma_W^{\Op_1}$ can be regarded to be roughly in equilibrium if 
\begin{align}
\label{eq:washout1}
  \frac{\Gamma_W^{\Op_1}}{H} &\equiv \frac{c}{n_\gamma H}\frac{T^{14}}{\Lambda^{10}}
	= c' \frac{\Lambda_\text{Pl}}{\Lambda}\left(\frac{T}{\Lambda}\right)^{9} \gtrsim 1,
\end{align}
where $c' = \pi^2 c/(\zeta(3) \,3.3 \sqrt{g_*}) \approx 0.3 \, c$ and the Planck scale $\Lambda_\text{Pl} = 1.2\times 10^{19}$~GeV. The Hubble parameter $H$ and the equilibrium photon density $n_\gamma$ are defined in Eqs.~\eqref{eq:hubble} and~\eqref{eq:pd}, respectively. A more accurate limit on the out-of-equilibrium temperature of the washout process for a successful baryogenesis scenario can be obtained by integrating Eq.~\eqref{eq:O3_be} to be given by
\begin{align}
\label{eq:washout2}
\hat{T}
	\simeq \left[
  9 T^9 \ln \left( \frac{d_{ \text{rec}}} {\eta^{\text{obs}}_{B}}  \right)
  + v^9 \right]^{\frac{1}{9}}\, ,
\end{align}
where $d_{ \text{rec}}\approx 1/27$ is the dilution factor due to entropy conservation when the Universe cools down from the temperature of baryon asymmetry generation $T=T^*$ to the recombination temperature $T=T_0$, such that $\eta_{\Delta B}(T_0)=g_s(T_0)/g_s(T_*)\, \eta_{\Delta B} (T_*)\simeq 1/27\, \eta_{\Delta B} (T_*)$, where $g_s$ is a function of temperature that maintains the relation $s= (2\pi^2/45) g_s T^3$, where $s$ is the entropy density. Furthermore, $v$ is the vacuum expectation value of the SM Higgs, and  $T$ is the out-of-equilibrium temperature in Eq.~\eqref{eq:washout1} is given by
\begin{align}
\label{eq:washout3}
T=\Lambda \left( \frac{1}{c'}\frac{\Lambda}{\Lambda_{\text{Pl}}}\right)^{\frac{1}{9}}\,.
\end{align}
In Fig.~\ref{fig:MIWO} left panel, we show the washout parameter $\Gamma_W/H$ as a function of temperature for the operator $\Op_1$ for different values of the EFT scale $\Lambda$ corresponding to the integrated out heavy new physics. As discussed in Sec.~\ref{sec:EFT}, the most stringent limits from $\nnbar$ oscillations constrain the NP scale to $\Lambda\ge 2.4\times 10^5 \text{ GeV}$. If in the future, $\nnbar$ oscillations would be observed and not involve any CP-violating interactions, they would imply a strong washout down to the scale indicated by the corresponding shaded area in Fig.~\ref{fig:MIWO} (left). The implication, on the other hand, becomes more visible in Fig.~\ref{fig:MIWO} (right), where we show the out-of-equilibrium temperature for the washout processes corresponding to $\Op_1$ as a function of the the EFT scale $\Lambda$ using Eq.~\eqref{eq:washout1} and Eq.~\eqref{eq:washout2}. An observation of $\nnbar$ oscillations around the scale of $\Lambda \approx 10^6 \text{ GeV}$, would imply a strong washout down to $\hat T \approx 1.4\times 10^5 \text{ GeV}$. Under the assumption of a pre-existing, generated asymmetry at a high scale, this would imply such a strong washout, that an asymmetry must be generated below this scale and above the current exclusion limits from the LHC. Hence, such a discovery would hint towards new physics possibly observable at future colliders, e.g. a $100~\mathrm{TeV}$ collider. 

Taking a completely agnostic approach towards the origin or flavour of the pre-existing asymmetry, as $\nnbar$ oscillations strictly involve first generation quarks only, the conclusions regarding the washout effects derived in this section are strictly applicable only in the case of a pre-existing baron asymmetry in the first generation quarks. In order to ensure the complete washout of a pre-existing asymmetry in all flavours, a complementary measurement of new physics arising from the operator in question involving second or third generation quarks (e.g. LHC searches, meson oscillations etc.) is needed besides an observation of $\nnbar$ oscillation, in the absence of flavour transitions. In order to study such a situation and its interplay with other experimental constraints, we will explore a simplified model (low-scale scenario) later in more detail. However, this is a conservative assumption as one would expect washout also in other flavours via spectator effects, as for instance discussed for leptogenesis in \cite{Dev:2017trv}.

\subsection{A comprehensive Boltzmann equation formalism for baryogenesis in models featuring $\nnbar$~oscillations}
While in the previous analysis, we have assumed that the new operator does not include any source of CP violation and only contributes to the washout of the baryon asymmetry, we want to refine our analysis by analysing a simplified set-up that allows for (a) including a source of CP violation in the NP operator and (b) different mass hierarchies within this operator. The latter is in particular important, as the the EFT approach presented in the previous subsection provides only an estimate of the washout within the validity of the EFT, i.e. when the masses of all new degrees of freedom are below its cutoff scale.

\begin{figure}[t]
\begin{center}
    \includegraphics[width = 0.45 \textwidth]{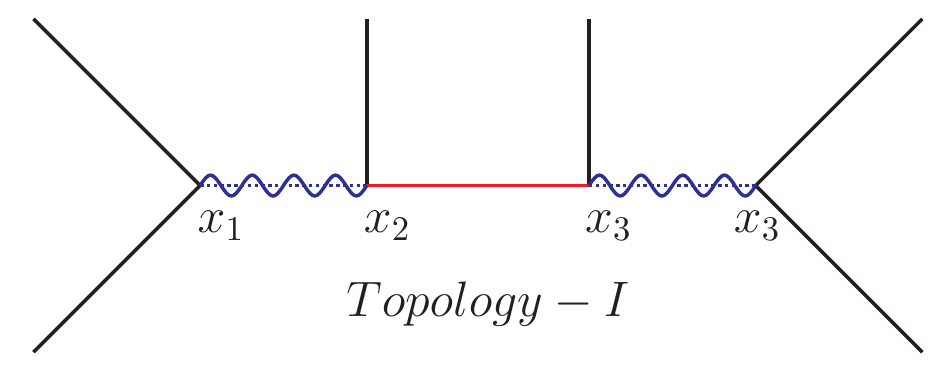}
     \includegraphics[width = 0.35 \textwidth]{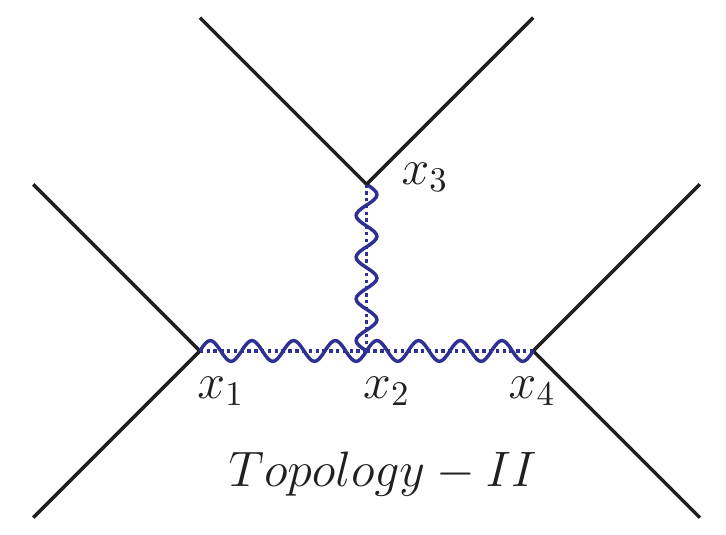}
\caption{The two basic tree-level topologies realising a dimension nine $\nnbar$ operator. Internal lines can be fermions (solid), or scalars / vectors (crossed wavy lines).
}
\label{fig:top}
\end{center}
\end{figure}

For these purposes, we are interested in the most generic topologies that such an operator could lead to. At tree level, the realisations for the short-range $\nnbar$ operators given in Eq.~\eqref{eq:operatorlist} can be classified in two possible topologies, which are shown in Fig.~\ref{fig:top}. In general, for topology I the internal mediators between vertices $x_1 -x_2$ and $x_3 -x_4$  can be either vector or scalar fields, while the particle between $x_2 -x_3$ must be a fermionic field. On the other hand, for topology II, all the internal mediators can be either scalars or vectors, with all possible combinations.
Topology I has been explored in the context of many specific model realisations in Refs.  \cite{Zwirner:1984is,Barbieri:1985ty,Mohapatra:1986bd,Lazarides:1986jt,Goity:1994dq,Babu:2001qr,Babu:2006wz,Allahverdi:2010im,Gu:2011ff,Gu:2011fp,Allahverdi:2013mza,Dev:2015uca,Dhuria:2015swa,Ghalsasi:2015mxa,Gu:2016ghu,Calibbi:2016ukt,Gu:2017cgp,Calibbi:2017rab,Allahverdi:2017edd} and recently, has been extensively discussed in the context of baryogenesis using a minimal simplified model set-up in~\cite{Grojean:2018fus}. The main focus of the remaining of this work will be on topology II, which has been proposed originally in Ref.~\cite{Mohapatra:1980qe} and has been realised in many UV complete and TeV scale models  \cite{Mohapatra:1982xz,Chang:1984qr,Babu:2006xc,Babu:2008rq,Baldes:2011mh,Babu:2012vc,Arnold:2012sd,Babu:2013yca,Patra:2014goa,Herrmann:2014fha}\footnote{See also Refs.~\cite{Bernal:2012gv,Asaka:2019ocw,Girmohanta:2020qfd} for some other exotic effective operator scenarios for  $\nnbar$~oscillation.}. While there are a few instances of studies for this scenario in the context of baryogenesis \cite{Babu:2012vc,Baldes:2011mh,Herrmann:2014fha}, a phenomenologically comprehensive and in depth exploration of the viable parameter space for baryogenesis still remained desirable. To this end, in this work, we explore both high- and low-scale (pre-electroweak) baryogenesis in the context of an observable $\nnbar$~oscillation lifetime while taking into account constraints from various complementary observables at the high-energy and high-intensity frontiers. Having already discussed the general EFT approach, in the following sections we consider a very general minimal set-up extending the SM with diquark scalar fields coupling to SM quark fields and a $B$ (and $B-L$) violating trilinear scalar coupling involving only the diquark scalar fields. This simplified set-up not only allows us to develop an in-depth prescription for the Boltzmann equation formalism but also makes the analysis very general and directly applicable to TeV scale and UV complete model realisations of topology II.  

%%%%%%%%%%%%%%%%%%%%%%%%%%%%%%%%%%%%%%%%%%%%%%%%%%%%%%%%%%%%%%%%%%%%%%%%%%%%%%%%%%%%%%%%%%%%%%%%%%%%%%%%%%%%%%%%%%%%%%%%%%%%%%%%%%%%%%%%%%%%%%%%%%%%%%%%%%%%%%%%%%%%%%%%%%%%%%%%%%%%%%%%%%%%%%%%%%%%%%%%%%%%%%%%%%%%%%%%%%%%
\subsubsection{The trilinear topology of $\nnbar$~oscillations: a simplified model with scalar diquarks}
\begin{figure}[t]
	\centering\includegraphics[width=0.7\textwidth]{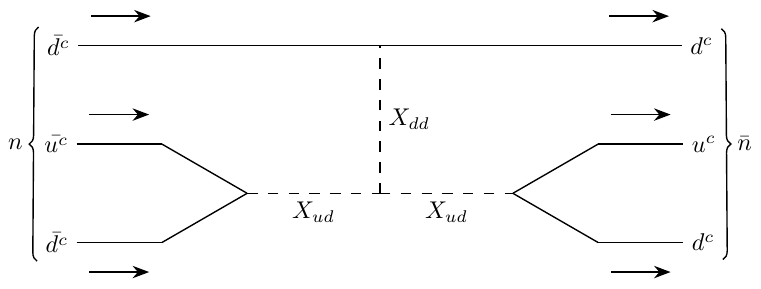}
	\caption{Diagram for $\nnbar$~oscillation with the diquark fields $X_{dd}$ and $X_{ud}$ descibed in the text.}
	\label{fig:diquarknnbar}
\end{figure}

{\label{subs:model}}
We consider the following Lagrangian including scalar diquarks given by
\begin{eqnarray}
  \label{lag:top2}
{\cal  L}_{II} &	=	&f^{dd}_{ij} X_{dd} \bar{d_{i}^c} \bar{d_{j}^c} + f^{uu}_{ij} X_{uu} \bar{ u_{i}^c} \bar{u_{j}^c} + \frac{f^{ud}_{ij}}{\sqrt 2} X_{ud}(\bar{u_{i}^c} \bar{d_{j}^c} + \bar{u_{j}^c}\bar{d_{i}^c}) \nonumber\\
&&+\lambda \xi
X_{dd} X_{ud} X_{ud}+{\tilde{\lambda}}\, \xi X_{uu} X_{dd} X_{dd} +{\rm h.c.} \, ,
\end{eqnarray}
where $\xi$ is a neutral complex scalar field, whose real part acquires a VEV and can be written as $\xi=v' + \frac{1}{\sqrt{2}} (S+ i \chi)$, with $v'\gg v$. Hereby, $\chi$ corresponds to the relevant goldstone mode associated with the symmetry broken by the VEV and is absorbed by the associated gauge boson. Before the $B-L$ symmetry is broken by $\xi$ acquiring a VEV, the new fields $X_{dd}$, $X_{uu}$, and $X_{ud}$ can be assigned consistently a baryon number $B=-2/3$ (and lepton number $L=0$), while $\xi$ carries $B-L=2$.  However, once $B-L$ is broken, the trilinear term will violate baryon number in units of $|\Delta B|=2$, as can be seen in the $\nnbar$~oscillation diagram in Fig.~\ref{fig:diquarknnbar}. The transformation properties of the scalar diquark fields under the SM gauge group and their associated baryon number charges are summarised in Tab.~\ref{tab:diquark_quantum_numbers}.
\begin{table}[h!]
	\begin{center}
	\begin{tabular}{l | c | c | c | c | c}
		\specialrule{.2em}{.3em}{.0em}
		\text{Field} & $SU(3)_C$ & $SU(2)_L$ & $U(1)_Y$ & $Q = T_{3L}+Y$ & $B$\\[1mm]
		\hline
		$X_{dd}$ & $\overline{6}$ or $3$ & $1$ & $+\frac{2}{3}$ & $+\frac{2}{3}$ & $-\frac{2}{3}$  \\[1mm]
		$X_{uu}$ & $\overline{6}$ or $3$ & $1$ & $-\frac{4}{3}$ & $-\frac{4}{3}$ & $-\frac{2}{3}$  \\[1mm]
		$X_{ud}$ & $\overline{6}$ or $3$ & $1$ & $-\frac{1}{3}$ & $-\frac{1}{3}$ & $-\frac{2}{3}$  \\[1mm]
		\specialrule{.2em}{.0em}{.3em}
	\end{tabular}
	\caption{Transformation properties and charges under the SM gauge group of the relevant scalar diquark fields.}
	\label{tab:diquark_quantum_numbers}
	\end{center}
\end{table}
Note that depending on the colour of the scalar diquarks, the associated Yukawa couplings are either symmetric or antisymmetric with respect to the flavour indices. If $X_{dd}$ ($X_{uu}$) transforms as a colour triplet under $SU(3)_c$ then  $f^{dd}_{ij}$ ($f^{uu}_{ij}$) must be flavour antisymmetric. On the other hand, if $X_{dd}$ ($X_{uu}$) transforms as a colour sextet then  $f^{dd}_{ij}$ ($f^{uu}_{ij}$) must be flavour symmetric.

Another point worth noting is that given the transformations of the scalar diquark field $X_{ud}$ under the SM gauge group it can also couple to a left-handed quark doublet: a colour sextet (triplet) $X_{ud}$ can couple to flavour (anti-)symmetric pair of $QQ$. However, in the presence of $X_{ud}$ couplings to both left- and right-handed quark pairs simultaneously, large chiral enhancements can be induced for flavour changing neutral current (FCNC)  $\Delta F=2$ operators (e.g. neutral meson mixing) and  $\Delta F=1$ operators (e.g $b\rightarrow s\gamma$), which in spite of being loop suppressed (with $X_{ud}$ in the loop) can lead to overwhelming rates disfavoured by the current stringent constraints from experiments for an $X_{ud}$ mass within the collider reach. Therefore, we assume that the couplings of $X_{ud}$ to a $QQ$ pair is negligible even if it gets generated due to radiative corrections.

Furthermore, if $X_{ud}$ and $X_{uu}$ transform as colour triplets, they can potentially also have leptoquark couplings. Leptoquark couplings when present simultaneously with diquark couplings lead to rapid proton decay for low $X_{ud}$ ($X_{uu}$) masses in the absence of any symmetry forbidding one of the couplings \cite{Babu:2012iv,Babu:2012vb,Hati:2018cqp}. In this work, we mainly focus on the case of colour sextet diquarks, implying flavour symmetric couplings $f^{dd}_{ij}$ ($f^{uu}_{ij}$) and absence of any leptoquark couplings (thereby avoiding any possibility of rapid proton decay). Interestingly, one can directly associate the Lagrangian in Eq.~\eqref{lag:top2} assuming colour-sextet scalar diquarks with a UV complete realisation as discussed in detail in App.~\ref{subs:UV}.

As we will later see in more detail, with respect to realising baryogenesis in this model, two regimes are of particular interest: \\
\textbf{High-scale scenario: }Here, we assume $m_{X_{uu}} \gg m_{X_{dd}} \gg m_{X_{ud}}$, in particular with $m_{X_{ud}} \sim\mathcal{O}(\text{TeV})$ and $m_{X_{dd}}\sim\mathcal{O}(10^{13-14}\text{GeV})$, while $m_{X_{uu}}\sim m_{\text{GUT}}$. Such a choice is naturally motivated by UV completions such as SO(10) to obtain gauge coupling unification, as discussed in App.~\ref{subs:UV}.\\
\textbf{Low-scale scenario: }We assume $m_{X_{uu}} \gg m_{X_{dd}} > m_{X_{ud}}$ is maintained but with both $m_{X_{ud}}$ and $m_{X_{dd}}$ not too far from $\sim\mathcal{O}(\text{TeV})$, while again $m_{X_{uu}}\sim m_{\text{GUT}}$. Even though such a scenario might require a more complex UV completion in order to address the colour vacuum stability as will be discussed in Sec.~\ref{sec:constraints}, this phenomenological scenario is particularly interesting as it allows for the possibility of two diquark states accessible at the LHC and future colliders. Also a number of ongoing low-energy experiments searching for baryon-number-violating processes e.g. dinucleon decay, are particularly relevant for such a scenario as will be discussed in Sec.~\ref{sec:constraints}.

For both scenarios above, the Lagrangian relevant for $\nnbar$~oscillations can be written in an effective form (at a scale below $m_{X_{dd}}$) as
\begin{eqnarray}
  \label{lag:top2eff}
{\cal  L}^{\text{eff}} 	=	\frac{f^{dd}_{ij} \lambda v'}{m_{X_{dd}}^2} X^{*}_{ud} X^{*}_{ud} \bar {d_{i}^c} \bar{d_{j}^c} + \frac{f^{ud}_{ij}}{\sqrt 2} X_{ud}(\bar{u_{i}^c} \bar{d_{j}^c} + \bar{u_{j}^c}\bar{d_{i}^c}),
\end{eqnarray}
where we have integrated out the heavy degrees of freedom. The relevant tree-level $\nnbar$~oscillation operator generated can be identified with $\Op^2_{RRR}$, defined in Eq.~\eqref{eqn:nnbar_obasis}. The effective Lagrangian for $\nnbar$~oscillation can be expressed, c.f. Eqs~\eqref{eq:operatorlist} and~\eqref{eq:oplisttilde}, as
\begin{eqnarray}
  \label{will:psl}
  \Lag^{\bar{n}\text{\--}n}_{\text{eff}}=\frac{ {(f^{ud}_{11})}^2  f^{dd}_{11} \lambda v'} { m_{X_{dd}}^2 m_{X_{ud}}^4} \mathcal{O}^2_{RRR} &=& \frac{ {(f^{ud}_{11})}^2  f^{dd}_{11} \lambda v'} { 4 m_{X_{dd}}^2 m_{X_{ud}}^4} \left(\Op_4 +\frac{3}{5} \tilde{\Op}_1\right)\nonumber\\
  &\approx & \frac{ 3\, {(f^{ud}_{11})}^2  f^{dd}_{11} \lambda v'} { 20\, m_{X_{dd}}^2 m_{X_{ud}}^4} \tilde{\Op}_1
\end{eqnarray}
where the last approximation follows from Eq.~\eqref{eq:vanO4} up to the uncertainties. For the numerical evaluation of the transition matrix element for operator $\tilde{\Op}_1(\mu)$, defined as $\tilde{\M}_1(\mu)\equiv \langle \bar{n}| \tilde{\Op}_1| n \rangle$ we take $\tilde{\M}_1(2\, \text{GeV})={\M}_1(2\, \text{GeV})$ given in Tab.~\ref{tab:gammar0} and to obtain the running effects in Eq.~\eqref{eq:operatorrunning}, we use the one- and two-loop $\overline{\text{MS}}$ anomalous dimensions $\gamma_i^{(0)}$ and $\gamma_i^{(1)}$, and the one-loop Landau gauge Regularisation-Independent-Momentum (RI-MOM) matching factor $r_i^{(0)}$, for $\tilde{\Op}_1$ given in Table~\ref{tab:gammar0}.

In a realistic model, such as the one discussed above, in addition to the $\nnbar$~oscillation operator $\tilde{\Op}_1$, a number of additional baryon number violating two-to-two scattering processes with NP particles (e.g. scalar diquark states) as the external states can give rise to washout processes. These processes become in particular important, when the new degrees of freedom feature a large hiearchy not captured within the validity of the previous EFT analysis. Therefore, taking the simplified model discussed above as a working example, we will explore the different relevant processes for CP violation and baryon asymmetry washout in full detail, accounting for different mass hierarchies of new physics.
 %%%%%%%%%%%%%%%%%%%%%%%%%%%%%%%%%%%%%%%%%%%%%%%%%%%%%%%%%%%%%%%%%%%%%%%%%%%%%%%%%%%%%%%%%%%%%%%%%%%%%%%%%%%%%%%%%%%%%%%%%%%%%%%%%%%%%%%%%%%%%%%%%%%%%%%%%%%%%%%%%%%%%%%%
\subsubsection{Derivation of the Boltzmann equation framework}
\begin{figure}[t]
	\begin{center}
	\includegraphics[height=3cm]{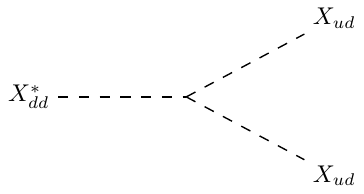}
	\includegraphics[height=3cm]{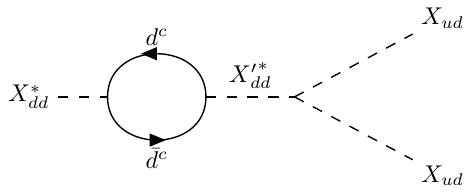}
	\caption{Tree-level and one-loop diagrams leading to a baryon asymmetry via $X_{dd}$ decays.}
	\label{fig:CPdecay}
	\end{center}
\end{figure}

{\label{subs:BEQ}}
We assume that the baryon asymmetry is generated by the $|\Delta B|=2$ decay $X_{dd} \rightarrow X^{*}_{ud} X^{*}_{ud}$ (once $\xi$ acquires a VEV) with the CP violation generated by the interference with loop diagrams involving the baryon number conserving decay mode $X_{dd} \rightarrow d^c d^c$, as shown in Fig.~\ref{fig:CPdecay}. It is straightforward to check that there is no vertex correction due to charge conservation. Only the self-energy diagram is present and it necessitates the introduction of an additional heavier $X_{dd}$ state, which we denote by $X'_{dd}$. To make our analysis as general as possible we consider the minimal relevant interactions of $X'_{dd}$ as 
\begin{eqnarray}
  \label{lag:Xddp}
{\cal  L}_{X'_{dd}} &	=	&f'^{dd}_{ij} X'_{dd}\bar{d_{i}^c} \bar{d_{j}^c} +\lambda' \xi
X'_{dd} X_{ud} X_{ud}+{\rm h.c.} \, .
\end{eqnarray}
The CP parameter $\epsilon$ is defined as \cite{Babu:2012vc}
\begin{eqnarray}
  \label{eq:eps}
\epsilon 	\equiv	\frac{\Gamma(X_{dd} \rightarrow X^{*}_{ud} X^{*}_{ud})-\Gamma(X_{dd}^* \rightarrow X_{ud} X_{ud})}{\Gamma^{\text{tot}}(X_{dd})}
	=	 \frac{1}{\pi} \text{Im}\left[ \frac{\text{Tr}\left[(f'^{dd})^\dagger f^{dd} \right] (\lambda' v')^* (\lambda v')}{\left|\lambda v'\right|^2}\right] \left( \frac{x}{1-x}\right) r\, ,
\end{eqnarray}
where  $\Gamma(X_{dd} \rightarrow X^{*}_{ud} X^{*}_{ud})$ and $\Gamma(X_{dd}^* \rightarrow X_{ud} X_{ud})$ are the decay widths for the decays $X_{dd} \rightarrow X^{*}_{ud} X^{*}_{ud}$ and $X_{dd}^* \rightarrow X_{ud} X_{ud}$, respectively. The total decay width  $\Gamma^{\text{tot}}(X_{dd})=\Gamma(X_{dd} \rightarrow X^{*}_{ud} X^{*}_{ud})+\Gamma(X_{dd} \rightarrow d^c d^c)$, $x$ is defined as $x\equiv m_{X_{dd}}^2/m_{X'_{dd}}^2$ and $r$ denotes the branching ratio for the decay mode $X_{dd} \rightarrow X^{*}_{ud} X^{*}_{ud}$.

Note that an alternative baryogenesis mechanism that can occur in this construction is the post-sphaleron baryogenesis via the decay of $S$ contained in $\xi$~\cite{Babu:2006xc,Babu:2008rq,Babu:2013yca} with $S$ being a real scalar field that can decay into six quarks ($S\rightarrow \bar{u^c} \bar{d^c} \bar{u^c} \bar{d^c} \bar{d^c} \bar{d^c}$) and antiquarks ($S\rightarrow u^c d^c u^c d^c d^c d^c$) leading to $|\Delta B|=2$. In such a scenario the relevant loop diagrams inducing the CP violation through interference employs $W^{\pm}$ loops\footnote{Note that this specific scenario is a particular realisation of an exception of the Nanopoulos-Weinberg theorem \cite{Nanopoulos:1979gx,Kolb:1979qa}, as has already been discussed in detail in Ref.~\cite{Babu:2006xc,Bhattacharya:2011sy,Babu:2013yca}.}, which necessitates flavour violation linking the CP violation directly to CKM mixing. In this scenario, it is therefore important to include loop contributions to $\nnbar$~oscillations in addition to tree level ones. A detailed discussion of the post-sphaleron baryogenesis mechanism is beyond the scope of the current work and will be presented elsewhere \cite{Fridell:2021aaa}.

Particle dynamics in the early Universe can be described using Boltzmann equations\cite{Griest:1990kh, Edsjo:1997bg, Giudice:2003jh}, which is well studied in particular for leptogenesis scenarios with right-handed Majorana neutrinos.
However, there are some crucial differences in this scenario, which must be taken into account to obtain a consistent description that we will discuss in the following. 

\begin{figure} [t]
	\begin{center}
	\hspace{-0.02\textwidth}$\underline{D_d^0}$\hspace{0.38\textwidth}$\underline{S_{s}^0}$\\[-4mm]
	\hspace{0.05\textwidth}\includegraphics[width=0.3\textwidth]{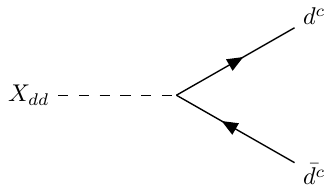}\hspace{0.05\textwidth}
	\raisebox{0mm}{\includegraphics[width=0.4\textwidth]{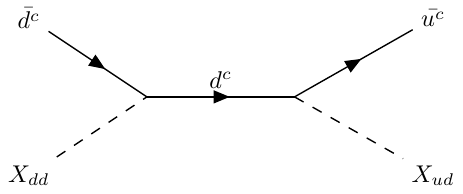}}\\[1mm]
	\hspace{-0.02\textwidth}$\underline{S_{ta}^0}$\hspace{0.38\textwidth}$\underline{S_{tb}^0}$\\
	\includegraphics[width=0.4\textwidth]{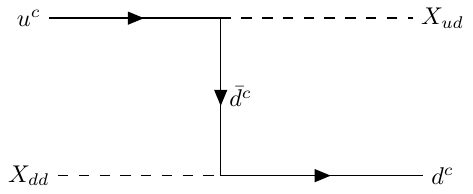}
	\includegraphics[width=0.4\textwidth]{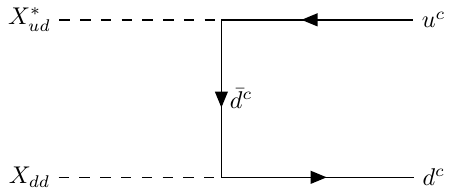}
	\caption{Baryon number conserving scattering and decay diagrams that involve the diquarks $X_{ud}$ and $X_{dd}$. The diagrams are here associated with the naming convention used in the Boltzmann equation, where the reverse processes are also employed.}
	\label{fig:gammaD1}
	\end{center}
\end{figure}

\begin{figure} [t]
\begin{center}
	\hspace{-0.02\textwidth}$\underline{D_d}$\hspace{0.38\textwidth}$\underline{S_s}$\\[-4mm]
	\hspace{0.05\textwidth}\includegraphics[width=0.3\textwidth]{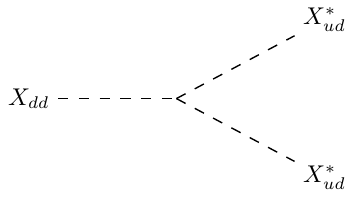}\hspace{0.05\textwidth}	\includegraphics[width=0.4\textwidth]{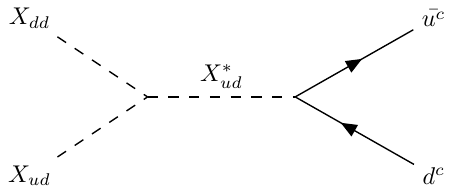}\\[-2mm]
	\hspace{-0.02\textwidth}$\underline{X_s}$\hspace{0.38\textwidth}$\underline{X_{t}}$\\
	\includegraphics[width=0.4\textwidth]{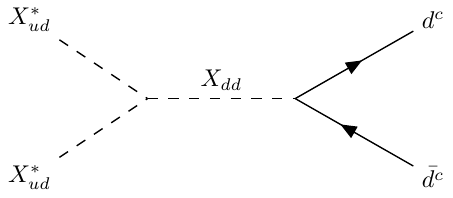}
	\includegraphics[width=0.4\textwidth]{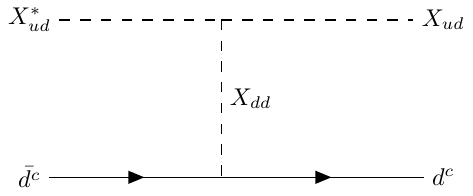}\\[5mm]
	\hspace{-0.02\textwidth}$\underline{S_{ta}}$\hspace{0.38\textwidth}$\underline{S_{tb}}$\\
	\includegraphics[width=0.4\textwidth]{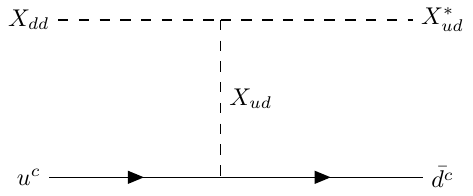}
	\includegraphics[width=0.4\textwidth]{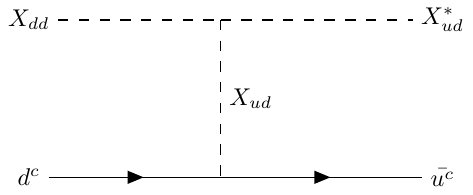}
	\caption{ Baryon number violating scattering and decay diagrams that involve the diquarks $X_{ud}$ and $X_{dd}$. The diagrams are here associated with the naming convention used in the Boltzmann equation, where the reverse processes are also employed.}
	\label{fig:gammaD2}
\end{center}
\end{figure}

For our simplified construction, cf. Eq.~\eqref{eq:bol_sd}, the relevant processes for baryogenesis can be classified as
\begin{eqnarray}{\label{eq:BEr_d1}}
|\Delta B|=0 &:& D_d^0= [X_{dd}\leftrightarrow d^c d^c] \nonumber\\
\phantom{|\Delta B|=0} &\phantom{:}& S_s^0=[X_{dd}\bar{d^c}\xleftrightarrow[s]{d^c} X_{ud} \bar{u^c}]
\quad S_{t_a}^0=[X_{dd} u^c \xleftrightarrow[t]{d^c} d^c X_{ud}  ] \quad S_{t_b}^0=[X_{dd} X_{ud}^* \xleftrightarrow[t]{d^c} d^c \bar{u^c}  ]    \nonumber\\
|\Delta B|=2 &:&   D_d= [X_{dd}\leftrightarrow X_{ud}^* X_{ud}^*] \nonumber\\
\phantom{|\Delta B|=0} &\phantom{:}&  X_s=[X_{ud}^* X_{ud}^*\xleftrightarrow[s]{X_{dd}} d^c d^c]
\quad X_t=[X_{ud}^* \bar{d^c}\xleftrightarrow[t]{X_{dd}} X_{ud} d^c ]
\nonumber\\
\phantom{|\Delta B|=2} &\phantom{:}&  S_s=[X_{dd} X_{ud} \xleftrightarrow[s]{X_{ud}^*} \bar{d^c} \bar{u^c}]  \quad S_{t_a}=[ X_{dd} u^c   \xleftrightarrow[t]{X_{ud}^*} X_{ud}^* \bar{d^c}] \quad S_{t_b}=[ X_{dd} d^c   \xleftrightarrow[t]{X_{ud}^*} X_{ud}^* \bar{u^c}  ],    \nonumber\\
\end{eqnarray}
where the Feynman diagrams for the processes are shown in Figs.~\ref{fig:gammaD1} and~\ref{fig:gammaD2}.
Now to compare this scenario with respect to the standard leptogenesis scenario let us note some key observations. 

\begin{figure}[t]
 	\begin{center}
 		\includegraphics[height=3.8cm]{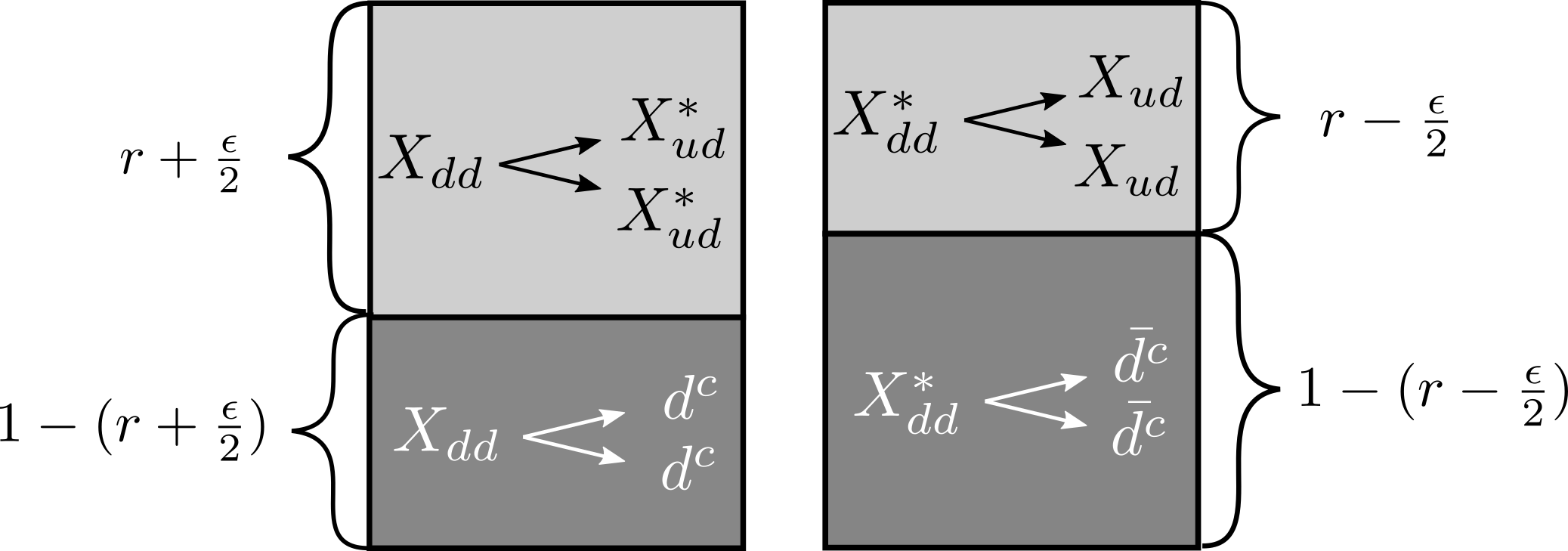}
 		\caption{Comparison of the different branching ratios of the $X_{dd}$ decay modes.}
 		\label{fig:XddGamma}
 	\end{center}
 \end{figure}

Since $X_{dd}$ can decay into either a pair of $X_{ud}$ or $d^c$ (after the $B-L$ breaking), one should include the change of the number density for both of these species in the Boltzmann equation for the baryon asymmetry. However, baryon number is only violated in the interaction $X_{dd} \leftrightarrow X^*_{ud} X^*_{ud}$ and $X^*_{dd} \leftrightarrow X_{ud} X_{ud}$. This can seen from the assignments in Tab.~\ref{tab:diquark_quantum_numbers}. Note, that in our analysis, we do not consider the possibility of a three-body decay involving this interaction term under the assumption that $m_\xi \gg m_{X_{dd}}$, leading to an early decoupling of such a decay mode. Due to this baryon number assignment, the decay modes $X_{dd}\leftrightarrow d^c d^c$ and $X_{dd}^{*}\leftrightarrow \bar{d^c} \bar{d^c}$ do not violate $B$, cf. Eq.~\eqref{lag:top2}. Therefore, in view of the thermal decoupling of $X_{dd} (X_{dd}^*)$, the decay of a pair of $X_{dd}$ and $X_{dd}^*$ can be considered as the analog of the heavy out-of-equilibrium decay of the Majorana fermion in the standard leptogenesis scenario. However, in contrast to the standard scenario, it is important to note that in presence of a CP asymmetry between $X_{dd}\leftrightarrow X_{ud}^* X_{ud}^*$ and $X_{dd}^{*}\leftrightarrow X_{ud} X_{ud}$ indirectly a CP asymmetry between $X_{dd}\leftrightarrow d^c d^c$ and $X_{dd}^{*}\leftrightarrow \bar{d^c} \bar{d^c}$ is generated. This is caused under the assumption of CPT invariance such that the total decay widths for $X_{dd}$ and $X_{dd}^*$ must be the same, as illustrated in Fig.~\ref{fig:XddGamma}.
 We note that being coloured particles, $X_{dd}$ and $X_{dd}^*$ can also be produced via two gluons $gg\rightarrow X_{dd} X_{dd}^*$, however such a mode is highly phase space suppressed at a temperature below  $m_{X_{dd}}$ (while the inverse process is doubly Boltzmann suppressed by $m_{X_{dd}}/T$), which is the interesting temperature for the out-of-equilibrium decay of $X_{dd}^{(*)}$ and its implications for baryogenesis. The process $gg\rightarrow X_{dd} X_{dd}^*$ generally dominates for $z< 1$, as is also demonstrated in~\cite{Herrmann:2014fha}. However, both the decay and the scattering washout channels dominate over any effect of $gg\rightarrow X_{dd} X_{dd}^*$ for $z> 1$, making this gauge scattering mode inconsequential for the final baryon asymmetry and mainly relevant for bringing the $X_{dd}$ density to equilibrium at the onset of baryogenesis. The reason for this is that the rate of $gg\rightarrow X_{dd} X_{dd}^*$ falls off much earlier than the washout via scatterings, due to the different dependence on the number density of $X_{dd}$ for $T<m_{X_{dd}}$. The gauge scattering features two $X_{dd}$ external legs, while the washout processes we consider have at most a single $X_{dd}$ external leg. Therefore, we do not include gauge scatterings in our analysis\footnote{This is in contrast to the type-II seesaw leptogenesis scenario, where the gauge interactions, e.g.\ $W^aW^a\rightarrow \Delta \Delta^*$, constitute one of the dominant interaction channels~\cite{Hambye:2005tk}. On the other hand, from Eq.~\eqref{eq:firstboltzmann}, it can be seen that a number of different scatterings affect the number density of $X_{dd}^{(*)}$ in addition to the gauge scatterings, which are more dominant than the gauge interactions $gg\rightarrow X_{dd} X_{dd}^*$ for $z\gtrsim 1$. Physically, the sub-dominance of gauge scatterings for the regime of baryon asymmetry generation can be understood by noticing the fact that the gauge scattering mode is subject to double phase space suppression of the number density of $X_{dd}(X_{dd}^*)$ for $T< m_{X_{dd}}$, as compared to scatterings like $S_{s(t)}^{(0)}$, which are only singly phase space suppressed with respect to $X_{dd}$.}.
 
 In the existing literature, both $X_{dd}\leftrightarrow X_{ud}^* X_{ud}^*$ and $X_{dd}\leftrightarrow d^c d^c$ modes have been taken into account to some extent, either while calculating the CP violation (see e.g. \cite{Babu:2012vc} and the references therein), or by including the Boltzmann equations for both $X_{ud}$ and $d^c$~\cite{Herrmann:2014fha}, however, a comprehensive formalism for the relevant Boltzmann equation for the baryon asymmetry is lacking. Here, we present a prescription for obtaining a consistent equation for the evolution of baryon asymmetry.
 
Based on the Boltzmann equations introduced in Sec.~\ref{subs:MIWO}, in particular Eqs.~\eqref{eq:BoltzmannY} and \eqref{eq:bol_sd} and the definition of the baryon asymmetry in Eq.~\eqref{eq:deltaBdef}, we can evaluate the final baryon asymmetry at the electroweak scale. After $X_{dd}^{(*)}$ goes out of equilibrium, both $X_{dd}\leftrightarrow X_{ud}^* X_{ud}^*$ and $X_{dd}\leftrightarrow d^c d^c$ modes (together with their CP conjugate modes) can generate a $B-L$ asymmetry and their interplay dictates the final baryon asymmetry. First, we discuss the necessary equations that describe the evolution of the out-of-equilibrium decay.

The number density of $X_{dd} (X_{dd}^*)$ per photon density at a given temperature can be obtained by solving the relevant Boltzmann equation given by  
\begin{eqnarray}\label{eq:bolxdd1}
2 zH(z)n_\gamma(z)\frac{d \eta_{X_{dd}^{(*)}}}{dz} &=& -D_d^0-\overline{D_d^0}-D_d-\overline{D_d}-S_s^0-\overline{S_s^0}-S_s-\overline{S_s}-S_t^0-\overline{S_t^0}-S_t-\overline{S_t}\, ,\nonumber\\
\end{eqnarray}
 where we note that the additional factor of 2 on the left-hand side accounts for the averaging factor since the right-hand side includes both $X_{dd}$ and $X_{dd}^*$ mediated processes. The different decay and scattering rates are defined in Eqs. \eqref{eq:bol_sd} and \eqref{eq:BEr_d1}, such that Eq.~\eqref{eq:bolxdd1} can further be expressed in terms of the rates in the following form
\begin{equation}
\begin{aligned}
\label{eq:firstboltzmann}
zH(z)n_\gamma(z)\frac{d \eta_{X_{dd}^{(*)}}}{dz} &=
-\left(\frac{\eta_{X_{dd}^{(*)}}}{\eta_{X_{dd}^{(*)}}^{\text{eq}}}-1\right)\left(\gamma_D^{X_{dd}^{(*)}}+\gamma_{S^0_s}+\gamma_{S_s}+\gamma_{S^0_{t_a}}+\gamma_{S_{t_a}}+\gamma_{S^0_{t_b}}+\gamma_{S_{t_b}}\right).
\end{aligned}
\end{equation}
The details of the prescription for obtaining the relevant density of scatterings $\gamma\equiv \gamma^{\text{eq}}$ is outlined in App.~\ref{app:boltzmann}. 

\begin{figure}[t]
 	\begin{center}
 		\includegraphics[height=3.2cm]{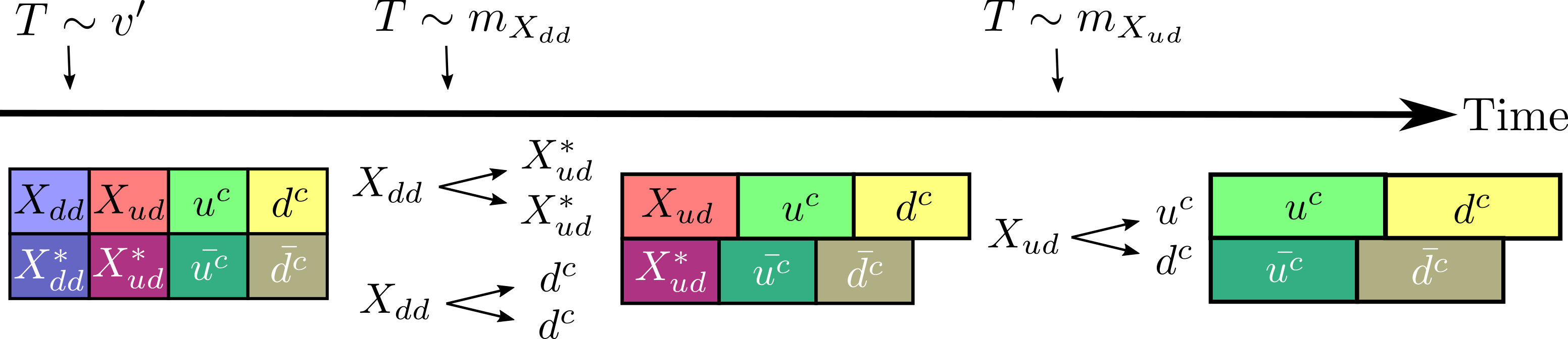}
 		\caption{A timeline (not to scale) that describes the chronology of events in the baryogenesis mechanism discussed in this section. Before the time of $B-L$ symmetry breaking ($T\gtrsim v'$), the particle species $X_{dd}$, $X_{ud}$, $u^c$ and $d^c$, as well as their corresponding antiparticles, are in thermal and chemical equilibrium. When $X_{dd}^{(*)}$ has decayed away ($T\lesssim m_{X_{dd}}$), an asymmetry has emerged in the relative number densities of $X_{ud}$, $u^c$, $d^c$ with their antiparticles. In the subsequent decay of $X_{ud}^{(*)}$ (at $T\lesssim m_{X_{ud}}$), this asymmetry is translated into a baryon asymmetry of the Universe stored in SM quarks.} 
 		\label{fig:TimeEvolve}
 	\end{center}
 \end{figure}

 We would like to note that even if we initially assume $\eta_{X_{dd}} = \eta_{X_{dd}^{*}}$, an asymmetry can be induced between $\eta_{X_{dd}^{*}}$ and $\eta_{X_{dd}}$ through the back reaction of $\eta_{X_{ud}^{*}}\eta_{X_{ud}^{*}}$ and $\eta_{X_{ud}}\eta_{X_{ud}}$ ($d^c d^c$ and $\bar{d^c} \bar{d^c}$) because of the asymmetry generated between $\eta_{X_{ud}}$ and $\eta_{X_{ud}^{*}}$ ($d^c$ and $\bar{d^c}$). Since during baryogenesis we are interested in the out-of-equilibrium decay of $X_{dd}^{(*)}$, such a secondary asymmetry is relevant for the Boltzmann equation for $\eta_{X_{dd}^{(*)}}$. However, such a contribution can be easily checked to be proportional to the generated baryon asymmetry $\eta_{\Delta B}$ which is very small compared to $\eta_{X_{dd}^{(*)}}$ for the temperature where the baryon asymmetry is dynamically generated (i.e.\ before the baryon asymmetry freezes out) c.f.\ Figs.~\ref{fig:1Dplots_highscale} and \ref{fig:1Dplots_midscale}. Therefore, we find it a good approximation to take $\eta_{X_{dd}} = \eta_{X_{dd}^{*}}$ through out the whole regime of baryogenesis. Now, to derive the Boltzmann equation for the baryon asymmetry one must consistently define the net baryon number density per photon density in terms of the number densities of the relevant species carrying SM baryon number that are in thermal equilibrium at the time of baryogenesis, see Eq.~\eqref{eq:def_B}. Again, we neglect $\frac{d}{dz}(\eta_{u_L}-\eta_{{\bar{u}_L}})$ and $\frac{d}{dz}(\eta_{d_L}-\eta_{{\bar{d}_L}})$ as $u_L$ and $d_L$ do not participate in any baryon-number-violating interactions generating or washing out the baryon asymmetry. Therefore, they can be decoupled from the Boltzmann equation\footnote{We note that even though $u_L$ and $d_L$ do not participate in any baryon number violating interactions, they can indirectly affect the baryon asymmetry through spectator processes, see e.g. \cite{Buchmuller:2001sr,Nardi:2005hs,Davidson:2008bu} for some relevant discussion. However, in the interest of simplification of analysis we ignore such secondary effects.}. 
Note that we include the number density of $X_{ud}$ in our definition of the final baryon asymmetry, which is valid in the regime when $X_{ud}$ is in thermal equilibrium. We have checked that this assumption holds for all scenarios presented in this work. Once $X_{ud}$ goes out of equilibrium one would need to write a new Boltzmann equation with the baryon number expressed in terms of only the SM quarks, taking into account the decay modes of $X_{ud}$. Since we are interested in the case where $m_{X_{dd}}> m_{X_{ud}}\gg m_{u,d}$, the baryon asymmetry generation freezes out by $z\equiv m_{X_{dd}}/T \sim \mathcal{O}(10)$. Therefore, for our analysis it will suffice to consider Eq.~\eqref{eq:def_B_X} with the assumption that the asymmetry generated in $X_{ud}-X_{ud}^*$ gets redistributed into SM quarks once $X_{ud}$ goes out of equilibrium (at a temperature below the baryon asymmetry generation freeze out), see Fig.~\ref{fig:TimeEvolve} for an illustration.
Hence, we arrive at the following definition,
\begin{equation}
  \label{eq:def_B_X}
  \eta_{\Delta B} \equiv \sum_{u,d} \frac{1}{3} [(\eta_{\bar{u^c}}-\eta_{u^c})+(\eta_{\bar{d^c}}-\eta_{d^c})+2 (\eta_{X_{ud}^*}-\eta_{X_{ud}})]\, ,
\end{equation}
where the sum over $(u,d)$ is over the number $N$ of generations in thermal equilibrium. 

 Then the Boltzmann equation for the baryon asymmetry per photon density can be obtained by differentiating Eq.~\eqref{eq:def_B_X}:
      \begin{eqnarray}\label{eq:bme1}
     z H n_\gamma   \frac{d \eta_{\Delta B}}{d\, z} =\frac{1}{3} z H n_\gamma \left[N \frac{d\, (\eta_{\bar{d^c}}-\eta_{{d^c}}) }{d\, z}+ N \frac{d\, (\eta_{\bar{u^c}}-\eta_{{u^c}}) }{d\, z}+ 2 \frac{d\, (\eta_{X_{ud}^*}-\eta_{X_{ud}}) }{d\, z}\right]\,.
      \end{eqnarray}
Now each term on the right-hand side of Eq.~\eqref{eq:bme1} corresponds to the standard Boltzmann equation describing the evolution of the number density of a particle species, (cf. Eqs. \eqref{eq:BoltzmannY} and \eqref{eq:bol_sd}) and is given by
 \begin{eqnarray}
zH(z)n_\gamma(z)\frac{d\eta_{X_{ud}^*}}{dz} &=& D_d-X_s-X_t-\overline{S_s}+S_t +\overline{S_s^0} +\overline{S_{t_a}^0} -S_{t_b}^0 \, , \label{eq:bebn1}\\
zH(z)n_\gamma(z)\frac{d\eta_{X_{ud}}}{dz} &=& \overline{D_d}-\overline{X_s}+X_t-S_s+\overline{S_t} +S_s^0 +S_{t_a}^0 -\overline{S_{t_b}^0}  \, , \label{eq:bebn2}\\
zH(z)n_\gamma(z)\frac{d\eta_{\bar{d^c}}}{dz} &=& \overline{D_d^0}+\overline{X_s}-X_t+S_s+S_{t_a}-\overline{S_{t_b}} -S_s^0 +\overline{S_{t}^0}  \, , \label{eq:bebn3}\\
zH(z)n_\gamma(z)\frac{d\eta_{d^c}}{dz} &=& D_d^0+X_s+X_t+\overline{S_s}+\overline{S_{t_a}}-S_{t_b} -\overline{S_s^0} +S_{t}^0  \, , \label{eq:bebn4} \\ 
zH(z)n_\gamma(z)\frac{d\eta_{\bar{u^c}}}{dz} &=& S_s-\overline{S_{t_a}}+S_{t_b} +S_s^0 -\overline{S_{t_a}^0} +S_{t_b}^0  \, , \label{eq:bebn5}\\
zH(z)n_\gamma(z)\frac{d\eta_{u^c}}{dz} &=& \overline{S_s}-S_{t_a}+\overline{S_{t_b}} +\overline{S_s^0} -S_{t_a}^0+\overline{S_{t_b}^0} \, , \label{eq:bebn6}  
\end{eqnarray}
where $S_t^0=S_{t_a}^0+S_{t_b}^0$ and $S_t=S_{t_a}+S_{t_b}$. Using Eqs.~(\ref{eq:bebn1}-\ref{eq:bebn6}) we can rewrite Eq.~\eqref{eq:bme1} as
      \begin{eqnarray}\label{eq:bme2}
     3 z H n_\gamma   \frac{d \eta_{\Delta B}}{d\, z} &=& 2( D_d-\overline{D_d})+ N(\overline{D_d^0} -D_d^0) -(N+2) \left[ (X_s -\overline{X_s})+2 X_t \right]\nonumber\\
	 &&+ 2 (N+1)\left[(S_s-\overline{S_s})+ (S_t-\overline{S_t})\right] 
	  -2\left[ (S_s^0-\overline{S}_s^0)+ (S_t^0-\overline{S}_t^0)\right]\, .
      \end{eqnarray}
Now let us discuss the relevant terms in some detail before employing the chemical potential relations relating all the species in chemical equilibrium. The contributions to the baryon asymmetry originating from decays can be parameterised in terms of the relevant decay rates as
\begin{eqnarray}
  \label{eq:decayrates}
 D_d &=& \gamma_D^{X_{dd}} \left[ {\left(r+\frac{\epsilon}{2}\right)} \frac{\eta_{X_{dd}}}{\eta_{X_{dd}}^{\text{eq}}}-{\left(r-\frac{\epsilon}{2}\right)} \frac{{\eta_{X_{ud}^*}}^2}{{\eta_{X_{ud}}^{\text{eq}}}^2}\right]\, ,\nonumber\\
  D_d^0 &=& \gamma_D^{X_{dd}} \left[ {\left(1-r-\frac{\epsilon}{2}\right)} \frac{\eta_{X_{dd}}}{\eta_{X_{dd}}^{\text{eq}}}-{\left(1-r+\frac{\epsilon}{2}\right)} \frac{{\eta_{d^c}}^2}{{\eta_{d^c}^{\text{eq}}}^2}\right]\, ,\nonumber\\
  \overline{D_d} &=&  \gamma_D^{X_{dd}^*} \left[ {\left(r-\frac{\epsilon}{2}\right)} \frac{\eta_{X_{dd}}}{\eta_{X_{dd}}^{\text{eq}}}- {\left(r+\frac{\epsilon}{2}\right)}  \frac{{\eta_{X_{ud}}}^2}{{\eta_{X_{ud}}^{\text{eq}}}^2}\right]\, ,\nonumber\\
 \overline{D_d^0} &=& \gamma_D^{X_{dd}^*} \left[ {\left(1-r+\frac{\epsilon}{2}\right)} \frac{\eta_{X_{dd}}}{\eta_{X_{dd}}^{\text{eq}}}-{\left(1-r-\frac{\epsilon}{2}\right)}  \frac{{\eta_{\bar{d^c}}}^2}{{\eta_{\bar{d^c}}^{\text{eq}}}^2}\right]\, ,
\end{eqnarray}
where $r$ corresponds to the average branching fraction for the $X_{dd}\rightarrow X_{ud}^* X_{ud}^*$ and $X_{dd}^* \rightarrow X_{ud} X_{ud}$ decay modes and $\epsilon$ corresponds to average $B$ asymmetry generated in the decay of a $X_{dd} (X_{dd}^*)$, as defined in Eq.~\eqref{eq:eps}. We summarise the branching ratios in Tab.~\ref{tbl:dm}.
\begin{table}[t]
	\begin{center}
{\begin{tabular}{l|c}
\hline
Process & Branching fraction\\
\hline
$X_{dd}\rightarrow X_{ud}^* X_{ud}^*$ & {$r+\frac{\epsilon}{2}$}\\
$X_{dd} \rightarrow d^c d^c$ &  {$1-(r+\frac{\epsilon}{2})$}\\
$X_{dd}^* \rightarrow X_{ud} X_{ud}$ &  {$r-\frac{\epsilon}{2}$}\\
$X_{dd}^* \rightarrow \bar{d^c} \bar{d^c}$ &  {$1-(r-\frac{\epsilon}{2})$}\\
\hline
\end{tabular}}
\caption{Two-body decay modes of $X_{dd}$ and the corresponding branching fractions. }
\label{tbl:dm}
\end{center}
\end{table}
We further assume CPT invariance implying the same total decay width for $X_{dd}$ and $X_{dd}^*$, and define $\gamma_D^{X_{dd}}=\gamma_D^{X_{dd}^*} \equiv \gamma_D^{\text{tot}}/2$, where $\gamma_D^{\text{tot}}$ is the total decay rate of the $X_{dd}$ and $X_{dd}^*$ pair.
Using the above parametrisation we obtain
\begin{eqnarray}
  \label{eq:drf}
 (D_d -\overline{D_d}) &=& \frac{\gamma_D^{\text{tot}}}{2} \left[\epsilon +\epsilon \frac{\eta_{X_{dd}}}{\eta_{X_{dd}}^{\text{eq}}} - r  
 \left(
 \frac {{\eta_{X^*_{ud}}}^2}{{\eta_{X_{ud}}^{\text{eq}}}^2}-
 \frac {{\eta_{X_{ud}}}^2}{{\eta_{X_{ud}}^{\text{eq}}}^2}
 \right)
 \right] \, ,\nonumber\\
 (D_d^0 - \overline{D_d^0}) &=& \frac{\gamma_D^{\text{tot}}}{2} \left[-\epsilon 
 -\epsilon \frac{\eta_{X_{dd}}}{\eta_{X_{dd}}^{\text{eq}}}
  - (1- r) 
 \left(
 \frac {{\eta_{\bar{d^c}}}^2}{{\eta_{\bar{d^c}}^{\text{eq}}}^2}-\frac {{\eta_{d^c}}^2}{{\eta_{\bar{d^c}}^{\text{eq}}}^2}
 \right)
 \right]\, .
\end{eqnarray}

Similar to the standard leptogenesis scenario, it is important to note that the $s$-channel scattering mediated by $X_{dd}$ must be calculated subtracting the CP-violating contribution due to the on-shell contributions in order to avoid double-counting, as this effect is already taken into account by successive (inverse) decays, $X_{ud}^* X_{ud}^* \leftrightarrow X_{dd} \leftrightarrow d^c d^c$ (real intermediate state subtraction). Therefore, we can write
\begin{eqnarray}\label{eq:bme6}
X_s &= \left(\frac{\eta_{X_{ud}^*}}{\eta_{X_{ud}}^{\text{eq}}}\right)^2 \left[\gamma_{X_{ud}^* X_{ud}^*\rightarrow d^c d^c}-\gamma_{X_{ud}^* X_{ud}^*\rightarrow d^c d^c}^{\text{on-shell}}\right]-\left(\frac{\eta_{d^c}}{\eta_{d^c}^{\text{eq}}}\right)^2 \left[\gamma_{d^c d^c\rightarrow X_{ud}^* X_{ud}^*}-\gamma_{d^c d^c\rightarrow X_{ud}^* X_{ud}^*}^{\text{on-shell}}\right]\, ,\nonumber\\
\overline{X_s} &= \left(\frac{\eta_{X_{ud}}}{\eta_{X_{ud}}^{\text{eq}}}\right)^2 \left[\gamma_{X_{ud} X_{ud}\rightarrow \bar{d^c} \bar{d^c}}-\gamma_{X_{ud} X_{ud}\rightarrow \bar{d^c} \bar{d^c}}^{\text{on-shell}}\right]-\left(\frac{\eta_{\bar{d^c}}}{\eta_{\bar{d^c}}^{\text{eq}}}\right)^2 \left[\gamma_{\bar{d^c} \bar{d^c}\rightarrow X_{ud} X_{ud}}-\gamma_{\bar{d^c} \bar{d^c}\rightarrow X_{ud} X_{ud}}^{\text{on-shell}}\right]\, ,
\end{eqnarray}
with
\begin{eqnarray}\label{eq:os}
\gamma_{X_{ud}^* X_{ud}^*\rightarrow d^c d^c}^{\text{on-shell}} &=& \gamma_{X_{ud}^* X_{ud}^*\rightarrow X_{dd}}\, Br(X_{dd} \rightarrow d^c d^c)
\simeq \left(r-r^2-\frac{\epsilon}{2}\right)\frac{\gamma_D^{\text{tot}}}{2} \, ,\nonumber\\
\gamma_{d^c d^c\rightarrow X_{ud}^* X_{ud}^*}^{\text{on-shell}} &=& \gamma_{d^c d^c\rightarrow X_{dd}} \, Br(X_{dd} \rightarrow X_{ud}^* X_{ud}^*)
\simeq \left(r-r^2+\frac{\epsilon}{2}\right)\frac{\gamma_D^{\text{tot}}}{2}\, ,\nonumber\\
\gamma_{X_{ud} X_{ud}\rightarrow \bar{d^c} \bar{d^c}}^{\text{on-shell}} &=& \gamma_{X_{ud} X_{ud}\rightarrow X_{dd}^*} \, Br(X_{dd}^* \rightarrow \bar{d^c} \bar{d^c})
\simeq \left(r-r^2+\frac{\epsilon}{2}\right)\frac{\gamma_D^{\text{tot}}}{2}\, ,\nonumber\\
\gamma_{\bar{d^c} \bar{d^c}\rightarrow X_{ud} X_{ud}}^{\text{on-shell}}&=& \gamma_{\bar{d^c} \bar{d^c}\rightarrow X_{dd}^*} \, Br(X_{dd}^* \rightarrow X_{ud} X_{ud})
\simeq \left(r-r^2-\frac{\epsilon}{2}\right)\frac{\gamma_D^{\text{tot}}}{2}\, ,
\end{eqnarray}
where the rightmost equalities are valid up to $\mathcal{O}(\epsilon)$.  Therefore, after simplifying Eq.~\eqref{eq:bme6} we obtain
\begin{eqnarray}\label{eq:bme7.0}
X_s-\overline{X_s} =\epsilon \gamma_D^{\text{tot}}+  \left(\gamma_{X_s}- ( r - r^2) \frac{\gamma_D^{\text{tot}}}{2}\right) \left[ \Bigg(\frac {{\eta_{X^*_{ud}}}^2}{{\eta_{X_{ud}}^{\text{eq}}}^2}-\frac {{\eta_{X_{ud}}}^2}{{\eta_{X_{ud}}^{\text{eq}}}^2}\Bigg)+ \Bigg(\frac {{\eta_{\bar{d^c}}}^2}{{\eta_{\bar{d^c}}^{\text{eq}}}^2}-\frac {{\eta_{d^c}}^2}{{\eta_{\bar{d^c}}^{\text{eq}}}^2}\Bigg) \right]\, .
\end{eqnarray}
Correspondingly, we obtain for the $t$-channel scattering mediated by $X_{dd}$ 
\begin{eqnarray}\label{eq:bme7.0.1}
X_t = \gamma_{X_t} \left( \frac {{\eta_{X^*_{ud}}}}{{\eta_{X_{ud}}^{\text{eq}}}} \frac {{\eta_{\bar{d^c}}}}{{\eta_{\bar{d^c}}^{\text{eq}}}}-\frac {{\eta_{X_{ud}}}}{{\eta_{X_{ud}}^{\text{eq}}}} \frac {{\eta_{d^c}}}{{\eta_{\bar{d^c}}^{\text{eq}}}}\right) \, ,
\end{eqnarray}
for the scattering processes with $X_{dd}$ in the initial or final
state (for $s$- and $t$-channel, respectively),
\begin{eqnarray}
	(S_s-\overline{S_s}) &=& \gamma_{S_s}
	\left[ 
	\left(\frac{\eta_{X_{dd}}}{\eta_{X_{dd}}^{\text{eq}}} \frac{\eta_{X_{ud}}}{\eta_{X_{ud}}^{\text{eq}}}- 
	\frac {{\eta_{\bar{d^c}}}}{{\eta_{\bar{d^c}}^{\text{eq}}}} 
	\frac {{\eta_{\bar{u^c}}}}{{\eta_{\bar{u^c}}^{\text{eq}}}}
	\right)
	-\left(\frac{\eta_{X_{dd}}}{\eta_{X_{dd}}^{\text{eq}}}
	\frac{\eta_{X^*_{ud}}}{\eta_{X_{ud}}^{\text{eq}}}- 
	\frac {{\eta_{d^c}}}{{\eta_{\bar{d^c}}^{\text{eq}}}} 
	\frac {{\eta_{u^c}}}{{\eta_{\bar{u^c}}^{\text{eq}}}}
	\right)
	\right] \, , \label{eq:bme8.0.1} \\
	(S_t-\overline{S_t}) 
	&=&(S_{t_a}-\overline{S}_{t_a})+(S_{t_b}-\overline{S}_{t_b})\nonumber\\ 
	&=&  \gamma_{S_{t_a}}
	\left[ \left( 
	\frac{\eta_{X_{dd}}}{\eta_{X_{dd}}^{\text{eq}}} 
	\frac{{\eta_{u^c}}}{{\eta_{\bar{u^c}}^{\text{eq}}}}-
	\frac{\eta_{X^*_{ud}}}{\eta_{X_{ud}}^{\text{eq}}} 
	\frac {{\eta_{\bar{d^c}}}}{{\eta_{\bar{d^c}}^{\text{eq}}}} 
	\right)
	-\left( 
	\frac{\eta_{X_{dd}}}{\eta_{X_{dd}}^{\text{eq}}} 
	\frac {{\eta_{\bar{u^c}}}}{{\eta_{\bar{u^c}}^{\text{eq}}}}-
	\frac{\eta_{X_{ud}}}{\eta_{X_{ud}}^{\text{eq}}}
	\frac {{\eta_{d^c}}}{{\eta_{\bar{d^c}}^{\text{eq}}}}  
	\right)\right]\nonumber \\
	&+&  \gamma_{S_{t_b}}
	\left[ \left(
	\frac{\eta_{X_{dd}}}{\eta_{X_{dd}}^{\text{eq}}} 
	\frac {{\eta_{d^c}}}{{\eta_{\bar{d^c}}^{\text{eq}}}}-
	\frac{\eta_{X^*_{ud}}}{\eta_{X_{ud}}^{\text{eq}}}
	\frac {{\eta_{\bar{u^c}}}}{{\eta_{\bar{u^c}}^{\text{eq}}}}
	 \right)
	-\left(
	 \frac{\eta_{X_{dd}}}{\eta_{X_{dd}}^{\text{eq}}}
	 \frac {{\eta_{\bar{d^c}}}}{{\eta_{\bar{d^c}}^{\text{eq}}}} -
	 \frac{\eta_{X_{ud}}}{\eta_{X_{ud}}^{\text{eq}}}
	 \frac{{\eta_{u^c}}}{{\eta_{\bar{u^c}}^{\text{eq}}}}
	\right)\right] \, , \label{eq:bme8.0.2}
\end{eqnarray}
and similarly for the quark mediated scatterings,
\begin{eqnarray}
	(S_s^0-\overline{S}_s^0) &=&   \gamma_{S_s^0}\left[
	\left( \frac {{\eta_{X_{dd}}}}{{\eta_{X_{dd}}^{\text{eq}}}} \frac {{\eta_{\bar{d^c}}}}{{\eta_{\bar{d^c}}^{\text{eq}}}} -\frac {{\eta_{X_{ud}}}}{{\eta_{X_{ud}}^{\text{eq}}}} \frac {{\eta_{\bar{u^c}}}}{{\eta_{\bar{u^c}}^{\text{eq}}}}    \right)-
	\left(  \frac {{\eta_{X_{dd}}}}{{\eta_{X_{dd}}^{\text{eq}}}} \frac {{\eta_{d^c}}}{{\eta_{\bar{d^c}}^{\text{eq}}}} - \frac {{\eta_{X_{ud}^*}}}{{\eta_{X_{ud}}^{\text{eq}}}} \frac {{\eta_{u^c}}}{{\eta_{\bar{u^c}}^{\text{eq}}}}    \right)
	\right] \, , \label{eq:bme10.0.1} \\
	(S_t^0-\overline{S}_t^0) &=&(S_{t_a}^0-\overline{S}_{t_a}^0)+(S_{t_b}^0-\overline{S}_{t_b}^0) \nonumber\\
	&=&  \gamma_{S_{t_a}^0}
	\left[\frac {{\eta_{X_{dd}}}}{{\eta_{X_{dd}}^{\text{eq}}}}
	\left( \frac {{\eta_{u^c}}}{{\eta_{\bar{u^c}}^{\text{eq}}}}- 
	\frac {{\eta_{\bar{u^c}}}}{{\eta_{\bar{u^c}}^{\text{eq}}}} \right)
	+\left( \frac {{\eta_{X_{ud}^*}}}{{\eta_{X_{ud}}^{\text{eq}}}} \frac {{\eta_{\bar{d^c}}}}{{\eta_{\bar{d^c}}^{\text{eq}}}}
	- \frac {{\eta_{X_{ud}}}}{{\eta_{X_{ud}}^{\text{eq}}}} \frac {{\eta_{d^c}}}{{\eta_{\bar{d^c}}^{\text{eq}}}}\right)
	\right] \nonumber \\
	&+& \gamma_{S_{t_b}^0}
	\left[ \frac {{\eta_{X_{dd}}}}{{\eta_{X_{dd}}^{\text{eq}}}}
	\left( \frac {{\eta_{X_{ud}^*}}}{{\eta_{X_{ud}}^{\text{eq}}}} -
	\frac {{\eta_{X_{ud}}}}{{\eta_{X_{ud}}^{\text{eq}}}} \right)
	+\left( \frac {{\eta_{\bar{d^c}}}}{{\eta_{\bar{d^c}}^{\text{eq}}}} \frac {{\eta_{u^c}}}{{\eta_{\bar{u^c}}^{\text{eq}}}}-\frac {{\eta_{d^c}}}{{\eta_{\bar{d^c}}^{\text{eq}}}} \frac {{\eta_{\bar{u^c}}}}{{\eta_{\bar{u^c}}^{\text{eq}}}}
	\right)
	\right]\, . \label{eq:bme10.0.2}
\end{eqnarray}
In order to solve the Boltzmann equations, we have to translate the different baryon densities on the right-hand side of Eq.~\eqref{eq:bme1} into $\eta_{\Delta B}$. 
As a first step, we express the ratio of the number density over the equilibrium density for different species in terms of the chemical potentials, recalling the approximation $\eta_a/\eta^{\eq}_a \approx e^{\mu_a/T} \approx 1 + \frac{\mu_a}{T}$ for a species $a$ with chemical potential $\mu_a$. Then we can express all relevant chemical potentials in terms of the chemical potential of a single species (following a similar prescription as in Sec.~\ref{subs:MIWO}) under the assumption that in addition to the SM Yukawa interactions and sphalerons, the $X_{ud}^{*}\leftrightarrow \bar{u^c} \bar{d^c}$ interactions are also in thermal equilibrium. We express all appearing chemical potentials for any particle species $a$ in terms of
\begin{equation}{\label{eq:def:x}}
\mu_a = x_a \mu_{X_{ud}^*}\,,
\end{equation}
which we summarise in App.~\ref{app:cp}. Finally, we arrive at the baryon asymmetry expressed in terms of one chemical potential $\mu_{X_{ud}^*}$
\begin{equation}
	 \frac {\eta_{\Delta B}}{\eta_{B}^{\text{eq}}}\equiv \frac{\eta_b-\bar \eta_b}{\eta_b^\eq} =  C_B\frac{\mu_{X_{ud}^*}}{T},
\end{equation}
where $C_B$ is given by (cf. App.~\ref{app:cp})
\begin{eqnarray}\label{eq:bme5}
C_B= \frac{\pi^2}{3\zeta(3)} \frac{6N+4 C_{X_{ud}^*}}{18N+4 C_{X_{ud}^*}} \, ,
 \end{eqnarray}
with $C_{X_{ud}^*} =6$ being the colour multiplicity of $X_{ud}^*$.
Hence, the combination of decay terms relevant for the evolution of baryon asymmetry in Eq.~\eqref{eq:bme2} is given by (cf. Eq.~\eqref{eq:drf})
\begin{eqnarray}\label{eq:bme3}
 2( D_d-\overline{D_d})+ N(\overline{D_d^0} -D_d^0)  = \frac{\gamma_D^{\text{tot}}}{2}\left[ (N+2) \epsilon\, \left( \frac{\eta_{X_{dd}}}{\eta_{X_{dd}}^{\text{eq}}}+1 \right)- \frac{4}{C_B}  \left\{ N (1- r ) x_{\bar{d^c}} +2 r  \right\} \frac {\eta_{\Delta B}}{\eta_B^{\text{eq}}}\right] \, ,
\end{eqnarray}
where $x_{\bar{d^c}}$ is given in Eq.~\eqref{app:eq:x}. Similarly, the various $s$- and $t$-channel scattering contributions in Eqs.~\eqref{eq:bme7.0}, \eqref{eq:bme7.0.1}, \eqref{eq:bme8.0.1}, \eqref{eq:bme8.0.2}, and \eqref{eq:bme10.0.1} can be expressed as
\begin{eqnarray}
X_s-\overline{X_s} &=& \epsilon \gamma_D^{\text{tot}}+ \frac{4}{C_B}(1+x_{\bar{d^c}}) \left[\gamma_{X_s}- ( r - r^2) \frac{\gamma_D^{\text{tot}}}{2}\right] \frac {\eta_{\Delta B}}{\eta_B^{\text{eq}}}\, , 
\label{eq:bme7}\\
X_t &=& \frac{2}{C_B} (1+x_{\bar{d^c}}) \gamma_{X_t} \frac {\eta_{\Delta B}}{\eta_B^{\text{eq}}}\, , \label{eq:bme7.1}\\
(S_s-\overline{S_s})+(S_t-\overline{S_t}) &=& -\frac{2}{C_B}\frac{\eta_{\Delta B}}{\eta_B^{\text{eq}}}\left[\tilde\gamma_{S_B}+ \tilde\gamma_{S_X} \frac{\eta_{X_{dd}}}{\eta_{X_{dd}}^{\text{eq}}}\right] \, ,
\label{eq:bme8}\\
	(S_s^0-\overline{S}_s^0)+(S_t^0-\overline{S}_t^0) &=&  \frac{2}{C_B}\frac{\eta_{\Delta B}}{\eta_B^{\text{eq}}}\left[\tilde\gamma_{S_B^0}+\tilde\gamma_{S_X^0} \frac{\eta_{X_{dd}}}{\eta_{X_{dd}}^{\text{eq}}} \right] \, ,
\label{eq:bme10}
\end{eqnarray}
where
\begin{eqnarray}
	\tilde\gamma_{S_B} &=& (x_{\bar{d^c}}+x_{\bar{u^c}}) \gamma_{S_s}+(x_{\bar{d^c}}+1) \gamma_{S_{t_a}}+(x_{\bar{u^c}}+1) \gamma_{S_{t_b}}\, ,\\
	\tilde\gamma_{S_X} &=& \gamma_{S_s}+x_{\bar{u^c}} \gamma_{S_{t_a}}+x_{\bar{d^c}} \gamma_{S_{t_b}}\, , \label{eq:bme9} \\
	\tilde\gamma_{S_B^0}&=& (-x_{\bar{u^c}}+1) \gamma_{S^0_s} +(x_{\bar{d^c}}+1) \gamma_{S^0_{t_a}}+(x_{\bar{d^c}}-x_{\bar{u^c}}) \gamma_{S^0_{t_b}}\, ,\\
	\tilde\gamma_{S_X^0}&=& x_{\bar{d^c}} \gamma_{S^0_s}-x_{\bar{u^c}} \gamma_{S^0_{t_a}}+ \gamma_{S^0_{t_b}}\, , \label{eq:bme11}
\end{eqnarray}
and $x_{\bar{u^c}}$ and $x_{\bar{d^c}}$ are given in Eq.~\eqref{app:eq:x}.

We then obtain the final form of the equation governing the evolution of the baryon asymmetry per photon density using Eqs. \eqref{eq:bme3}, \eqref{eq:bme7}, \eqref{eq:bme7.1}, \eqref{eq:bme8} and \eqref{eq:bme10} with Eq.~\eqref{eq:bme2} given by
\begin{equation}
\begin{aligned}
\label{eq:secondboltzmann}
 3 zH(z)n_\gamma(z)\frac{d \eta_{B}}{dz} &= \frac{\gamma_D^{\text{tot}}}{2}\left[ (N+2) \epsilon\, \left( \frac{\eta_{X_{dd}}}{\eta_{X_{dd}}^{\text{eq}}}-1 \right)- \frac{4}{C_B}  \left\{ N (1- r ) x_{\bar{d^c}} +2 r \right\} \frac {\eta_{\Delta B}}{\eta_B^{\text{eq}}}\right] 
 - \frac{4}{C_B}\frac {\eta_{\Delta B}}{\eta_{B}^{\text{eq}}}\\
 & \times \left[ (N+2)(1+x_{\bar{d^c}}) \frac{\gamma_X^{\text{sub}}}{2}
 +\left\{ (N+1) \tilde\gamma_{S_B}+\tilde\gamma_{S_B^0}\right\}
  +\frac{\eta_{X_{dd}}}{\eta_{X_{dd}}^\eq} \left\{ (N+1)\tilde\gamma_{S_X}+\tilde\gamma_{S_X^0}
  \right\}
 \right]
\end{aligned}
 \end{equation}
 where
\begin{eqnarray}
 \gamma_X^{\text{sub}}\equiv 2 \gamma_{X_s}+ 2 \gamma_{X_t}- \left( r - r^2\right) \gamma_D^{\text{tot}}\,
\end{eqnarray}
is the on-shell contribution subtracted scattering rate with the on-shell part given by $\gamma_{X_s}^{\text{on-shell}}=( r - r^2) \frac{\gamma_D^{\text{tot}}}{2} +\mathcal{O}(\epsilon)$, as can be verified from Eqs. \eqref{eq:os}. 
Note that after solving Eq.~\eqref{eq:secondboltzmann}, one has to include a factor $d_\gamma\approx 1/27$ for the dilution of the photon density in order to obtain the final baryon asymmetry at the recombination epoch $T_0$. We included this dilution factor for the later presented numerical analysis to obtain the final $\eta_{\Delta B}(T_0)$. 
For solving the Boltzmann equations for the high-scale and low-scale scenario defined in section~\ref{subs:model} we make the following assumptions which we summarise below.

\smallskip

\smallskip
{\bf High-scale scenario:}
\begin{itemize}
    \item Given that the SM top Yukawa coupling is in equilibrium for $T\gg 10^{13}$ GeV, and bottom as well as tau Yukawa couplings come in equilibrium just below $10^{13}$ GeV, we assume that all these Yukawas interactions as well as gauge interactions are in chemical equilibrium during the whole range of temperature relevant for high scale baryogenesis. This implies that the chemical potential relations due to transfer of the baryon asymmetry from $SU(2)_L$ singlet quarks to $SU(2)_L$ doublet quarks in equilibrium (c.f. Appendix \ref{app:cp}) by imposing the relevant chemical potential constraints due to Yukawa interactions. For a first exploration, we restrict ourselves to the single flavour approximation by assuming that the (inverse) decay of $X_{dd}^{(*)}$ and $X_{ud}^{(*)}$ dominantly occurs along the third generation quarks in flavour space. The respective chemical potential relations then correspond to the case $N=1$ in Appendix~\ref{app:cp}. In the future, it would be interesting to perform a detailed analysis including potential flavour correlations.

    \item Under the assumption that a single flavour of quarks $b$ and $t$, as well as the gauge interactions, are in equilibrium, the chemical potential relation $2\mu_{Q_3}-\mu_{\bar{u^c}_3}-\mu_{\bar{d^c}_3}=0$ is enforced, therefore making the chemical potentials of these single flavour quarks no longer affected by the QCD sphaleron constraint $\sum_i^3 2\mu_{Q_i}-\mu_{\bar{u^c}_i}-\mu_{\bar{d^c}_i}=0$. However, the partial equilibriation of Yukawa interactions can lead to a change of the final yield by little (e.g.\ this effect is $\sim 10\%$~\cite{Garbrecht:2014kda} for leptogenesis), which we do not take into account as it has little bearing in our final parameter space conclusions. For simplicity, we assume that the electroweak sphalerons are also in equilibrium together with the QCD sphalerons from the beginning of baryogenesis.
\end{itemize}

\smallskip

{\bf Low-scale scenario:}
\begin{itemize}
    \item We assume that the Yukawa couplings for all three generation of quarks as well as the gauge interactions are in equilibrium during the baryon asymmetry generation. The species $X_{ud}^*$, and the process $X_{ud}^*\leftrightarrow \bar{u^c}+\bar{d^c}$ are also assumed to be in equilibrium for all three generations. We further  assume that both QCD sphalerons and electroweak sphalerons are also in equilibrium. In this case the chemical potential relation $2\mu_{Q_i}-\mu_{\bar{u_i}^c}-\mu_{\bar{d_i}^c} = 0$ for $i={1,2,3}$ is enforced, which takes into account the effect of transfer of asymmetry from singlet to doublet quarks. The final chemical potential relations correspond to the case $N=3$ in the Appendix~\ref{app:cp}.
        \item To simplify the analysis we consider the case of flavour universal and diagonal couplings of $X_{dd}$ and $X_{ud}$ to three quark generations, implying that $X_{ud}^{(*)}$ decays about equally to all flavours and produces an equal asymmetry in all flavours. In the general case, the Boltzmann equation in this scenario is a matrix equation in flavour space, with the baryon-to-photon density, CP-asymmetry, decay rates and the washout rates all generalised to matrices in the flavour space. Starting from a physical basis where the off-diagonal elements are suppressed one can trace the matrices to obtain the sum over flavour space. In the absence of flavour correlations, one can solve for the final asymmetry by solving each flavour separately and adding the solutions. However, given our assumption of equal asymmetry generation and washout in all flavours, it suffices to simply multiply the final asymmetry by a factor 3, with the assumption that any induced flavour off-diagonal contributions are negligible and $\text{Tr}\left([\gamma_D] [\eta_L]\right)=\text{Tr}[\gamma_D] \text{Tr}[\eta_L]$ is a good approximation, where the square brackets done matrices in flavour space. For more details, we refer to e.g.~\cite{Abada:2006fw,Domcke:2020quw} discussing potential flavour effects and their implications, which is beyond the scope of this work.  
         \item To present the numerical results for the low-scale scenario in section~\ref{sec:results}, we allow for the possibility that $X_{dd}$ and $X_{ud}$ Yukawa couplings $f_{ud}$ and $f_{dd}$ are hierarchical among themselves (e.g. we consider the two benchmark cases $f_{ud}=f_{dd}$ and $f_{ud}= 10 f_{dd}$), we assume that $f_{ud}$ and $f_{dd}$ are universal across all three generations.
\end{itemize}

Before discussing in detail the parameter space that leads to the observed baryon asymmetry, we will discuss the relevant phenomenological constraints.

\section{Phenomenological constraints}\label{sec:constraints}
Diquarks are subject to different phenomenological constraints, both from experimental observables as well as theoretical conditions. The relevant constrains are particularly strong for diquark states that are not much heavier than the electroweak symmetry breaking scale. For instance, the LHC probes already $\mathcal{O}$(TeV) mass scales and provides high precision measurements of the Flavour Changing Neutral Current (FCNC) processes in mesonic observables. Moreover, dinucleon decays, which provide a complimentary probe to $\nnbar$ oscillations, are also of particular interest with the expected improvements on future experimental sensitivities. Furthermore, considerations of a colour preserving vacuum provides useful constraints on the $B-L$ breaking scale and mass hierarchy between diquark states, when more than one of the diquarks are light. In this section, we provide an overview of all relevant constraints for our study of the baryogenesis parameter space and will comment on additional constraints, which can be relevant in scenarios beyond the studied ones.

%%%%%%%%%%%%%%%%%%%%%%%%%%%%%%%%%%%%%%%%%%%%%%%%%%%%%%%%%%%%%%%%%%%%%%%%%%%%%%%%%%%%%%%%%%%%%%%%%%%%%%%%%%%%%%%%%%%%%%%%%%%%%%%%%%%%%%%%%%%%%%%%%%%%%%%%%%%%%%%%%%%
\subsection{Direct LHC searches}{\label{subs:lhc}}
Scalar diquarks have been studied extensively in the context of collider searches in the literature, see e.g. \cite{Mohapatra:2007af,Atag:1998xq,Cakir:2005iw,Chen:2008hh,Han:2009ya,Gogoladze:2010xd,Berger:2010fy,Han:2010rf,Baldes:2011mh,Richardson:2011df,Karabacak:2012rn,Kohda:2012sr,Chivukula:2015zma,Zhan:2013sza,Liu:2013hpa,Pascual-Dias:2020hxo}. At the LHC or future colliders they can be produced through the annihilation of a pair of quarks via an $s$-channel resonance decaying into two quarks producing a dijet final state. In principle, it is also possible to produce a pair of scalar diquarks e.g.\ through gluon-gluon fusion. For diquark couplings of $\mathcal{O}(0.1)$ and higher, the resonant production cross section dominates over the pair production~\cite{Richardson:2011df}. For smaller couplings, the pair production cross section can potentially dominate over the resonant production (the resonant production contains one power of the diquark coupling, while the gluon fusion pair production only includes gauge couplings)~\cite{Giudice:2011ak}. As limits from the current collider searches for the pair production mode is only available till roughly TeV scale~\cite{ATLAS:2017jnp}, while the resonant production search limits reach $\mathcal{O}(10)$ TeV~\cite{Sirunyan:2018xlo}, we consider only the resonant production, providing a very good estimate of the current LHC reach.

Since we assume in our simplified framework that $X_{uu}$ is very heavy, it will be beyond the collider reach; however, $X_{ud}$ being around few TeV of mass is actively probed by collider searches. The accessibility of $X_{dd}$ at colliders depends on the respective baryogenesis scenario (with a mass around the GUT scale in the high-scale scenario or around a few TeV for the low-scale scenario).

Depending on the generation of the quarks, we can distinguish between a resonant dijet or a top+jet signature. The partonic differential cross section for the latter ($u_i d_j \rightarrow X_{ud} \rightarrow t d_k$) is given by
\begin{eqnarray}
 \frac{ d \hat{\sigma}( u_i d_j \rightarrow X_{ud} \rightarrow t d_k)}
 {d \cos \theta^*}
 = \frac{ |f_{ij}^{ud}|^2 |f_{3k}^{ud}|^2}{8 \pi \hat{s}}
   \frac{(\hat{s}- m_t^2)^2}
    {(\hat{s}-m_{X_{ud}}^2)^2
  + m_{X_{ud}}^2 \Gamma_{X_{ud}}^2 },
\end{eqnarray}
where we have neglected all quark masses except the top quark mass. Hereby, $\theta^*$ denotes the scattering angle and $\hat{s}$ the center-of-mass energy of the partons. As will be discussed later, spin and colour multiplicity factors depending on the spin and colour representation of the initial and final states need to be added correspondingly. The total decay width $\Gamma_{X_{ud}}$ of the scalar diquark $X_{ud}$ is given by the sum of its partial decay widths,
\begin{eqnarray}
 \Gamma (X_{ud} \to u_i d_j)|_{u_i\neq t}
 &=& \frac{C_{X_{ud}}}{8 \pi}   |f_{ij}^{ud}|^2 \; m_{{X_{ud}}},   \nonumber \\
 \Gamma (X_{ud} \to t d_j)
 &=& \frac{C_{X_{ud}}}{8 \pi}  |f_{3j}^{ud}|^2 \; m_{{X_{ud}}}
   \left(  1- \frac{m_t^2}{m_{{X_{ud}}}^2} \right)^2\, ,
\end{eqnarray}
where $C_{X_{ud}}$ is the colour multiplicity of $X_{ud}$.

Similarly, one can obtain the relevant partonic cross section for the resonant dijet signature ($d_i d_j \rightarrow X_{dd} \rightarrow d_k d_l$),
\begin{eqnarray}
 \frac{ d \hat{\sigma}(d_i d_j \rightarrow X_{dd} \rightarrow d_k d_l)}
 {d \cos \theta^*}
 = \frac{|f_{ij}^{dd}|^2 |f_{kl}^{dd}|^2}{16 \pi }
   \frac{\hat{s}}
    {(\hat{s}-m_{X_{dd}}^2)^2
  + m_{X_{dd}}^2 \Gamma_{X_{dd}}^2 },
\end{eqnarray}
with the relevant partial decay width,
\begin{eqnarray}
 \Gamma (X_{dd} \to d_i d_j)
 &=& \frac{C_{X_{ud}}}{16 \pi}   |f_{ij}^{dd}|^2 \; m_{{X_{dd}}}.
\end{eqnarray}
Given the partonic scattering cross section, the experimentally measured hadronic cross section (e.g. at the LHC) can be obtained by employing the relevant parton distribution functions (PDFs) $f(x)$ and summing over all partons~\cite{Harris:2011bh}
\begin{equation}
  \sigma = \sum_{ij}{\int{dx_1dx_2f_i(x_1,\mu_F^2)f_j(x_2,\mu_F^2)\hat{\sigma}_{ij}\left(\alpha_s(\mu_R^2),\frac{Q^2}{\mu_F^2},\frac{Q^2}{\mu_R^2}\right)}} \, ,
\end{equation}
where $Q$ is the characteristic scale of the interaction, e.g. the invariant mass in a two-to-two partonic scattering. Furthermore, $\mu_F$ corresponds to the factorisation scale, which factorises the non-perturbative contributions from the short-distance hard scattering and $\mu_R$ is the scheme-dependent renormalisation scale. Note that the scales $\mu_{F,R}$ are parameters determined for a fixed-order calculation, while in general the total cross section should be independent of these scales at any given order of the perturbative expansion. The Parton Distribution Functions (PDFs) can be taken into account by introducing the parton luminosity factor defined as~\cite{Harris:2011bh}
\begin{equation}
  \frac{dL_{ij}}{d\tau} = \int_0^1\int_0^1{dx_1dx_2f_i(x_1)f_j(x_2)\delta(x_1x_2-\tau)},
\end{equation}
where
\begin{equation}
  \tau = x_1x_2 = \frac{\hat{s}}{s}.
\end{equation}
In the above relations the initial partons are carrying fractions $x_1$ and $x_2$ of the hadron momentum and the invariant mass of the two-parton system is defined as $\hat{s}\equiv x_1 x_2 s$, with $\sqrt{s}$ being the energy of the colliding hadrons in center-of-mass frame. Furthermore, constraints are also imposed on the rapidities $\overline{y}$ of the final state partons observed at the collider experiments as jets. It is particularly convenient to express the parton luminosity in terms of the variables $\tau$ and $\bar{y}$ as
\begin{equation}
  \frac{dL_{ij}(\bar{y}_{min},\bar{y}_{max})}{d\tau} = \int_{\bar{y}_{min}}^{\bar{y}_{max}}{f_i\left(\sqrt{\tau}e^{\bar{y}}\right)f_j\left(\sqrt{\tau}e^{-\bar{y}}\right)d\bar{y}} \, ,
\end{equation}
where we use the identity $dx_1dx_2 = \frac{\partial(\tau,\bar{y})}{\partial{x_1,x_2}}d\tau d\bar{y} = d\tau d\bar{y}$ and the rapidity can be expressed as a function of the momentum fractions as $\bar{y}=1/2 \ln (x_1/x_2)$ in the center-of-mass frame. Consequently, the total cross section can be expressed in terms of the parton luminosity factor and the partonic cross section as
\begin{equation}
  \sigma_{had} = \sum_{i,j}{\int{\frac{d\tau}{\tau}\left[\frac{dL_{ij}}{d\hat{s}}\right]\left[\hat{s}\,\hat{\sigma}_{ij}\right]}}.
\end{equation}
\begin{figure}[t!]
	\centering
	\mbox{\hspace{-4mm}
\includegraphics[height=5.0cm]{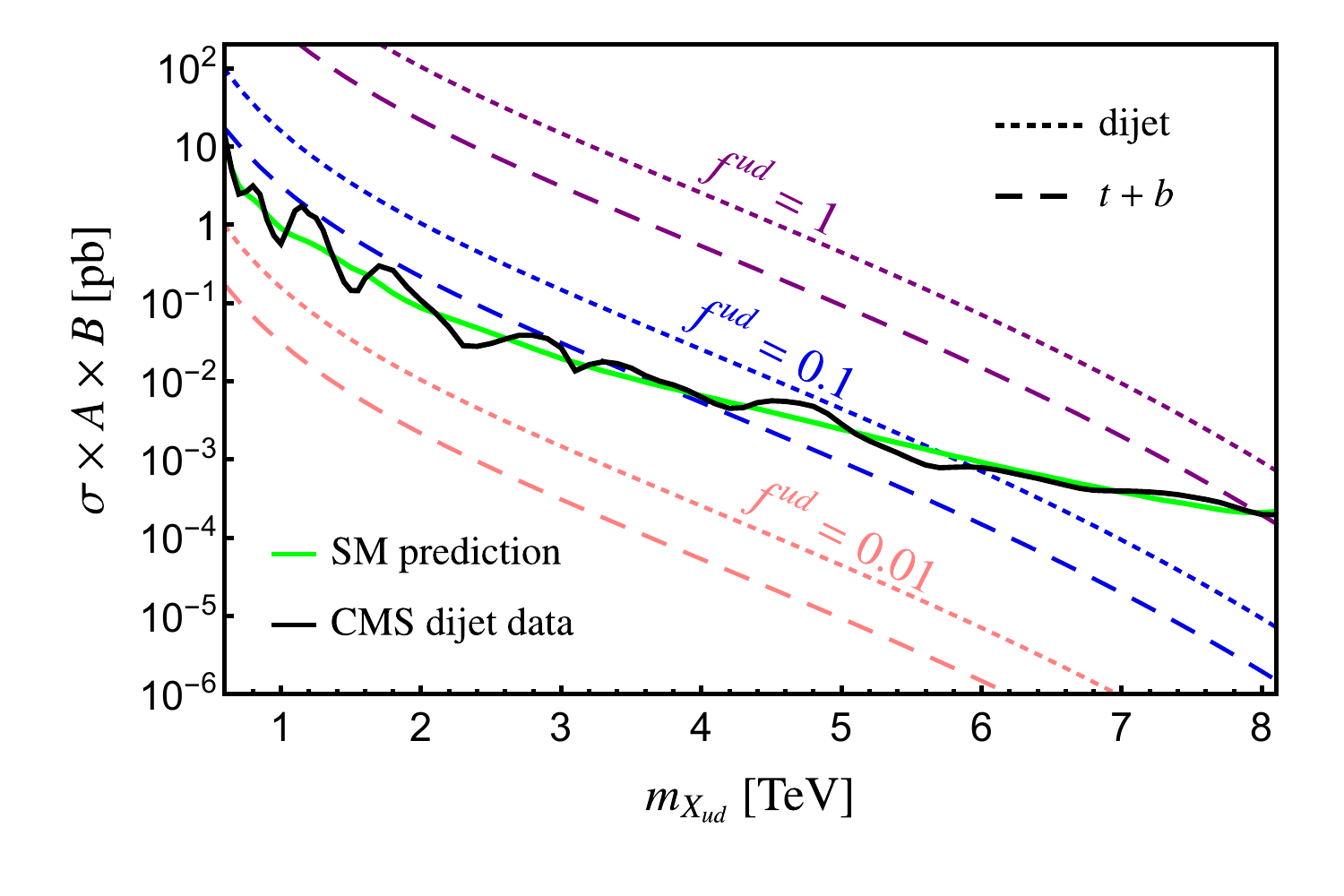}\hspace{-7mm}
\includegraphics[height=5.0cm]{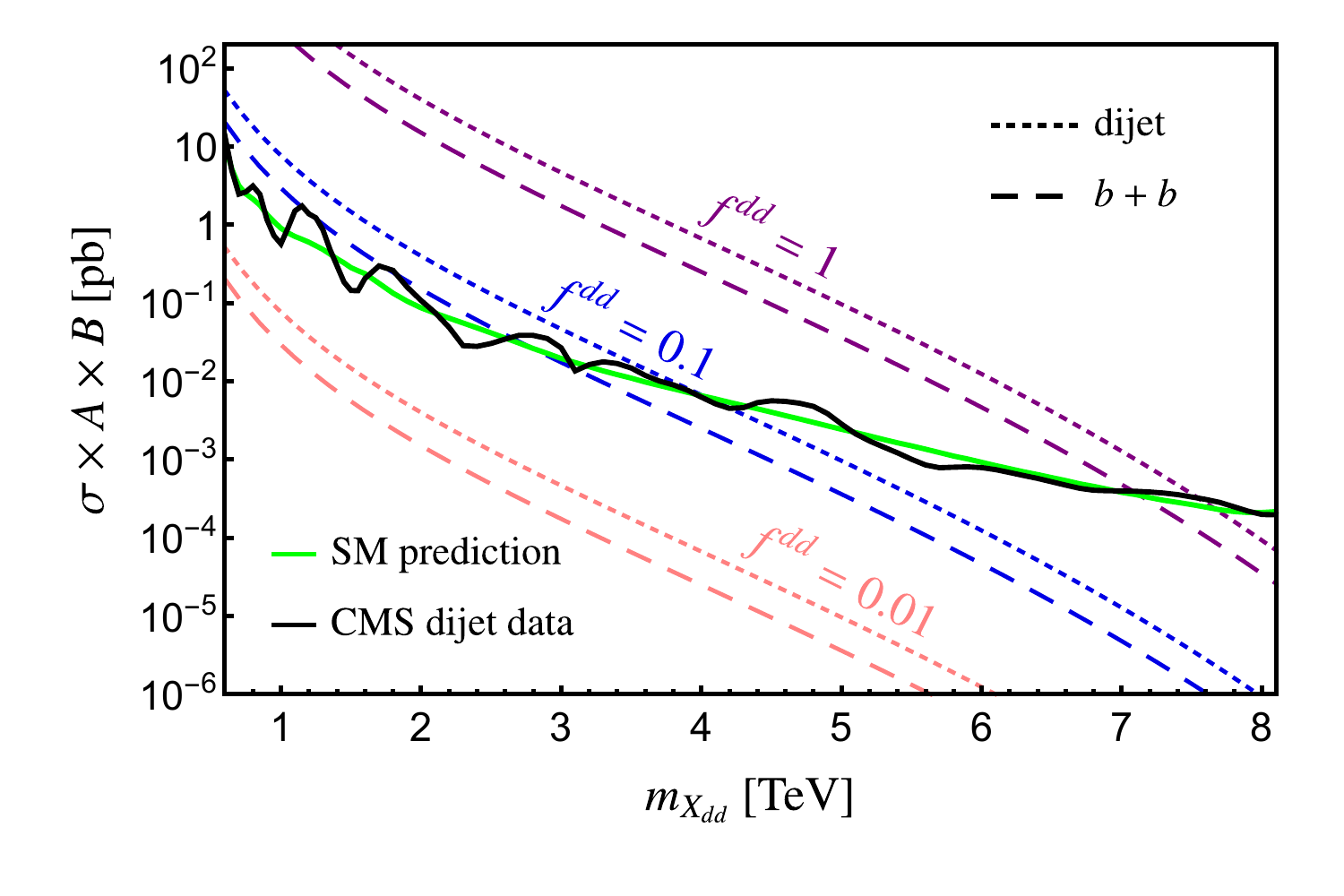}}\\[-5mm]
\mbox{\hspace{-4mm}
\includegraphics[height=5.0cm]{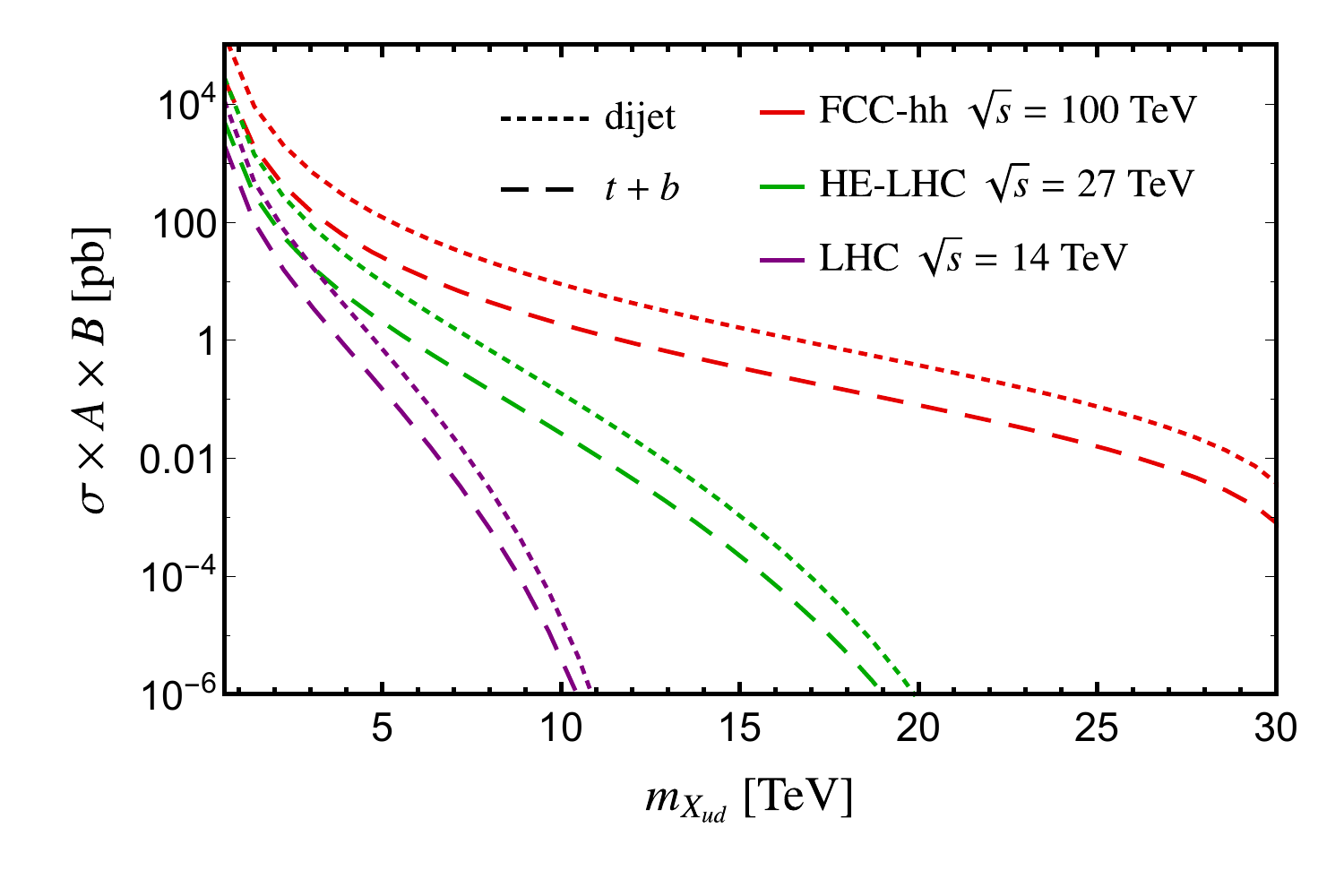}\hspace{-7mm}
\includegraphics[height=5.01cm]{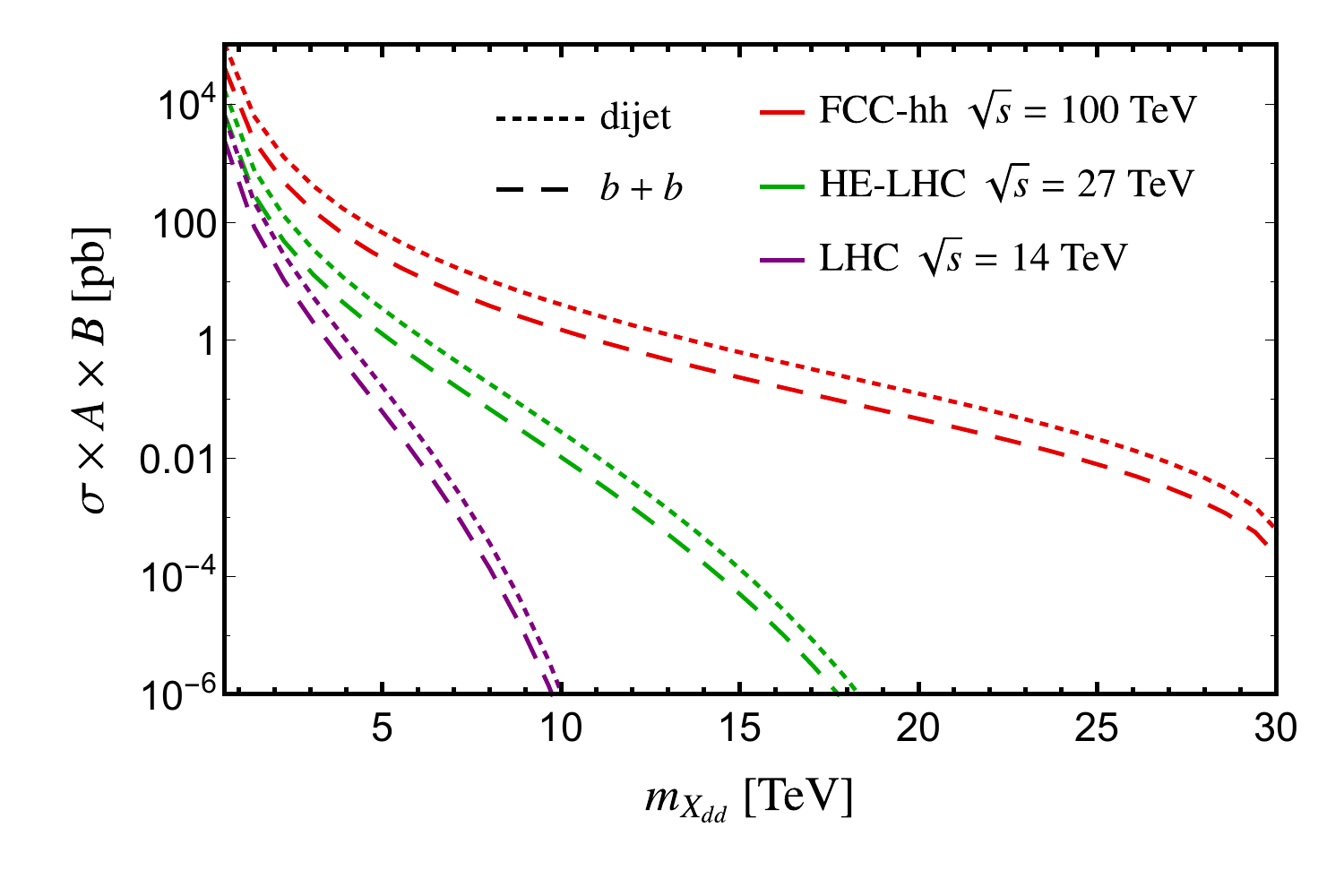}}\vspace{-7mm}
\label{fig:CMS}
\caption{Cross section of the scattering $u+d\to X_{ud}\to \text{dijet}~(t + \text{jet})$ (left column) and $d+d\to X_{dd}\to \text{dijet}~(b + b)$ (right column), multiplied by the corresponding experimental acceptance $\mathcal{A}$ and branching ratio BR, as a function of the diquark mediator mass. In the top left (right), the cross sections for three different values of the diquark coupling $f^{ud}$ $(f^{dd})$ at 13 TeV center-of-mass energy at LHC are shown, for $f=1.0$, $0.1$, and $0.01$, corresponding to purple, blue, and pink lines, respectively. The short-dashed lines correspond to the total cross section into dijet final state, and the long-dashed lines correspond to final state with third generation quarks. In black and green are the experimental data and corresponding SM prediction for a dijet final state search at the CMS experiment with 13 TeV center-of-mass energy and an integrated luminosity of 36 fb$^{-1}$. In the bottom row, the projected cross sections for $f=1$ at LHC, HE-LHC and FCC-hh are shown.}
\end{figure}
Even though an exact calculation of the cross section or the decay widths should include all possible Feynman diagrams, for all practical purposes, the experimental searches are principally focused on narrow resonances, expecting resonant peaks on a smoothly falling dijet mass spectrum corresponding to a $s$-channel decay mode of the resonance. Such a cross section for a resonance decaying via $s$-channel can be approximated by a Breit-Wigner form
\begin{equation}
  \hat{\sigma}(m)\left(a+b\rightarrow X\rightarrow c+d\right) = 16\pi \times \mathcal{N}\times\frac{\Gamma(a+b\rightarrow X)\times\Gamma(X\rightarrow c+d)}{\left(m^2-m_X^2\right)^2+m_X^2\Gamma^2_X},
\end{equation}
where $\Gamma_X$ and $m_X$ correspond to the total width and the mass of the resonance, and $m\simeq \sqrt{\hat{s}}=\sqrt{\tau s}$. The partial widths $\Gamma(a+b\rightarrow X)$ and $\Gamma(X\rightarrow c+d)$ correspond to the creation of the resonance from specific initial states and the decay of the resonance to the specific final states, respectively. The different multiplicity factors are taken into account through $\mathcal{N}$, defined as
\begin{equation}
  \mathcal{N} \equiv \frac{N_{S_X}}{N_{S_a}N_{S_b}}\frac{C_X}{C_a C_b},
\end{equation}
with $N_{S_X}$ and $N_{S_{a,b}}$ representing the spin multiplicities of the resonance and the initial state particles, respectively (e.g. for scalar diquarks $N_{S_X}=1$ and $N_{S_{a,b}}=2$ for initial state quarks). $C_X$ and $C_{a,b}$ are the relevant colour multiplicities (e.g. for colour sextet diquarks $C_X=6$ and for initial state quarks $C_{a,b}=3$). It is relevant to note here that the cross section above is obtained after integrating over $\cos\theta^*$, which in practice is constrained by the kinematics of the experimental searches. Therefore it is convenient to express the Breit-Wigner partonic cross section as
\begin{equation}
  \hat{\sigma}(m)=\frac{16\pi\times\mathcal{N}\times\mathcal{A} \times \text{BR}_{\text{par}} \times \Gamma_X^2}{\left(m^2-m_X^2\right)^2+m_X^2\Gamma^2_X},
\end{equation}
where $\text{BR}_{\text{par}}$ corresponds to the branching fraction of the partonic subprocess, and $\mathcal{A}$ is the experimental acceptance factor after the $\cos\theta^*$ cut. The Breit-Wigner partonic cross section can further be simplified in the narrow-width approximation ($\Gamma_X \ll m_X$)
\begin{equation}
\frac{1}{\left(m^2-m_X^2\right)^2+m_X^2\Gamma^2_X}\approx\frac{\pi}{m_X\Gamma_X}\delta(m^2-m_X^2) \, ,
\end{equation}
leading to the hadronic cross section 
\begin{equation}
  \sigma_{\text{had}}(m_X) = 16\pi^2\times\mathcal{N}\times\mathcal{A} \times \text{BR}_{\text{par}}\times\left[\frac{dL(\bar{y}_{min},\bar{y}_{max})}{d\hat{s}}\right]_{\hat{s}=m^2_X}\times\frac{\Gamma_X}{m_X}\, ,
\end{equation}
which, for given possibilities of initial ($a,b$) and final  ($c,d$) state partons, can be expressed as
\begin{eqnarray}
  \sigma_{\text{had}}^{\text{tot}}(m_X)&=&  16\pi^2\times\mathcal{N}\times\mathcal{A}
 \times \sum_{ab} (1 + \delta_{ab}) \text{BR} (X\to ab) \left[\frac{dL_{ab}(\bar{y}_{min},\bar{y}_{max})}{d\hat{s}}\right]_{\hat{s}=m^2_X} \nonumber \\
 &&  \times \sum_{cd} \text{BR} (X\to cd) \times\frac{\Gamma_X}{m_X}\, ,
\end{eqnarray}
where the factor $(1 + \delta_{ab})$ accounts for the possibility of two identical incoming partons (which gets compensated by a factor $1/2$ in the partial width of the final state phase space $\Gamma(X\to ab)$ for identical partons). We follow the prescription of Ref. \cite{Sirunyan:2018xlo} for the estimation of the acceptance parameter: defined as $\mathcal{A}=\mathcal{A}_\Delta \mathcal{A}_\eta$, with $\mathcal{A}_\Delta$ being the acceptance by requiring $\left|\Delta \eta\right| < 1.3$ for the dijet system and $\mathcal{A}_\eta$ is the acceptance factor due to also requiring $\left|\eta\right| < 2.5$ for each jet, separately. Assuming the decay of the scalar diquark resonances to be isotropic, one has $\mathcal{A}_\Delta=0.57$ for all masses and $\mathcal{A}_\eta \simeq 1$ \cite{Sirunyan:2018xlo}. Note that in case of a single top or anti-top quark in the final state, the top quark can decay before hadronising. This provides the possibility of tagging it to reconstruct the invariant mass of the diquark resonance. To provide some crude benchmark estimates for the $t+b$ final state we use the top tagging efficiency $\sim 0.3$ and bottom tagging efficiency $\sim 0.8$ for the decay channels with a $t$ or $b$ quark in the final state \cite{Sirunyan:2017ukk}.

Using CTEQ6L1 \cite{Pumplin:2002vw} PDFs and computing the PDF luminosity functions using the ManeParse package \cite{Clark:2016jgm}, we show in Fig.~\ref{fig:CMS} (top left) the cross sections of the scattering $u+d\to X_{ud}\to \text{dijet}$ and $u+d\to X_{ud}\to t + b$ as a function of the diquark mass $m_{X_{ud}}$ for three different benchmark values of the coupling $f^{ud}$. Given that the current best limits on the $t+b$ final state searches \cite{Sirunyan:2017ukk} are at best of the order of a dijet final state search, the limits considering the $t+b$ final state are less stringent as compared to that of dijet searches. While we still show the $t+b$ ($b+b$) final state in our plots, we consider only the dijet constraints in the following. Fig.~\ref{fig:CMS} (top right) shows the corresponding cross sections of the scattering processes $d+d\to X_{dd}\to \text{dijet}$ and $d+d\to X_{dd}\to b + b$ as functions of the diquark mass $m_{X_{dd}}$ for three different benchmark values of the coupling $f^{ud}$. Again, we show the exclusive two bottom quark final states separately, due to the possibility of tagging the $b$ quarks in light of recent experimental improvements in $b$-tagging. For reference, we indicate the current experimental limits on dijet searches from the CMS collaboration \cite{Sirunyan:2018xlo} as well as the SM prediction.

The current search limits already exclude parts of the parameter space for mass ranges as high as 8 TeV, which is expected to be improved significantly with more data and future colliders e.g. HE-LHC with 27 TeV center-of-mass energy and FCC-hh with 100 TeV center-of-mass energy \cite{Abada:2019ono,Benedikt:2018csr}. In Fig.~\ref{fig:CMS} bottom row we show a comparison of expected cross sections at LHC, HE-LHC and FCC-hh for coupling $f^{ud}$ ($f^{dd}$) of order unity. Given the expected several $\text{ab}^{-1}$ luminosity from future LHC upgrades, the collider searches will play a key role in probing the parameter space of baryogenesis complimenting the $\nnbar$~oscillation searches.

%%%%%%%%%%%%%%%%%%%%%%%%%%%%%%%%%%%%%%%%%%%%%%%%%%%%%%%%%%%%%%%%%%%%%%%%%%%%%%%%%%%%%%%%%%%%%%%%%%%%%%%%%%%%%%%%%%%%%%%%%%%%%%%%%%%%%%%%%%%%%%%%%%%%%%%%%%%%%%%%%%%%%%%%%
\subsection{Meson oscillations}{\label{subs:fcnc}}
FCNC processes such as neutral meson oscillations and flavour changing non-leptonic meson decays provide stringent constraints on the masses and couplings of the scalar diquark states \cite{Giudice:2011ak}. Since we assume $X_{uu}$ to be very heavy\footnote{For the case where $X_{uu}$ is sufficiently light to be of phenomenological interest, $D^0-D^0$ can provide tight constraints on the masses and couplings of $X_{uu}$, see e.g. Ref.~\cite{Dorsner:2010cu}.}, we mainly focus here on constraints on $X_{dd}$ and $X_{ud}$. 

\begin{figure}[t!]
	\centering\raisebox{2.5mm}{
	\includegraphics[width=0.4\textwidth]{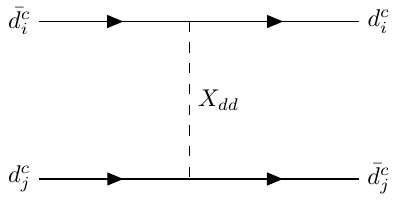}}
	\includegraphics[width=0.4\textwidth]{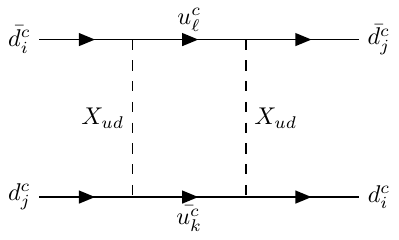}
	\caption{Tree- (left) and loop-level (right) meson oscillation diagrams mediated by the diquarks $X_{dd}$ or $X_{ud}$.}
	\label{fig:meson_osc}
\end{figure}

The contributions to neutral meson mixing, $M-\overline{M}$ with $M=B_s, B_d, K$, can be described as
\begin{equation}
	\langle M |H_{\text{eff}}|\overline{M} \rangle = \sum_j (C_j^{\text{SM}}+C_j^{\text{NP}}) \langle M |\Op_j |\overline{M} \rangle\, ,
\end{equation}
where $\Op_j$ are the effective four-quark operators and $C_j^{\text{SM (NP)}}$ denotes the associated Wilson coefficients for the SM (NP).

For the SM, the relevant operator for the $|\Delta F|=2$ process $B_q-\overline{B_q}$ is given by
\begin{equation}
	\Op_{V_L}=(\bar{q} \gamma_\mu P_L b)(\bar{q} \gamma^\mu P_L b)\, ,
\end{equation}
with the associated SM Wilson coefficient~\cite{Lenz:2010gu}
\begin{equation}
C^{\text{SM}}_{V_L} =\frac{G_{F}^2}{4\pi^2}m_W^2\hat{\eta}_BS_0(x_t)(V_{tb}V_{td(s)}^*)^2 \, ,
\end{equation}
where $\hat{\eta}_B = 0.8393$ accounts for QCD-corrections, and $S_0(x_t) = 2.35$ is the Inami-Lim function for the SM top quark box diagram, with $x_t = \overline{m}_t^2/m_W^2$.

For $\Delta S=2$ kaon oscillations  $K-\overline{K}$, the relevant SM operator is given by
\begin{equation}
	\Op_{V_L}=(\bar{d} \gamma_\mu P_L s)(\bar{d} \gamma^\mu P_L s)\, ,
\end{equation}
with the corresponding SM Wilson coefficient~\cite{Lenz:2010gu}
\begin{equation}
C^{\text{SM}}_{V_L} = \frac{G_{F}^2}{4\pi^2}m_W^2 \Big(\eta_{tt} S_0(x_t)(V_{ts}V_{td}^*)^2 + \eta_{cc} S_0(x_c)(V_{cs}V_{cd}^*)^2 + 2\eta_{ct} S_0(x_c,x_t)V_{ts}V_{td}^*V_{cs}V_{cd}^* \Big),
\end{equation}
where $\eta_{tt} = 0.5765$, $\eta_{cc} = 1.39~(1.29\text{ GeV}/\overline{m}_c)^{1.1}$, and $\eta_{ct}=0.47$ take into account QCD corrections. The Inami-Lim functions including the charm quark are given by $S_0(x_c)\approx x_c$, $S_0(x_c,x_t)\approx -x_c\log(x_c) + 0.56x_c$, with $x_c=\overline{m}_c^2/m_W^2$.

The exchange of the diquarks $X_{dd}$ and $X_{ud}$ gives rise to the following operator \cite{Babu:2013yca}
\begin{equation}
	\Op_{V_R}=(\bar{d}_j \gamma_\mu P_R d_i)(\bar{d}_j \gamma^\mu P_R d_i)\, ,
\end{equation}
where $(d_i,d_j)=(s,b), (d,b), (d,s)$ for $B_s, B_d, K$ oscillations, respectively. With $X_{dd}$ being a colour sextet, it contains flavour symmetric couplings to a down quark pair and can therefore mediate $\Delta F=2$ neutral meson mixing at tree level as well as at one-loop level. However, in the particular case where the coupling matrix $f^{dd}$ in Eq.~\eqref{lag:top2} is diagonal, the one loop contributions vanish, as can be quickly verified by considering the one-loop diagram in Fig.~\ref{fig:meson_osc} and replacing $u$ by $d$ and $X_{ud}$ by $X_{dd}$ in the loop. The exchange of $X_{dd}$ at tree-level can be related to the Wilson coefficient
\begin{eqnarray}\label{eq:osxdd}
C^{\text{NP}}_{V_R}=-\frac{1}{2}\frac{f^{dd}_{ii} (f^{dd}_{jj})^{*}}{m_{X_{dd}}^{2}} \, .
\end{eqnarray}
Furthermore, $X_{ud}$ cannot contribute via a tree-level contribution to the meson mixing. However, in contrast to $X_{dd}$, it can induce neutral meson mixing at one-loop level even if one starts with a diagonal structure for $f^{ud}$ in Eq.~\eqref{lag:top2} (written in the flavour basis), for instance in the presence of a right-handed analog of the CKM mixing matrix (see e.g. discussion in Refs.~\cite{Senjanovic:2016vxw,Senjanovic:2018xtu}). 
The exchange of the $X_{ud}$ in the one-loop box diagrams lead to \cite{Babu:2013yca}
\begin{eqnarray}\label{eq:osxud}
    C^{\text{NP}}_{V_R}= \frac{3}{256\pi^{2}} \frac{1}{m_{X_{ud}}^{2}} \left[\widehat{f}^{ud}_{ki}(\widehat{f}^{ud}_{kj})^*\right]^2 \, ,
   \end{eqnarray}
where we have used the fact that $f^{ud}_{ij}$ is symmetric and we have defined $\widehat{f}^{ud}_{ij}=(V_R)_{ik} f^{ud}_{kj}$, with $V_R$ being the right-handed quark mixing matrix diagonalising the right-handed quark charged current (similar to the CKM matrix for left-handed currents).
\begin{table}[t]
	\centering
	\begin{tabular}{l | c | c  }
		\specialrule{.2em}{.3em}{.0em}
		\text{Observable} & Diagram & Constraint \\[1mm]
		\hline
		 $\Delta_{B_{s}}$ & tree-level & $\left|f^{dd}_{22} (f^{dd}_{33})^{*}\right| \lesssim 3.1\times 10^{-4} \, \left(m_{X_{dd}}/\text{TeV}\right)^2$ \cite{Bona:2007vi} \\[1mm]
		                  & one-loop  & $\left|\sum_k \widehat{f}^{ud}_{k2} (\widehat{f}^{ud}_{k3})^{*}\right| \lesssim 0.36 \, \left(m_{X_{ud}}/\text{TeV}\right)$ \cite{Bona:2007vi} \\[1mm]
						  \hline
	     $\Delta_{B_{d}}$ & tree-level  & $\left|f^{dd}_{11} (f^{dd}_{33})^{*}\right| \lesssim 1.5\times 10^{-5} \, \left(m_{X_{dd}}/\text{TeV}\right)^2$ \cite{Bona:2007vi} \\[1mm]
		                  & one-loop  & $\left|\sum_k \widehat{f}^{ud}_{k3} (\widehat{f}^{ud}_{k1})^{*} \right| \lesssim 0.08 \, \left(m_{X_{ud}}/\text{TeV}\right)$ \cite{Bona:2007vi} \\[1mm]
						   \hline
		     $\Delta m_K$ & tree-level  & $\left|f^{dd}_{11} (f^{dd}_{22})^{*}\right| \lesssim 2.2\times 10^{-6} \, \left(m_{X_{dd}}/\text{TeV}\right)^2$ \cite{Charles:2015gya}  \\[1mm]
		                  & one-loop  & $\left|\sum_k \widehat{f}^{ud}_{k2} (\widehat{f}^{ud}_{k1})^{*}\right| \lesssim 0.03 \, \left(m_{X_{ud}}/\text{TeV}\right)$ \cite{Charles:2015gya} \\[1mm]
		\specialrule{.2em}{.0em}{.3em}
	\end{tabular}
	\caption{Latest constraints on the diquark couplings from $B_s^0-\overline{B_s^0}$, $B_d^0-\overline{B_d^0}$, and $K^0-\overline{K^0}$ oscillations using 95\% CL experimental limits.}
	\label{tab:flav_con}
\end{table}

Following the prescription of \cite{Charles:2015gya,Lenz:2010gu,Bona:2007vi,Hati:2018fzc}, we define the ratio of the total contribution to the SM for $M-\overline{M}$ oscillations as
\begin{equation}
\Delta_M =  \frac{\langle M |H_{\text{eff}}| \overline{M}\rangle}{\langle M |H_{\text{eff}}^\text{SM}| \overline{M}\rangle}=1+\frac{C^{\text{NP}}}{C^{\text{SM}}}\, ,
\end{equation}
where $\Delta_M$ is an experimentally determined complex parameter that depends on the respective meson $M$. It is experimentally constrained to $\Delta_{B_{s(d)}} = 1.11^{+0.96}_{-0.48} (1.05^{+1.0}_{-0.52})$ for the $B_s^0-\overline{B_s^0}$ and $B_d^0-\overline{B_d^0}$ oscillations, and to $\text{Re}(\Delta_K) = 0.93^{+1.14}_{-0.43}$ and $\text{Im}(\Delta_K) = 0.92^{+0.39}_{-0.26}$ for the $K^0-\overline{K^0}$ oscillations\cite{Bona:2007vi,Charles:2015gya}. Using the above 95\% CL experimental limits, we derive the relevant constraints summarised in Tab.~\ref{tab:flav_con}. In addition to the neutral meson oscillations, the diquark coupling can also give rise to a number of non-leptonic rare meson decays at tree- or loop-level, however the relevant constraints obtained are comparably weaker \cite{Giudice:2011ak,Babu:2013yca}.

Generally, the latest experimental data on neutral meson oscillations provide some of the most stringent constraints on the $X_{dd}$ scalar diquark couplings with masses around the TeV scale (see Tab.~\ref{tab:flav_con}). Therefore, the constraints from neutral meson oscillations will be of great importance in synergy with the direct collider searches for probing low-scale baryogenesis scenarios.

%%%%%%%%%%%%%%%%%%%%%%%%%%%%%%%%%%%%%%%%%%%%%%%%%%%%%%%%%%%%%%%%%%%%%%%%%%%%%%%%%%%%%%%%%%%%%%%%%%%%%%%%%%%%%%%%%%%%%%%%%%%%%%%%%%%%%%%%%%%%%%%%%%%%%%%%%%%%%%%%%%%
\subsection{Dinucleon decay}{\label{subs:dinucleon}}

\begin{figure}[t] 
\centerline{
   \includegraphics[height=2.7in]{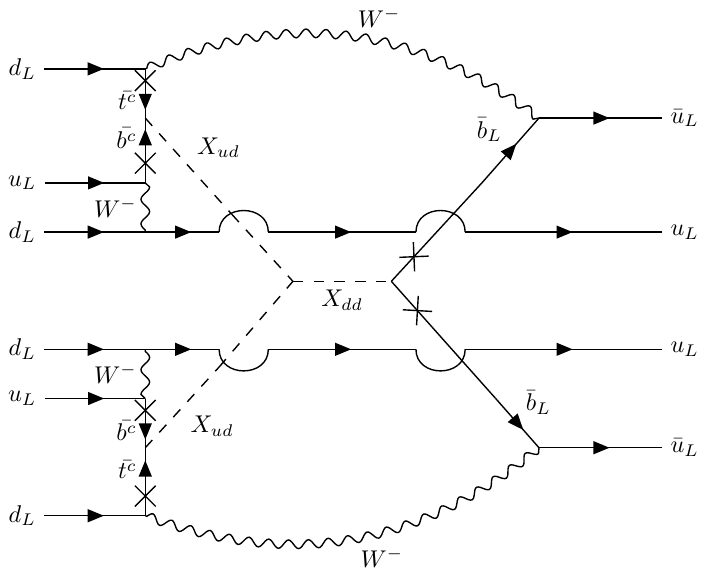}
   }
\caption{A representative diagram giving rise to the dinucleon decay mode $nn \to \pi^{0}\pi^{0}$ at two-loop level with the scalar diquarks dominantly coupling to third generation quarks.}
\label{dnd}
\end{figure}

In addition to $\nnbar$ oscillations, the dinucleon decays can also provide useful constraints for diquark couplings to the SM quarks. For the diquark couplings to the first generation of SM quarks both $\nnbar$~oscillations and dinucleon decay, e.g. $nn \to \pi^{0}\pi^{0}$, can occur at tree-level\footnote{Note that we take the neutral decay mode $nn\to\pi^0\pi^0$ as an example case. For other modes such as $pp\to\pi^+\pi^+$ our discussion can be straightforwardly generalised.}. Given the availability of better numerical estimates for the well studied transition matrix elements for $\nnbar$~oscillations as compared to potentially large hadronic uncertainties for dinucleon decay matrix elements, the former provides more reliable constraints in this case. However, for the scenario where the diquark couples dominantly to the third generation SM quarks, the dinucleon decay modes, e.g. $nn \to \pi^{0}\pi^{0}$ (induced at the two-loop level), provide a more stringent constraint as compared to the $\nnbar$~oscillation (induced at the three-loop level)\footnote{We note that, in Ref.~\cite{Berezhiani:2018xsx} it has been pointed out that some EFT operators which are odd under charge conjugation ($C$), parity $(P_z)$ (and $CP_z$) can lead to dinucleon decay, but not $\nnbar$ transition. In our simplified model the relevant operator correspond to the $C$, $(P_z)$, and $CP_z$ even case, which can a priory lead to both dinucleon decay as well as $\nnbar$ transition simultaneously.}. In particular, for the scenario of low-scale baryogenesis, where both $X_{dd}$ and $X_{ud}$ masses lie around TeV scale, the dinucleon decays are of particular interest if the diquark couples dominantly to the third generation SM quarks. Even though we will mainly focus on the simplest case of flavour diagonal and universal couplings of diquarks to SM quarks for the study of baryogenesis, below we briefly discuss the case where the dinucleon decay mode $nn \to \pi^{0}\pi^{0}$ can occur at two-loop level, when the diquark couples dominantly to the third generation SM quarks. A representative Feynman diagram inducing such a mode is shown in Fig.~\ref{dnd}. Given that in our simplified model the diquarks can exclusively couple to right-handed quarks one would require mass insertions for top and bottom quarks in the loops which we shall ignore in the following estimate in the interest of simplification and due to the large uncertainties involved in estimating the transition matrix elements. The relevant dinucleon decay rate is given by~\cite{Goity:1994dq}
	\begin{equation}
	\label{eq:dnd}
	\Gamma \simeq \frac{9}{32 \pi}\frac{\left|\mathcal{C}_{\text{DN}}\right|^2}{m_n^2} |\langle\pi^0 \pi^0|\mathcal{O}_{\text{DN}}^{15}|nn \rangle|^2 \rho_n,
	\end{equation}
where $\rho_n\simeq 0.25\, \text{fm}^{-3}$ is the nucleon density, and $m_n$ the neutron mass. The Wilson coefficient $\mathcal{C}_{\text{DN}}$ induced at the two-loop level is given by
	\begin{equation}
	\label{eq:dndwc}
\mathcal{C}_{\text{DN}}=\left(\frac{\lambda  (f^{ud}_{33})^2 f^{dd}_{33} |V_{ub}|^{4}|V_{td}|^{2}{g_{2}}^{8}}{32}\right) \frac{v' m_t^2}{m_W^4}\,I\,,
	\end{equation}
	where under the simplifying assumption of vanishing external momenta the relevant two-loop integral $I$ can be written as
	\begin{equation}
		\label{eq:tli}
	I= \int \frac{d^4 k_1}{(2\pi)^4}\int \frac{d^4 k_2}{(2\pi)^4}\, \prod_{i=1}^{2}
\left(\frac{1}{k_i^2-m_t^2} \frac{1}{k_i^2-m_b^2}
 \frac{1}{k_i^2-m_{X_{ud}}^2} \frac{1}{k_i^2-m_W^2} \right)
	\frac{1}{(k_1+k_2)^2-m_{X_{dd}}^2}\, .
	\end{equation}
Using partial fraction and rescaling all the masses by the scalar mass $m_{X_{dd}}$, the integral $I$ can further be expressed as \cite{Sierra:2014rxa,Bolton:2019bou}
		\begin{eqnarray}
		\label{eq:tl2}
	I&=&  \frac{1}{(2\pi)^8} \frac{1}{(m_t^2-m_b^2)^2} \frac{1}{(m_{X_{ud}}^2-m_W^2)^2} \frac{1}{m_{X_{dd}}^2} [(J_{t,X_{ud};t,X_{ud}}+J_{t,X_{ud};b,W}+J_{t,W;t,W}+J_{t,W;b,X_{ud}}\nonumber\\
	&& 
	+J_{b,X_{ud};t,W}+J_{b,X_{ud};b,X_{ud}}+J_{b,W;t,X_{ud}}+J_{b,W;b,W})-(J_{t,X_{ud};t,W}+J_{t,X_{ud};b,X_{ud}}+J_{t,W;t,X_{ud}}\nonumber\\
	&&
	+J_{t,W;b,W}+J_{b,X_{ud};t,X_{ud}}+J_{b,X_{ud};b,W}+J_{b,W;t,W}+J_{b,W;b,X_{ud}}        )]\, ,
		\end{eqnarray}
where the integral $J_{\alpha,\beta;\mu,\nu}$ is defined as
		\begin{equation}
		\label{eq:tl3}
J_{\alpha,\beta;\mu,\nu}= \int d^4k_1 \int d^4k_2
  \frac{1}{(k_1^2-r_\alpha)(k_1^2-t_{\beta})(k_2^2-r_\mu)(k_2^2-t_{\nu})([k_1+k_2]^2-1)}\ ,
		\end{equation}
with $r_{\{ \alpha,\mu \}} =(m_{F_{\{ \alpha,\mu \}}}/m_{X_{dd}})^2$ for fermionic and $t_{\{ \beta,\nu \}}=(m_{B_{\{ \beta,\nu \}}}/m_{X_{dd}})^2$ for bosonic masses. Note that the momenta $k_{\{1,2\}}$ are rescaled with respect to Eq.~\eqref{eq:tli}.
Now, the integral $J_{\alpha,\beta;\mu,\nu}$ can be reduced in terms of the standard integral as
\begin{equation}\label{eq:tl4}
J_{\alpha,\beta;\mu,\nu} =
  \frac{\pi^4}{(t_{\beta}-r_\alpha)(t_{\nu}-r_\mu)}
  \left[ -{ g}(t_{\beta},t_{\nu})
    +{ g}(r_\alpha,t_{\nu})
    +{ g}(t_{\beta},r_\mu)
    -{ g}(r_\alpha,r_\mu)\right]\ ,
\end{equation}
where in general $D = 4 +\epsilon$ dimensions we have
		\begin{equation}
		\label{eq:tl5}
{g}(a,b)= \mu^{\epsilon} \int d^n k_1 \int d^n k_2 \frac{1}{(k_1^2-a)(k_2^2-b)[(k_1+k_2)^2-1]}
 \ ,
		\end{equation}
which includes an infinite and a finite piece \cite{Sierra:2014rxa}. After collecting the different terms in Eq.~\eqref{eq:tl2}, the divergent pieces in $g(a,b)$ drop out, leaving the finite piece
\begin{eqnarray}
	\label{eq:tl6}
 g(a,b) & = & \frac{a}{2}\ln a \ln b +
\sum_{\pm} \pm \frac{a(1-a)+3ab+2(1-t)x_{\pm}}{2\omega}
\\ \nonumber
& \times & \left[\text{Sp}(\frac{x_{\pm}}{x_{\pm}-a})
              -\text{Sp}(\frac{x_{\pm}-a}{x_{\pm}})
              +\text{Sp}(\frac{b-1}{x_{\pm}})
              -\text{Sp}(\frac{b-1}{x_{\pm}-a})  \right]\ ,
\end{eqnarray}
with the argument of the Spence's function containing the parameters
\begin{eqnarray}
	\label{eq:xpm}
x_{\pm}=\frac{1}{2}(-1+a+b \pm \omega), & & \omega=\sqrt{1+a^2+b^2-2(a+b+ab)}\ .
\end{eqnarray}

  In order to estimate the dinucleon decay rates numerically for the relevant matrix element of the dimension-15 operator $\mathcal{O}_{\text{DN}}^{15}$ in Eq.~\eqref{eq:dnd}, we take $|\langle\pi^0 \pi^0|\mathcal{O}_{\text{DN}}^{15}|nn \rangle|\sim \Lambda^{11}$, where $\Lambda$ corresponds to a scale between $\Lambda_{\text{QCD}}$ and $m_n$. To be conservative we take the most stringent possible constraint corresponding to $\Lambda=m_n$.
  
\begin{figure}[t]
  	\centering
  	\includegraphics[height=8cm]{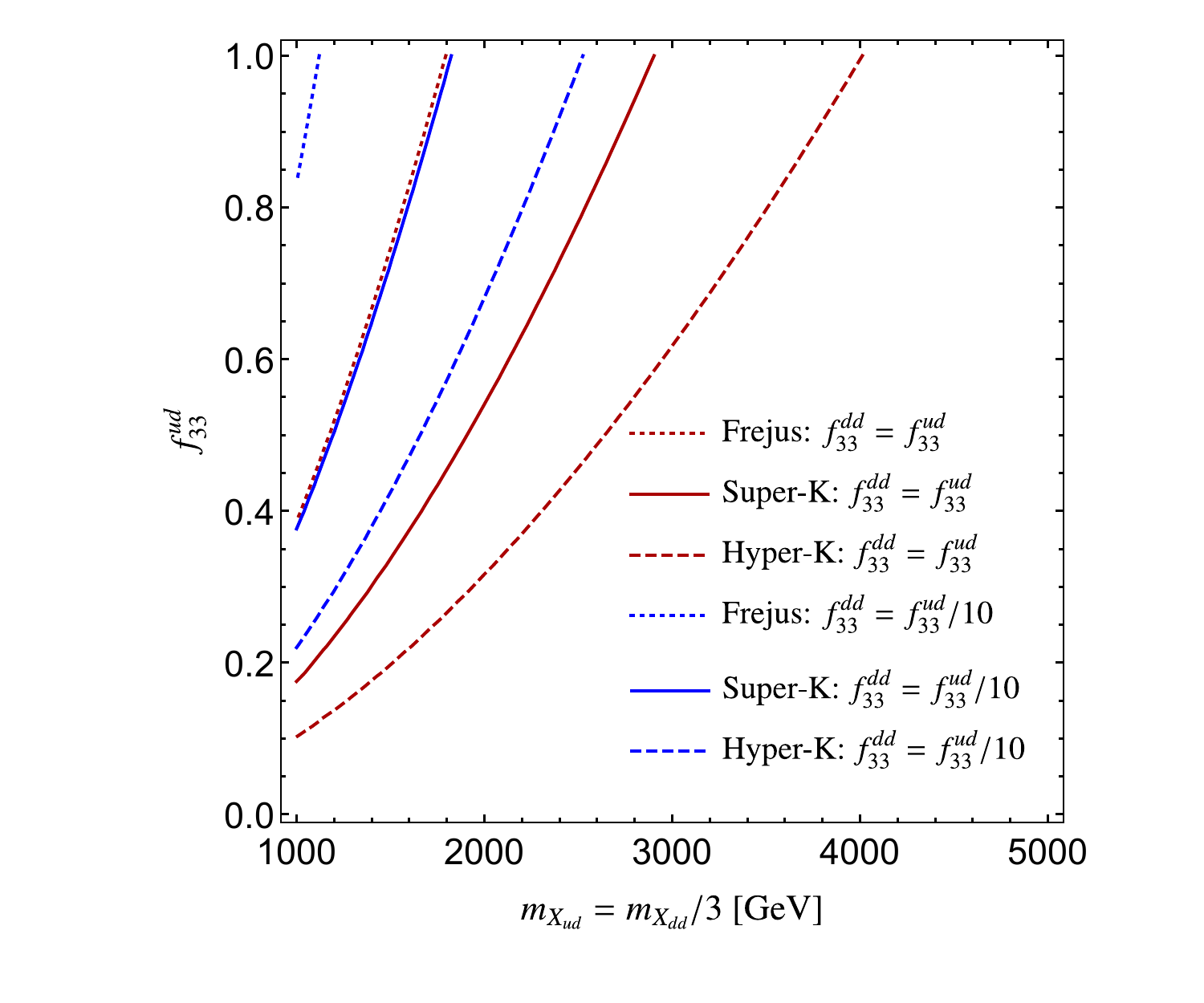}
  	\caption{Contours in the coupling-mass plane demonstrating the potential reach of different dinucleon decay experiments. The red lines correspond to the case $f^{dd}=f^{ud}$, the blue lines to $f^{ud}=10f^{dd}$.}
  	\label{fig:dinucleonplot}
  \end{figure}

 The most stringent current experimental constraint on the $nn \to \pi^{0}\pi^{0}$  partial lifetime due to Super-Kamiokande is $\tau > 4.04\times10^{32}$ years \cite{Gustafson:2015qyo}, which improved over the earlier limit from the Frejus experiment of $\tau > 3.4\times10^{30}$ years \cite{Berger:1991fa}. The future experiments, such as Hyper-Kamiokande, is expected to improve on the current results by up to an order of magnitude \cite{Takhistov:2016eqm}. In Fig.~\ref{fig:dinucleonplot} we show the relevant current and future experimental constraints on the parameter space spanned by the coupling and diquark masses. The current constraints are relevant for masses of $X_{ud}$ and $X_{dd}$ around a few TeV. For more robust constraints, a more accurate estimation of the relevant matrix elements is desirable.
%
%%%%%%%%%%%%%%%%%%%%%%%%%%%%%%%%%%%%%%%%%%%%%%%%%%%%%%%%%%%%%%%%%%%%%%%%%%%%%%%%%%%%%%%%%%%%%%%%%%%%%%%%%%%%%%%%%%%%%%%%%%%%%%%%%%%%%%%%%%%%%%%%%%%%%%%%%%%%%%%%%%%
\subsection{Colour preserving vacuum}{\label{subs:cpv}}
From a phenomenological point of view, in the absence of a specified UV completion, the effective trilinear coupling of the form $\mu X_{dd} X_{ud} X_{ud}$ (e.g. induced by the vacuum expectation value of $\xi$ in Eq.~\eqref{lag:top2}) can be constrained by the requirement of colour preserving vacua.
\begin{figure}[t] % fig 3
\centering
  \mbox{\includegraphics[height=1.5in]{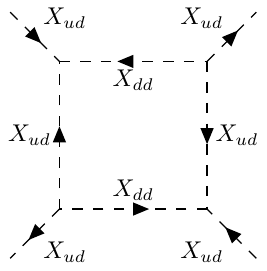}\hspace{8mm}
  \includegraphics[height=1.5in]{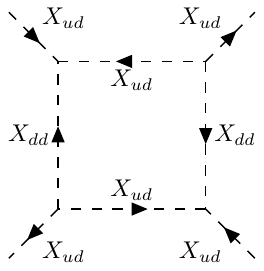}}\\[5mm]

\mbox{\includegraphics[height=1.5in]{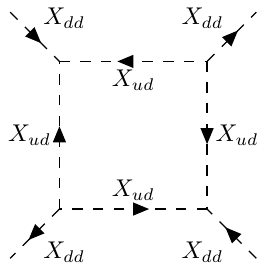}\hspace{8mm}
\includegraphics[height=1.5in]{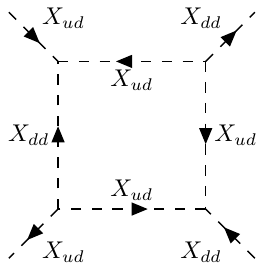}}
   
\caption{Diagrams giving rise to the effective quartic interaction terms
of the fields $X_{ud}$ and $X_{dd}$.}
\label{vpc}
\end{figure}
As shown in Fig.~\ref{vpc}, the effective trilinear coupling of the form $\mu X_{dd} X_{ud} X_{ud}$ leads to effective quartic interactions for the $X_{ud}$ and $X_{dd}$ fields given by
\begin{equation} \label{Lphi4}
-{\cal L}_{\mathrm{eff} } = \lambda_{\mathrm{eff}} (X_{ud})^2 (X_{ud}^\dagger)^2 \ + \
\lambda'_{\mathrm{eff}} (X_{dd})^2 (X_{dd}^\dagger)^2 \ + \
\lambda''_{\mathrm{eff}} (X_{ud} X_{ud}^\dagger) (X_{dd} X_{dd}^\dagger)~.
\end{equation}
%.
The relevant induced effective corrections are given by \cite{Babu:2002uu}
\begin{eqnarray}
\lambda_{\mathrm{eff}} & \sim & -\frac{1}{2 \pi^2} \frac{\mu^4}{(m_{X_{dd}}^2 - m_{X_{ud}}^2)^2}
\left[
 \frac{m_{X_{dd}}^2 + m_{X_{ud}}^2}{m_{X_{dd}}^2 - m_{X_{ud}}^2}
    \hbox{log} \left(\frac{m_{X_{dd}}^2}{m_{X_{ud}}^2}\right)
\ - \ 2 \right] \, ,\label{Lambdaeff1} \\
\lambda'_{\mathrm{eff}} & \sim & -\frac{1}{4 \pi^2}\ \frac{\mu^4}{6 m_{X_{ud}}^4} \, , \label{Lambdaeff2}\\
\lambda''_{\mathrm{eff}} & \sim & -\frac{1}{\pi^2}\ \frac{\mu^4 m_{X_{dd}}^2}{
    2(m_{X_{dd}}^2-m_{X_{ud}}^2)^3} \left[ \frac{m_{X_{dd}}^2}{m_{X_{ud}}^2} - \frac{m_{X_{ud}}^2}{m_{X_{dd}}^2}
    - 2 \hbox{log}\left( \frac{m_{X_{dd}}^2}{m_{X_{ud}}^2} \right) \right] \,. \label{Lambdaeff3}
\end{eqnarray}
In the particular limit in which two masses are similar ($m_{X_{dd}} \sim m_{X_{ud}} $) the following relation holds
$$
2 \lambda_{\mathrm{eff}} \ \approx \ \lambda''_{\mathrm{eff}} \ \approx \ 4\lambda'_{\mathrm{eff}}
 \ = \ -\frac{1}{\pi^2}\ \frac{\mu^4}{6 m_{X_{ud}}^4} .
$$
With the effective couplings being negative (in the absence of a UV embedding), the tree-level Lagrangian must contain terms similar to the effective quartic couplings $\lambda_{\mathrm{eff}}, \lambda'_{\mathrm{eff}}, \lambda''_{\mathrm{eff}}$, which are positive and greater in their absolute value  to ensure a stable colour preserving vacuum. 

Moreover, the requirement that the theory has to be perturbative ($\lambda_{\mathrm{eff}}, \lambda'_{\mathrm{eff}},
\lambda''_{\mathrm{eff}} < 1$)
imposes the following constraints on the $\mu$ parameter:
\begin{equation}{\label{eq:cpvac}}
\mu  \ < \  \left\{
\begin{array}{ll}
 m_{X_{ud}} \times ( 6 \pi^2)^{1/4} & \hbox{for} \ m_{X_{dd}} \sim m_{X_{ud}} \, , \\
 m_{X_{ud}} \times ( 24 \pi^2)^{1/4} & \hbox{for} \ m_{X_{dd}} > m_{X_{ud}} \, ,
\end{array}
\right.
\end{equation}
which provide important benchmark constraints fixing the hierarchy between $m_{X_{ud}}$, $m_{X_{dd}}$ and $\mu=\lambda v'$ in Eq.~\eqref{lag:top2}. Consequently we fix $\lambda v' \sim m_{X_{dd}}$ to a few times $m_{X_{dd}}$ for the subsequent numerical study of the baryogenesis scenarios where the hierarchy $m_{X_{dd}} > m_{X_{ud}}$ is maintained but with both $m_{X_{ud}}$ and $m_{X_{dd}}$ not too far from each other to ensure a stable colour preserving vacuum.

For the high-scale scenario on the other hand, a given UV completion must take into account the relevant Higgs multiplet in which the scalar fields are embedded. In the presence of a field at a particularly low scale, one must assure that the relevant interactions involving that field in the low-energy theory preserves the vacuum with a perturbative coupling. For example, when $m_{X_{ud}}$ is around few TeV and $m_{X_{dd}}$ is around the GUT scale, then for a low-energy theory at a temperature $T<m_{X_{dd}}$ one can integrate out $X_{dd}$, leaving all its effects encoded in the couplings of $X_{ud}$. In order to make sure that in such a scenario the low-energy theory is in the perturbative regime, one requires $\lambda_{\mathrm{eff}}< 1$ in Eq.~\eqref{Lambdaeff1}. This is ensured by the condition
\begin{equation}
	\mu  \sim \mathcal{O}\left( m_{X_{dd}} \right)\quad \hbox{for} \quad m_{X_{dd}} \gg T > m_{X_{ud}} \, .
\end{equation}
Besides, the couplings of the heavy state $X_{uu}$ with $X_{dd}$ can effectively ensure that the other radiative corrections $\lambda'_{\mathrm{eff}}$ and $\lambda''_{\mathrm{eff}}$ in Eqs.~\eqref{Lambdaeff2} and \eqref{Lambdaeff3} do not exceed the tree-level terms~\cite{Babu:2012vc}. 

To conclude, for the high-scale scenario the conditions for a colour preserving vacuum can be naturally satisfied in a UV completion where the diquark field $X_{ud}$ lies around TeV scale and the other diquark fields $X_{dd}$ and $X_{uu}$ are significantly heavy with masses around the GUT scale. On the other hand, for the phenomenologically driven low-scale scenario, where $X_{ud}$ and $X_{dd}$ masses are not fixed by a UV completion, one requires a very mild hierarchy between  $m_{X_{ud}}$, $m_{X_{dd}}$ and $\lambda v'$ to satisfy the conditions from  colour preserving vacua: $\lambda v' \sim \Op(\text{few})\times m_{X_{dd}} \sim \Op(\text{few})\times m_{X_{ud}}$. Therefore, we use the benchmark choices $\lambda v'=1.2 m_{X_{dd}}$ and $m_{X_{dd}}=3 m_{X_{ud}}$ for our subsequent numerical analysis of this type of scenario to satisfy the conditions given in Eq.~\eqref{eq:cpvac}.
%%%%%%%%%%%%%%%%%%%%%%%%%%%%%%%%%%%%%%%%%%%%%%%%%%%%%%%%%%%%%%%%%%%%%%%%%%%%%%%%%%%%%%%%%%%%%%%%%%%%%%%%%%%%%%%%%%%%%%%%%%%%%%%%%%%%%%%%%%%%%%%%%%%%%%%%%%%%%%%%%%%%%%%
\subsection{Comments on other possible constraints}
In addition to the constraints discussed above, $X_{dd}$ diquark couplings can also contribute to the neutron electric dipole moment (in the presence of complex phases in diquark couplings) and to electroweak precision observables (e.g. related to $Z\rightarrow d^c \bar{d^c}$) as discussed in~\cite{Giudice:2011ak}. In the presence of $X_{dd}$ couplings to both left- and right-handed down-type quark pairs these loop observables can in particular be sizeable due to chiral enhancement and in particular when heavy quarks can be in the loop due to flavour antisymmetric couplings. However, given that in our scenario $X_{dd}$ couples only to the right-handed quarks and flavour symmetrically, one would require additional quark mass insertions to realise such observables leading to both chirality and loop suppressed contributions.

\section{Results and discussion}\label{sec:results}
Using the formalism described in Sec.~\ref{sec:boltzmann}, in this section we present our findings regarding the viability of successful baryogenesis for the simplified model described in Sec.~\ref{subs:model}, and confront the results with the experimental constraints discussed in the previous section. Hereby, we distinguish two distinct scenarios, which we will refer to as the high- and low-scale scenarios. Assuming the mechanism that generates the BAU is the out-of-equilibrium decay of the $X_{dd}$ diquark, as described in Sec.~\ref{sec:boltzmann}, both the high- and low-scale scenario can lead to a baryon asymmetry close to the observed value, but with very different detection prospects. We furthermore define, for both scenarios, a model-dependent CP-violation parameter $\epsilon$, originating from the interference between the tree-level decay of $X_{dd}$ and the 1-loop decay with $X_{dd}'$, another generation of $X_{dd}$, as introduced in Eq.~\eqref{eq:eps}, and simplified here by selecting as a benchmark scenario\footnote{Note that we assume in writing Eq.~\eqref{eq:eps2} that one of the couplings is purely imaginary  to maximise the CP violation without any loss of generality.}, $f_{dd}'=f_{dd}$ and $\lambda'=\lambda$, such that
\begin{eqnarray}
	\label{eq:eps2}
	\epsilon = \frac{1}{\pi}\text{Tr}\left[(f_{dd})^\dagger f_{dd} \right]\left( \frac{x}{1-x}\right)r\, ,
\end{eqnarray}
where $x=(m_{X_{dd}}/m_{X_{dd}}^\prime)^2$, and $r$ is the branching ratio for $X_{dd}\to X_{ud}^*X_{ud}^*$. Hereby, to simplify the forthcoming baryogenesis parameter space analysis, we will assume a flavour diagonal and universal structure for (in general 3$\times$3 coupling matrices) $f^{ud(dd)}_{ij}$ (neglecting any flavour effects for baryogenesis) and henceforth will denote them as $f_{ud(dd)}$ for brevity.
The maximal value of $\epsilon$ is $\epsilon=2 r$ and is the most optimistic scenario, for the other cases we select as a benchmark scenario $x=0.2$.
\begin{table}[t!]
	\centering
	\begin{tabular}{l | l | c | c }
		\specialrule{.2em}{.3em}{.0em}
		$T\text{(GeV)}$ & Process & High-scale & Low-scale \\[1mm]
		\hline
		{$\gg 10^{13}$} & Gauge interactions + $t$ Yukawa  & {\textcolor{newgreen}{\ding{51}}} &   \textcolor{newgreen}{\ding{51}}  \\[1mm]
		\hline
		$> 10^{13}$ & QCD sphaleron & \textcolor{newgreen}{\ding{51}} & \textcolor{newgreen}{\ding{51}}  \\[1mm]
		\hline
		$\sim 10^{13}$ & EW sphaleron + $b$ Yukawa & \textcolor{newgreen}{\ding{51}} &   \textcolor{newgreen}{\ding{51}}  \\[1mm]		
		\hline
		$\sim 10^{12}$ & $\tau$ Yukawa & \textcolor{newgreen}{\ding{51}}  & \textcolor{newgreen}{\ding{51}}  \\[1mm]
		\hline
		$\sim 10^{12}$ & $c$ Yukawa & \textcolor{red}{\ding{55}} & \textcolor{newgreen}{\ding{51}}  \\[1mm]
		\hline
		$\sim 10^{10}$ & $\mu$, $s$ Yukawa & \textcolor{red}{\ding{55}} & \textcolor{newgreen}{\ding{51}}  \\[1mm]
		\hline
		$\ll 10^{8}$ & $u$, $d$, $e$ Yukawa & \textcolor{red}{\ding{55}} & \textcolor{newgreen}{\ding{51}} \\[1mm]
		\specialrule{.2em}{.0em}{.3em}
	\end{tabular}
	\caption{Approximate temperatures at which different SM processes come into equilibrium in the early Universe \cite{Garbrecht:2014kda}. Green ticks mark the processes that are assumed to be in equilibrium in the high- and low-scale scenarios, and red crosses mark the interactions that are assumed to not be in equilibrium.}
	\label{tab:Tscale}
\end{table}
For the numerical evaluation of the final baryon asymmetry in two different baryogenesis scenarios, we take into account the appropriate SM processes in equilibrium, as tabulated in Tab.~\ref{tab:Tscale}. The green ticks show the SM processes which are taken into account at the corresponding temperature while solving the Boltzmann equations. During the computation of the baryon asymmetry in the high-scale baryogenesis scenario, we assume that only the Yukawa couplings of the third generation quarks are in chemical equilibrium, while all three generations are assumed to be in chemical equilibrium in the low-scale baryogenesis scenario. All quark generations are considered to be in thermal equilibrium in both scenarios. Furthermore, we also assume that during high-scale baryogenesis, only the decay of $X_{dd}$ to third generation quarks are relevant, since other two lighter generations do not come into equilibrium with the thermal bath of the Universe until much later than the temperature relevant for high-scale baryogenesis. On the other hand, for the low-scale baryogenesis scenario, we consider the decay of $X_{dd}$ to all three generations of quarks. In the following, we proceed to discuss the two scenarios in more detail and present the numerical results of the baryogenesis mechanism of Sec.~\ref{sec:boltzmann}.\\

\begin{figure}[ht!]
\centering
\includegraphics[height=7.8cm]{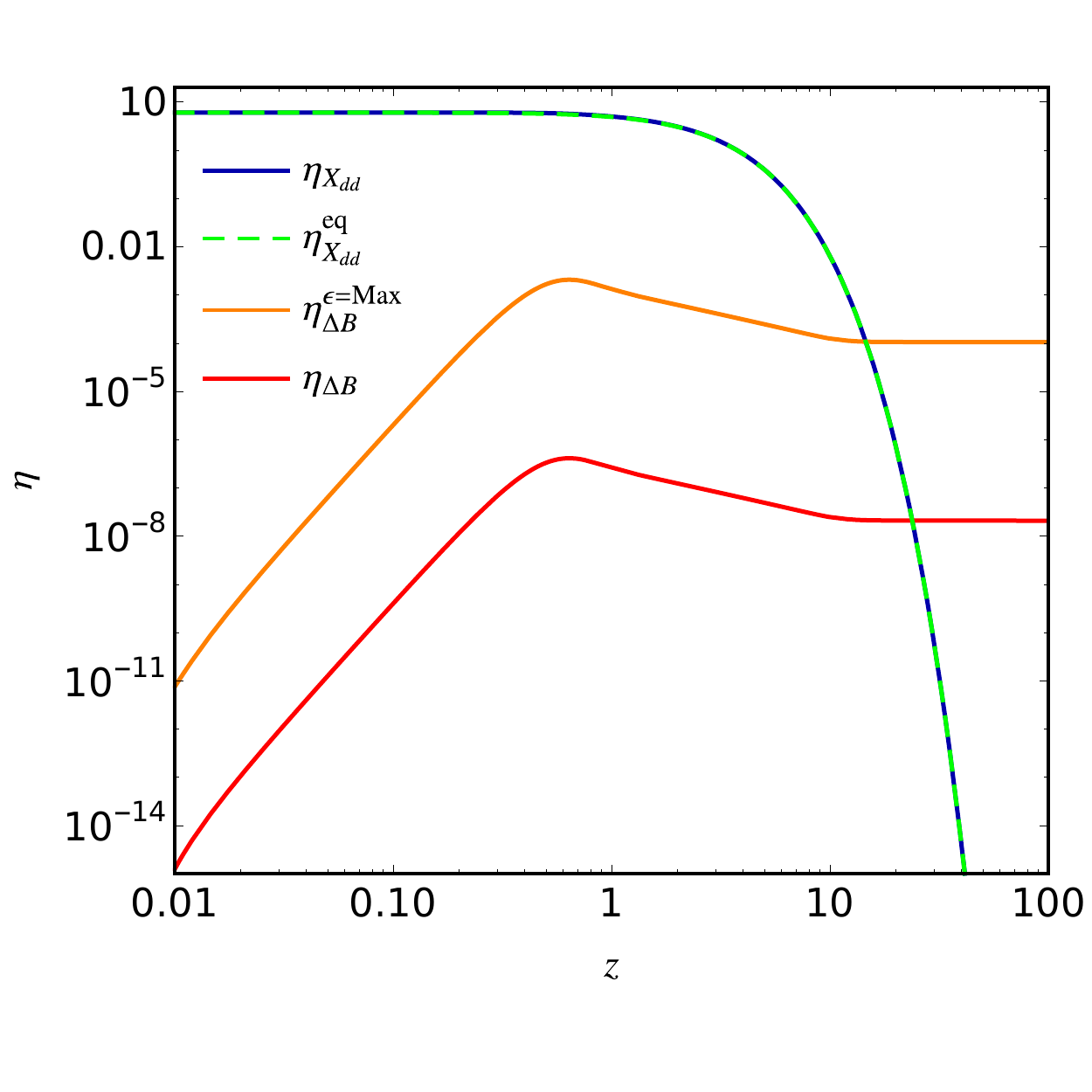}\hspace{-1mm}
\includegraphics[height=7.8cm]{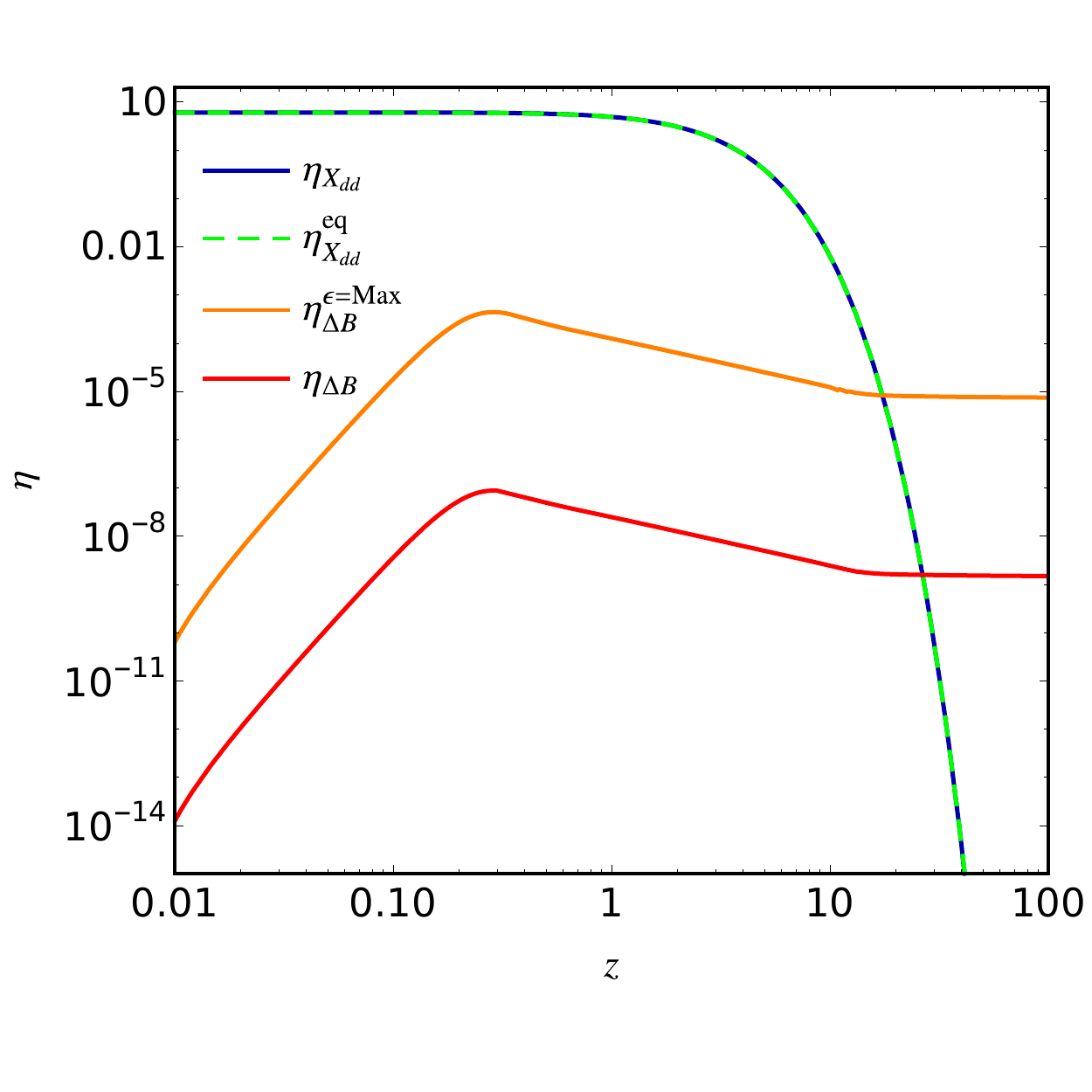}\\[-10mm]
\includegraphics[height=7.8cm]{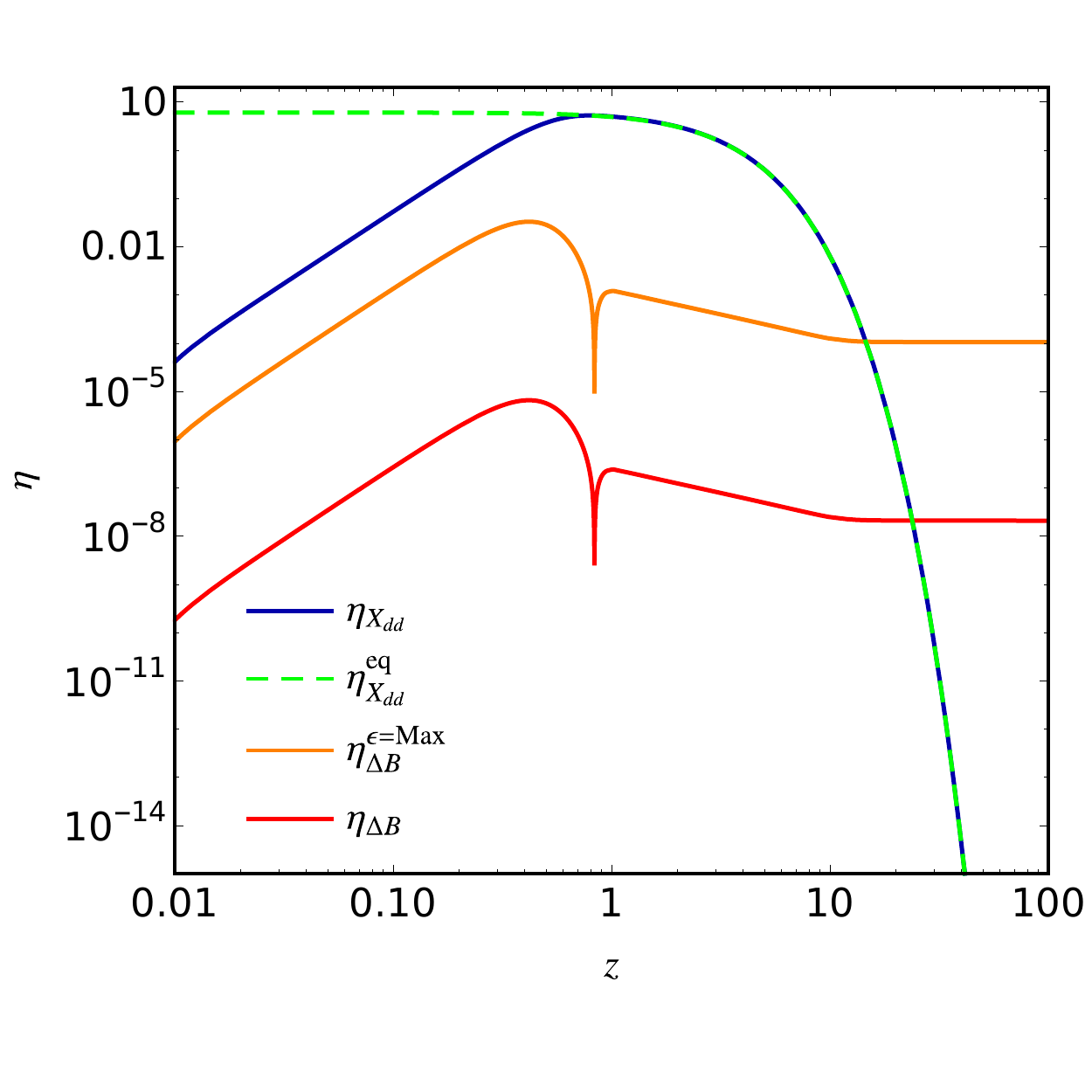}\hspace{-1mm}
\includegraphics[height=7.8cm]{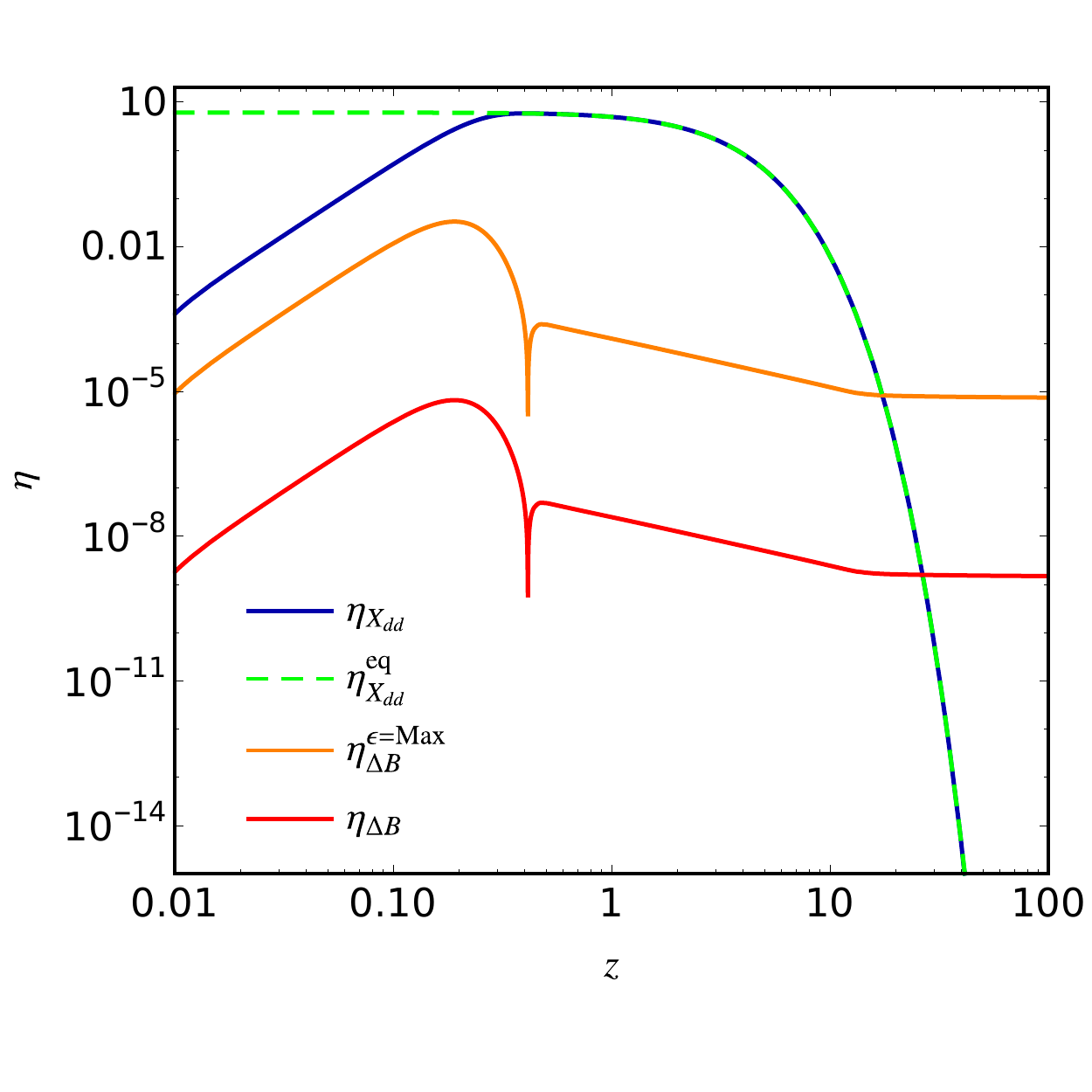}
\caption{{\bf High-scale scenario: }Evolution of baryon asymmetry $\eta_{\Delta B}$ with respect to $z=m_{X_{dd}}/T$ with $f_{ud}=f_{dd}=0.05$, $\lambda v'=1.5 m_{X_{dd}}$, and $m_{X_{ud}}=5$ TeV, for both the maximal value of $\epsilon$ (orange line) and the value $\epsilon\approx 4.0\times 10^{-4}$ given by Eq.~\eqref{eq:eps2} with $x=0.2$ (red line). The blue line shows the number density of $X_{dd}$ (normalised to the photon number density) and the dashed green line shows the equilibrium number density of $X_{dd}$. {\it Top row: }Number density evolution with $X_{dd}$ starting in equilibrium. {\it Bottom row: }No initial abundance of $X_{dd}$. {\it Left column:} $m_{X_{dd}}= 10^{14} \mathrm{ GeV}$. {\it Right column:} $m_{X_{ud}} = 10^{13}$ GeV. }
\label{fig:1Dplots_highscale}
\end{figure}
\begin{figure}[ht!]
	\centering
	\mbox{
		\includegraphics[height=7.7cm]{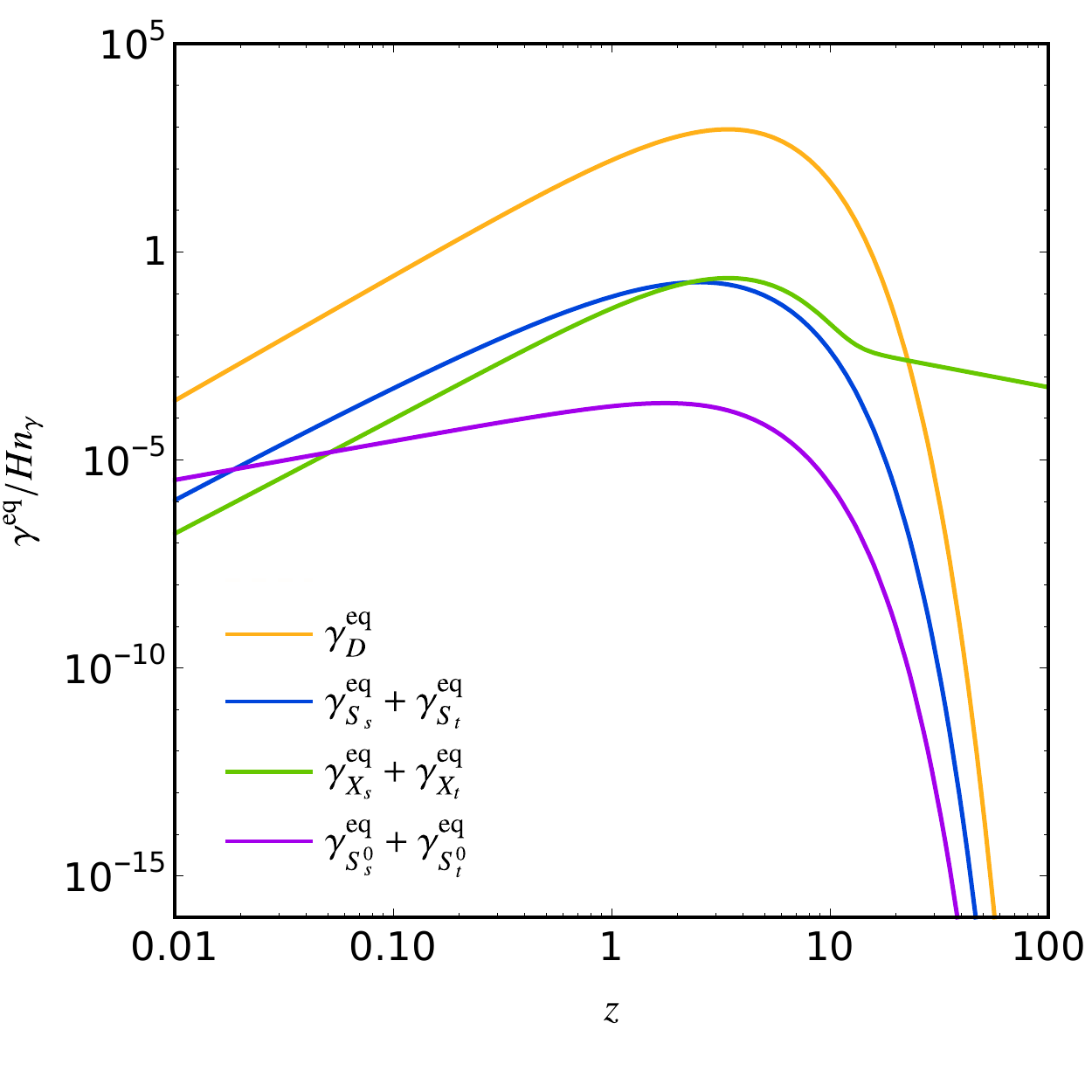}}
	\caption{Equilibrium decay and scattering rates with respect to $z=m_{X_{dd}}/T$ and $f_{ud}=f_{dd}=0.05$ for the high-scale scenario with $m_{X_{dd}}=10^{14}$~GeV, $m_{X_{ud}}=5$~TeV and $\lambda v' = 1.5m_{X_{dd}}$. The solid orange line shows the decay rate of $X_{dd}$, while the solid (dashed) blue, green, and purple lines show the scattering rates $S_s$ ($S_t$), $X_s$ ($X_t$), and $S_s^0$ ($S_t^0$) respectively. For the notation see Sec.~\ref{sec:boltzmann}. }
	\label{fig:rates_highscale}
\end{figure}

\begin{figure}[ht!]
	\centering
	\includegraphics[height=3.0in]{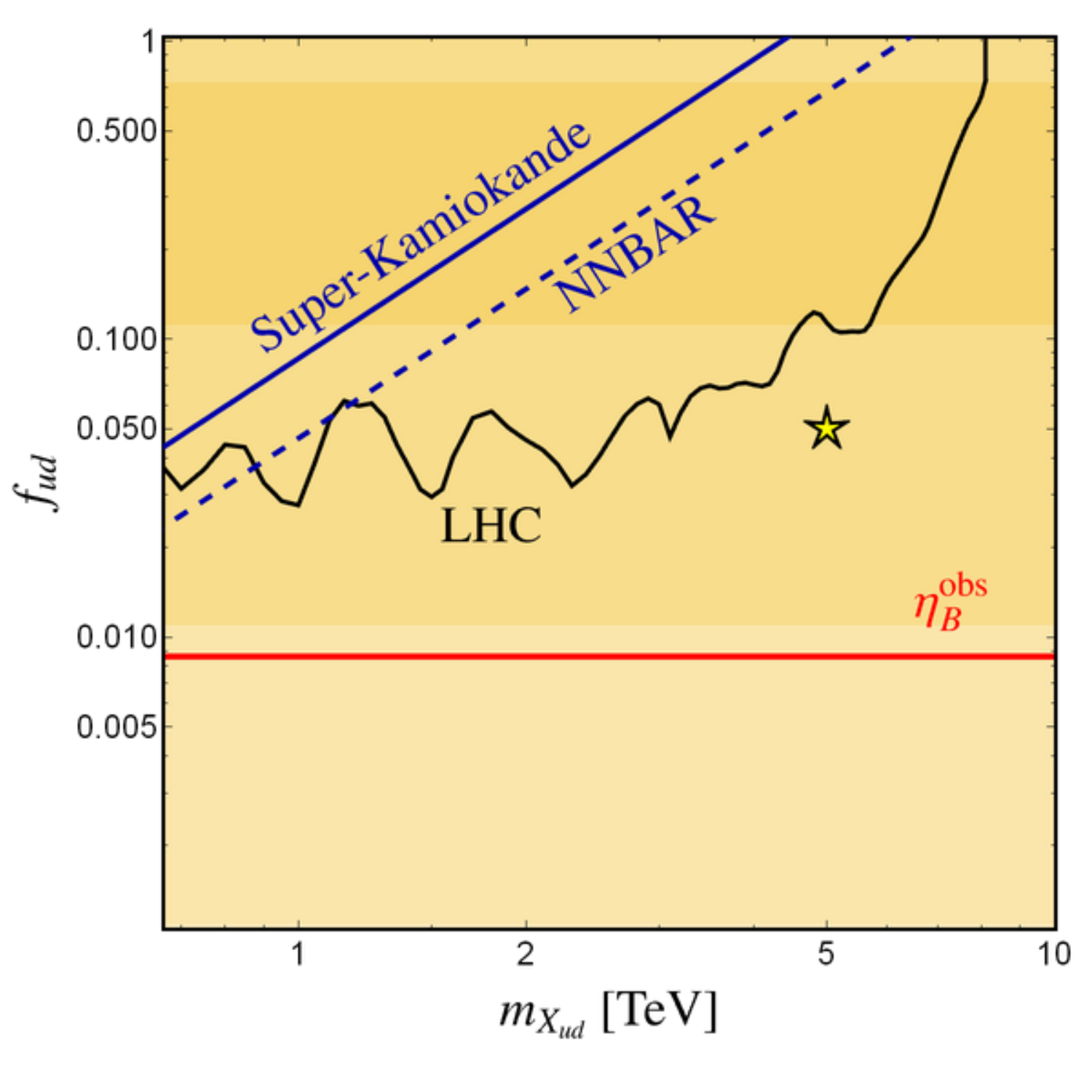}\hspace{-1mm}
	\includegraphics[height=3.0in]{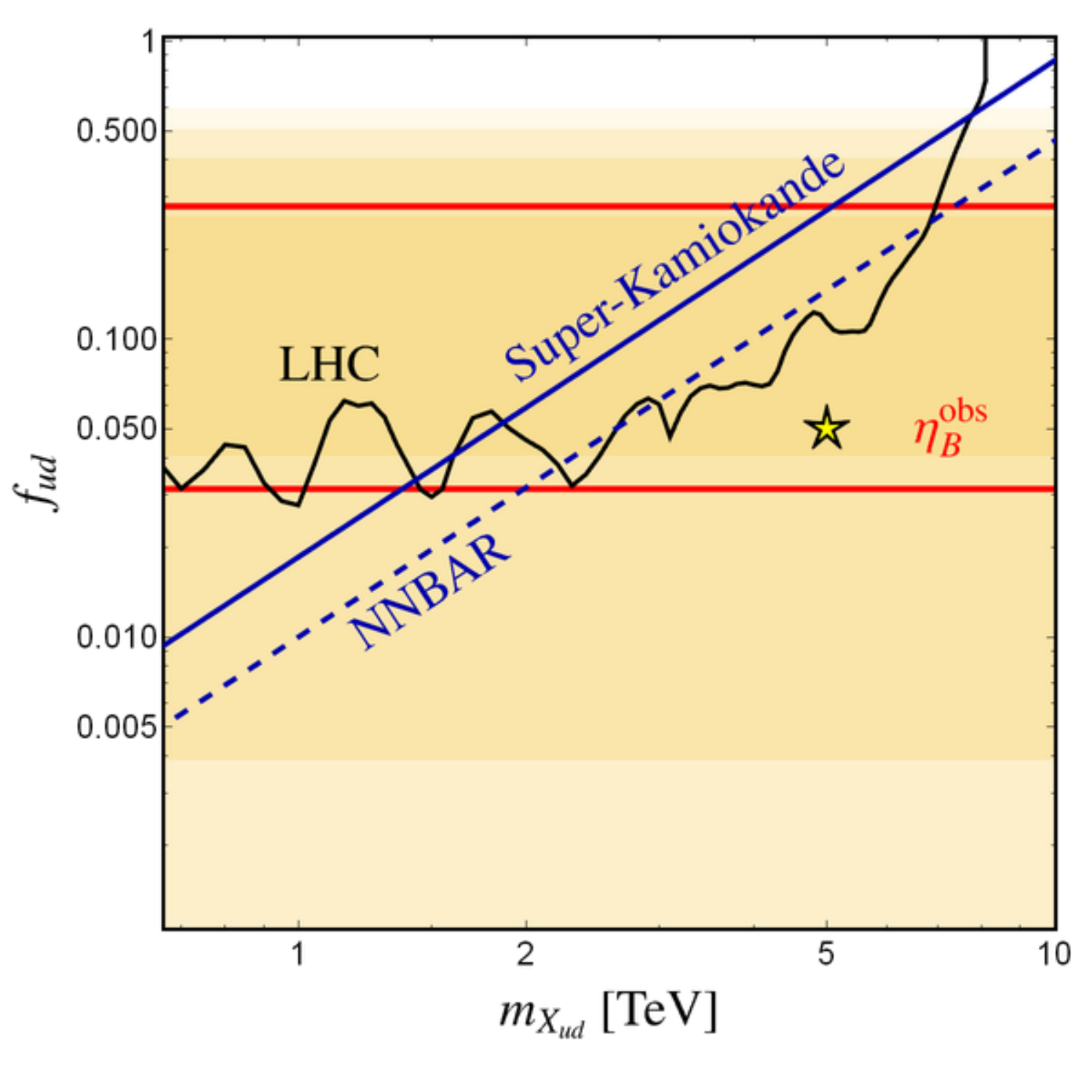}\\[-4mm]
	\includegraphics[height=3.0in]{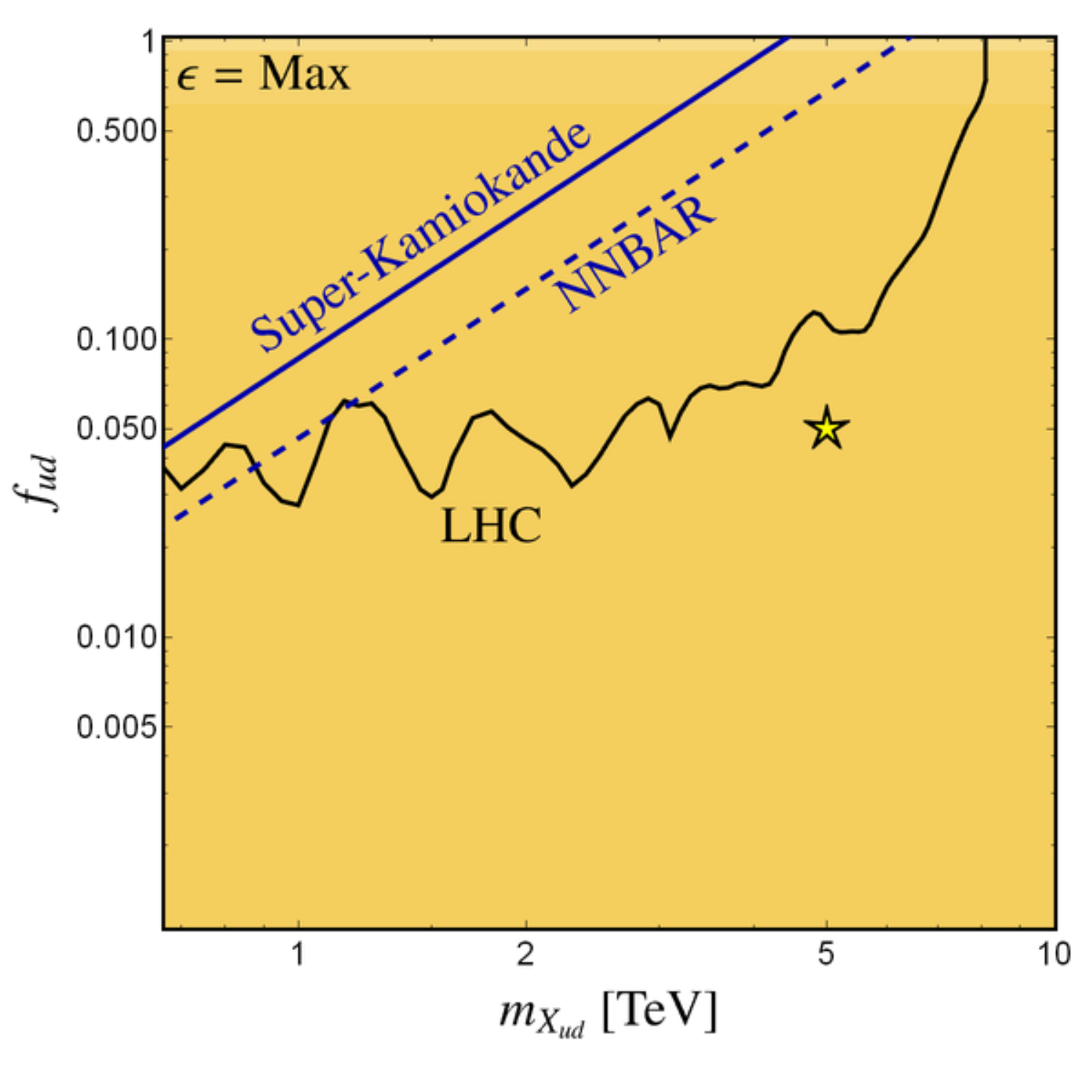}\hspace{-1mm}
	\includegraphics[height=3.0in]{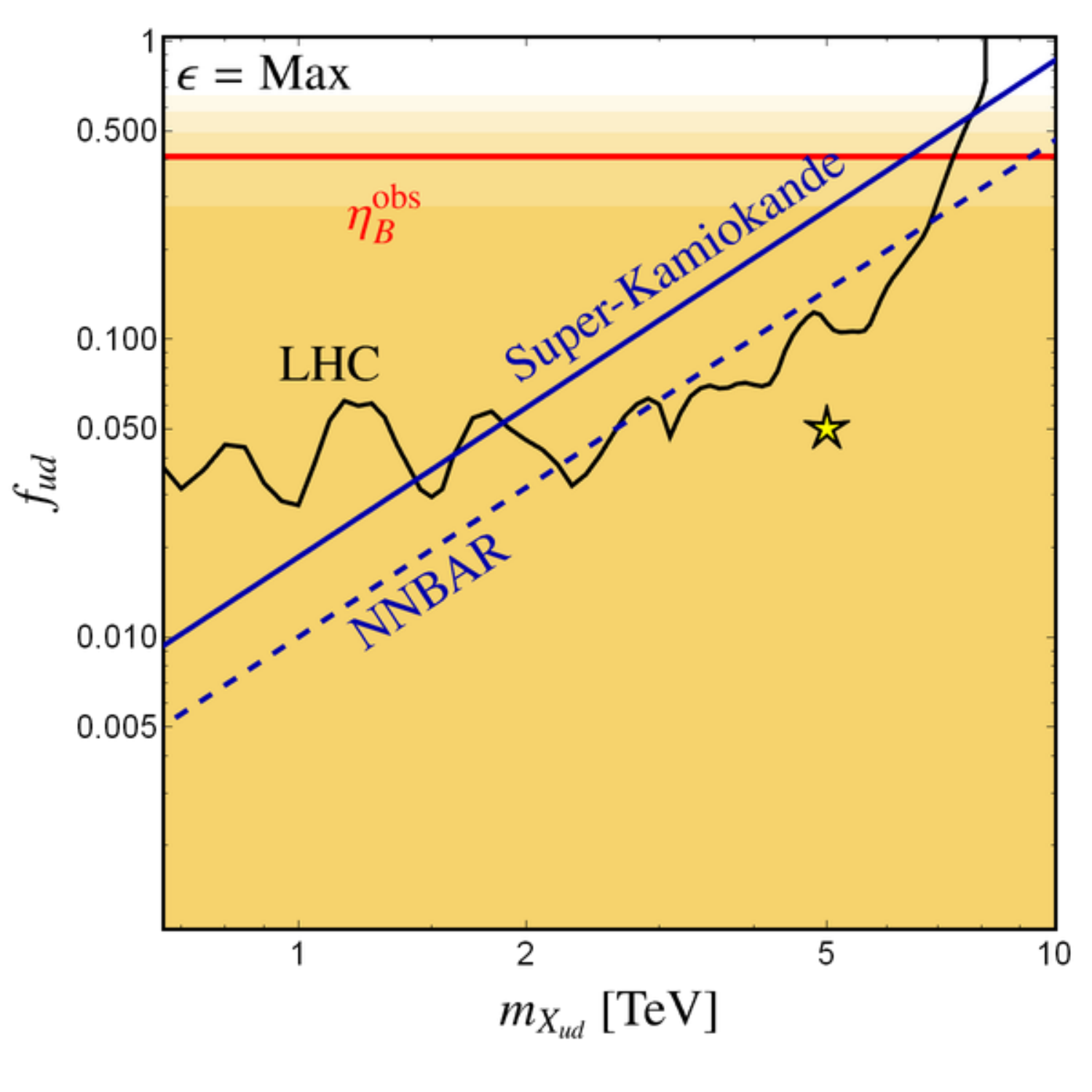}\\
	\includegraphics[scale=0.79]{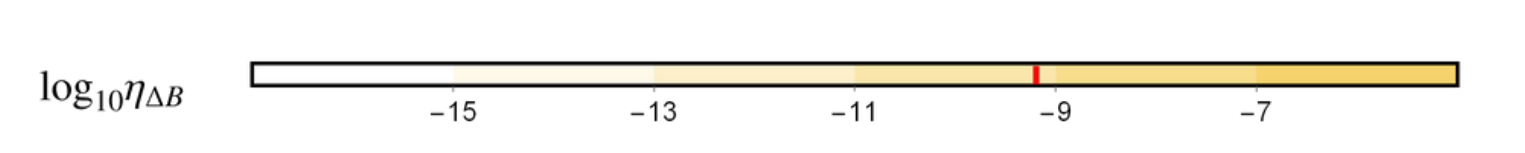}	
	\caption{{\bf High-scale scenario: }Yellow shading indicates the number density of baryon number~$\eta_{\Delta B}$ (normalised to photon number density) at the time of electroweak symmetry breaking in the $m_{X_{ud}}$\---$f_{ud}$ plane under the assumption of $f_{ud}=f_{dd}$ and $\lambda v'=1.5 m_{X_{dd}}$. \textit{ Top row:} The CP-violating parameter $\epsilon$ (as defined in Eq.~\eqref{eq:eps2}) is chosen with $x=0.2$ . \textit{Bottom row:} The maximum value $\epsilon = 2r$ is used. \textit{Left column:} $m_{X_{dd}}=10^{14}$ GeV. \textit{Right column:} $m_{X_{dd}}=10^{13}$ GeV. The red line marks the observed value for the baryon asymmetry $\eta_{B}^\text{obs}=6.2\times 10^{-10}$, the blue lines correspond to exclusion limits from $\nnbar$~oscillation set by Super Kamiokande (solid) and future exclusion limits set by NNBAR (dashed). The black lines correspond to the exclusion limit from dijet production at the LHC based on CMS-searches with $\sqrt{s}=13$ TeV and $36$ fb$^{-1}$ integrated luminosity.}
	\label{fig:highscale}
\end{figure}

{\bf High-scale scenario:} In the high-scale baryogenesis scenario, the $X_{dd}$ diquark has a mass $\mathcal{O}(10^{13\--14})$ GeV, while the mass of $X_{ud}$ is around $\mathcal{O}(10^{3\--4})$ GeV. From a theoretical perspective, the large hierarchy between the masses of $X_{dd}$ and $X_{dd}$ is naturally motivated by gauge coupling unification in the context of a GUT embedding of the model as discussed in App.~\ref{subs:UV}. Having a mass near the GUT scale, $X_{dd}$ would remain inaccessible at the direct collider searches; however, conspiring together with a TeV scale $X_{ud}$ it can still lead to an observable rate of $\nnbar$ oscillations, therefore making the scenario phenomenologically exciting. For instance, taking $\lambda$ $\approx$ $\mathcal{O}(1)\times (m_{X_{dd}})/v'$ and $f_{ud,dd}$ $\approx$ $\mathcal{O}(0.1\--1)$ together with the above discussed combination of masses, one finds that the $\nnbar$~oscillation rate is very close to the current experimental bound. On the other hand, since $m_{X_{ud}}$ is $\mathcal{O}(10^{3\--4})$ GeV, it can potentially be directly produced at the LHC and future colliders providing complementarity between collider searches and $\nnbar$~oscillation in constraining parts of the parameter space for a successful high-scale baryogenesis scenario.

For illustration, we present an example of the typical dynamics of the baryon asymmetry generation for the high-scale baryogenesis scenario in Fig.~\ref{fig:1Dplots_highscale}. We show the evolution of $\eta_{\Delta B}$ with respect to $z$, for the benchmark choices $f_{ud}=f_{dd}=0.05$, $m_{X_{ud}}=5$ TeV and $\lambda v'=1.5 m_{X_{dd}}$. The plots in the top row correspond to the case of zero initial abundance of $X_{dd}$, the plots in bottom row corresponds to the case of an initial thermal abundance of $X_{dd}$. The plots in the left and right columns correspond to the choices $m_{X_{dd}}=10^{14}$ GeV and $m_{X_{dd}}=10^{13}$ GeV, respectively. Furthermore, a final baryon asymmetry greater than the one observed in the Universe is achieved for a maximal CP violation $\epsilon=2 r$, as well as for the case where $\epsilon$ is given by Eq.~\eqref{eq:eps2} with $x=0.2$. The evolution that occurs for $m_{X_{dd}}= 10^{14}$ GeV (left column) and $m_{X_{dd}}=10^{13}$ GeV (right column) are similar, with the difference that the asymmetry generation starts earlier and gets more severely washed out in the latter case, since both decay rates and washout processes are less suppressed with respect to the Hubble rate. Whether the initial number density of $X_{dd}$ is in equilibrium or not has little effect on the final baryon asymmetry, as can be seen by comparing the top and bottom plots in Fig.~\ref{fig:1Dplots_highscale}. 

To illustrate the evolution of the relevant decay and scattering processes leading to the evolution of the baryon asymmetry with $z$ in the high-scale scenario, we show Fig.~\ref{fig:rates_highscale} the decay rates and different scattering rates for the same choice of parameters as in the plots in the left column of Fig.~\ref{fig:1Dplots_highscale}. We note that in the high-scale scenario, quark and $X_{ud}$ mediated washout processes $S_{s(t)}^{(0)}$ are subdominant and the $X_{dd}$ mediated $X_{s(t)}$ processes provide the most dominant washout for $z\gtrsim 10$. This can be understood as follows. The quark and $X_{ud}$ mediated scatterings $S_{s(t)}^{(0)}$ contain heavy $X_{dd}$ as one of its external legs and leads to large Boltzmann (or phase space) suppression of these scattering rates for $T\ll m_{X_{dd}}$. On the other hand, the $X_{dd}$ mediated $X_{s(t)}$ scatterings are much less-severely suppressed due to $X_{dd}$ mass in the propagator for $T\ll m_{X_{dd}}$. One particular caveat to the above discussion is the situation when $f_{ud}>f_{dd}$ and the coupling $f_{ud}$ is large, $\Op(1)$. In such a case the washout processes $S_t$, $S_s$, $S_t^0$ and $S_s^0$ processes can still be quite effective for large $z$ and can potentially be comparable to $X_{s(t)}$. As a benchmark, as long as the coupling $f_{dd}$ does not exceed values of $\mathcal{O}(0.1)$ the dominant washout processes $X_s$ and $X_t$ remain modest leading to the generation of a sizeable final baryon asymmetry. With the washout processes being modest in the high-scale scenario set-up for $f_{dd}\lesssim \mathcal{O}(0.1)$, a baryon asymmetry close to the observed value can be found over a wide parameter range of phenomenological interest, as shown in Fig.~\ref{fig:highscale}. This statement holds for both cases, a vanishing initial $X_{dd}$ number density, and a non-zero initial $X_{dd}$ number density in equilibrium.

In Fig.~\ref{fig:highscale}, we show the parameter space of the high-scale baryogenesis scenario in the $m_{X_{ud}}$\---$f_{ud}$ plane with $f_{ud}=f_{dd}$. The generated final baryon asymmetry is shown in yellow with deeper shades corresponding to higher yields. The red lines indicate the observed value\footnote{The bottom left plot in Fig.~\ref{fig:highscale} contains no red line, since the whole parameter space that is shown leads to an asymmetry greater than the observed one.}. Note that in Fig.~\ref{fig:highscale}, we take the diquark couplings $f_{ud}$ and $f_{dd}$ to be equal, ranging from $f_{ud}=f_{dd}=10^{-3}$ to unity. In general $f_{ud}$ and $f_{dd}$ couplings could be treated as independent parameters; however, we note that for a large hierarchy between $f_{ud}$ and $f_{dd}$, the connection of baryogenesis to the relevant experimental observables gets weakened. This is because on the one hand the LHC direct searches for $X_{ud}$ and $\nnbar$~oscillation rate largely depend on a high value for the coupling $f_{ud}$, on the other hand the baryon asymmetry depends to a large degree on $f_{dd}$. Therefore, we find that fixing the ratio of $f_{ud}$ and $f_{dd}$ to unity in Fig.~\ref{fig:highscale} illustrates the interplay between the final baryon asymmetry and the reach of experiments more clearly. While in Fig.~\ref{fig:highscale} the tri-linear scalar diquark coupling is taken to be $\lambda v'=1.5m_{X_{dd}}$, we have found that a lower or higher value can also impact the final baryon asymmetry. A lower value for $\lambda v'$ would generally result in a larger baryon asymmetry at the cost of lowering the experimental reach of the $\nnbar$ oscillations, in contrast to a higher value which can run into problems with the constraints from colour preserving vacua (see Sec.~\ref{subs:cpv}). However, if we relax the constraints from colour preservation of the vacuum, then a higher value of the tri-linear coupling $\lambda v'$ would lead to an increased decay and washout rate, leading to an asymmetry generation at later times and a more effective washout, resulting in a lower final asymmetry. In all four plots of Fig.~\ref{fig:highscale}, the baryon asymmetry is generally lower for very high values of $f_{ud}$  (i.e. when this coupling is $\mathcal{O}(1)$) as the consequence of an increased washout from $X_s$ and $X_t$. This effect is more pronounced in the right plots, as $m_{X_{dd}}$ is an order of magnitude lower as compared to the left plots.  For lower values $f_{ud}$ $\lesssim$ $\mathcal{O}(0.1)$, the asymmetry again gets smaller for smaller couplings in the top plots of Fig.~\ref{fig:highscale}, while it remains rather constant in the bottom plots. This effect is due to the CP-asymmetry parameter $\epsilon$ being smaller for small couplings in the top plots\footnote{The CP-asymmetry parameter $\epsilon$ does not depend on $f_{ud}$, but rather $f_{dd}$. However, since $f_{ud} = f_{dd}$ in Fig.~\ref{fig:highscale}, $\epsilon$ changes along the vertical direction in the top plots by the variation of $f_{ud}$.}. This is in contrast to the plots in the bottom row, where the maximal CP asymmetry $\epsilon=2 r$ is used.

The reach of $\nnbar$ oscillations and collider experiments, which provide important constraints for the high-scale baryogenesis parameter space, do however, depend on the mass scales $m_{X_{dd}}$ and $m_{X_{ud}}$. As can be seen in the left column of Fig.~\ref{fig:highscale}, for $m_{X_{dd}}=10^{14}$ GeV, the LHC proves to be a better probe than $\nnbar$~oscillation experiments in parts of the parameter space. In contrast, for a slightly smaller mass, $m_{X_{dd}}=10^{13}$ GeV, (Fig.~\ref{fig:highscale}, right column) $\nnbar$ oscillations almost completely dominate in sensitivity. Furthermore, considering that both the LHC and $\nnbar$~oscillation rate depend non-trivially on the smallness of $m_{X_{ud}}$ and the generated baryon asymmetry is largely insensitive to small changes in $m_{X_{ud}}$, a large part of the parameter space in which baryogenesis is successful (e.g. when $m_{X_{ud}}\gg\mathcal{O}(10)$ TeV) remains unreachable by any current or near future experiments.

Note that the final baryon asymmetry shown in Fig.~\ref{fig:highscale} was calculated under the assumption that $X_{dd}$ and $X_{ud}$ dominantly decay into to third generation quarks, while the $\nnbar$ oscillation requires $X_{dd}$ and $X_{ud}$ couplings to first generation quarks. In general both, the SM quark Yukawa coupling matrix and the $X_{dd}$ (and $X_{ud}$) Yukawa coupling matrix need not to be diagonal in the same basis. Moreover, the inverse decays and washout processes can lead to the generation of an asymmetry in all flavours even if the initial asymmetry is generated in only one of the flavours. If an initial asymmetry is generated in a the third flavour of quarks together with an equal and opposite asymmetry in an unflavoured $X_{ud}$ (comparable to the Higgs asymmetry generated in the leptogenesis~\cite{Dev:2017trv}), then through inverse decay or washout processes the $X_{ud}$ asymmetry will induce asymmetry in all three quark flavours realising a thermal contact among flavours such that asymmetry in one flavour induces an asymmetry in the other flavours. Therefore, the asymmetry that will "leak" into the first generation will be subject to washout effects from the first generation Yukawa couplings of  $X_{dd}$ and $X_{ud}$ inducing $\nnbar$ oscillations. In a full two-flavour treatment of the high-scale scenario we expect that the first generation interactions relevant for $\nnbar$ oscillation operators will therefore at least partially washout the asymmetry along the perpendicular direction to the third generation quarks.

To this end, the synergy of LHC searches for a $tb$ vs.\ dijet final state searches with the $\nnbar$ oscillation will provide valuable hint towards the underlying flavour dynamics of the high-scale scenario. A comprehensive analysis of these interesting flavour effects is beyond the scope of the present work and will be addressed in the future.

In summary, the high-scale scenario is a case where baryogenesis can be successful, while at the same time, the baryon-number-violating mechanism can be potentially discovered by $\nnbar$~oscillation experiments. Moreover, the $X_{ud}$ diquark is close to the current reach of the LHC. In case a signal is observed in either of these experiments, the other experiment can be used to set further constraints, perhaps rejecting or confirming this mechanism as the one responsible for baryogenesis.\\

\begin{figure}[htb!]
	\centering
	\mbox{
		\includegraphics[height=7.7cm]{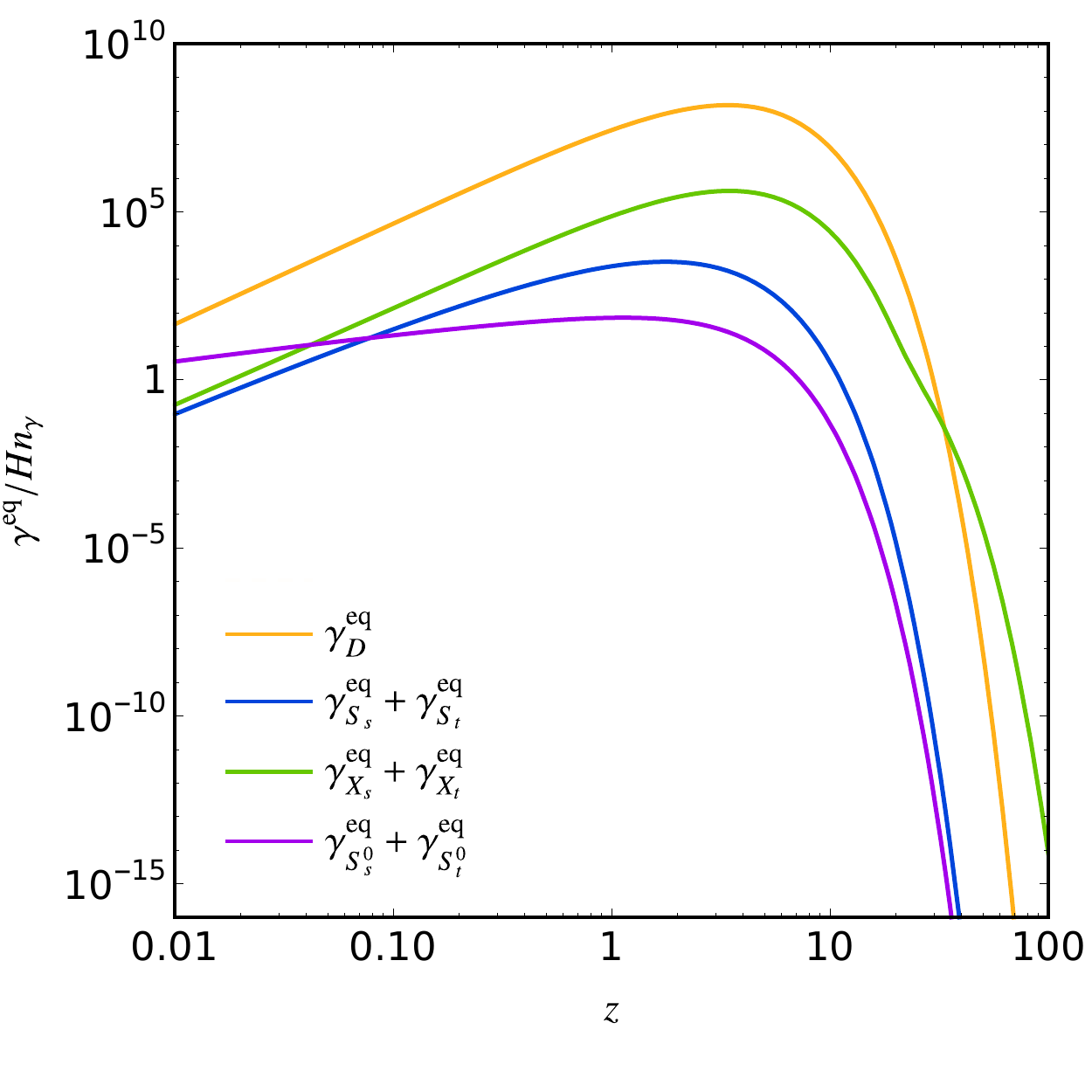}}
	\caption{Equilibrium decay and scattering rates with respect to $z=m_{X_{dd}}/T$ and $f_{ud}=f_{dd}=0.05$ for the low-scale scenario with $m_{X_{dd}}=3\times10^8$~GeV, $m_{X_{ud}}=10^8$~GeV, and $\lambda v' = 1.2 m_{X_{dd}}$. The solid orange line shows the decay rate of $X_{dd}$, while the solid (dashed) blue, green, and purple lines show the scattering rates $S_s$ ($S_t$), $X_s$ ($X_t$), and $S_s^0$ ($S_t^0$) respectively. }
	\label{fig:rates_lowscale}
\end{figure}

%%%%%%%%%%%%%%%%%%%%%%%%%%%%%%%%%%%%%%%%%%%%%%%%%%%%%%%%%%%%%%%%%%%%%%%%%%%%%%%%%%%%%%%%%%%%%%%%%%%%%%%%%%%%%%%%%%%%%%%%%%%%%%%%%%%%%%%%%%%%%%%%%%%%%%%%%%%%%%%%%%%%%%%%%%%%%
{\bf Low-scale scenario:} In the low-scale baryogenesis scenario, we consider the case where the diquark $X_{dd}$ mass is a few times the mass of $X_{ud}$, with $m_{X_{ud}}$ varying in the range $\mathcal{O}(10^{3\-- 9})$ GeV. The phenomenological choice of a mild hierarchy between $m_{X_{dd}}$ and $m_{X_{ud}}$ is primarily driven by the constraints from colour preserving vacua as discussed in Sec.~\ref{sec:constraints}.  
\begin{figure}[ht!]
	\centering
	\mbox{
		\includegraphics[height=7.8cm]{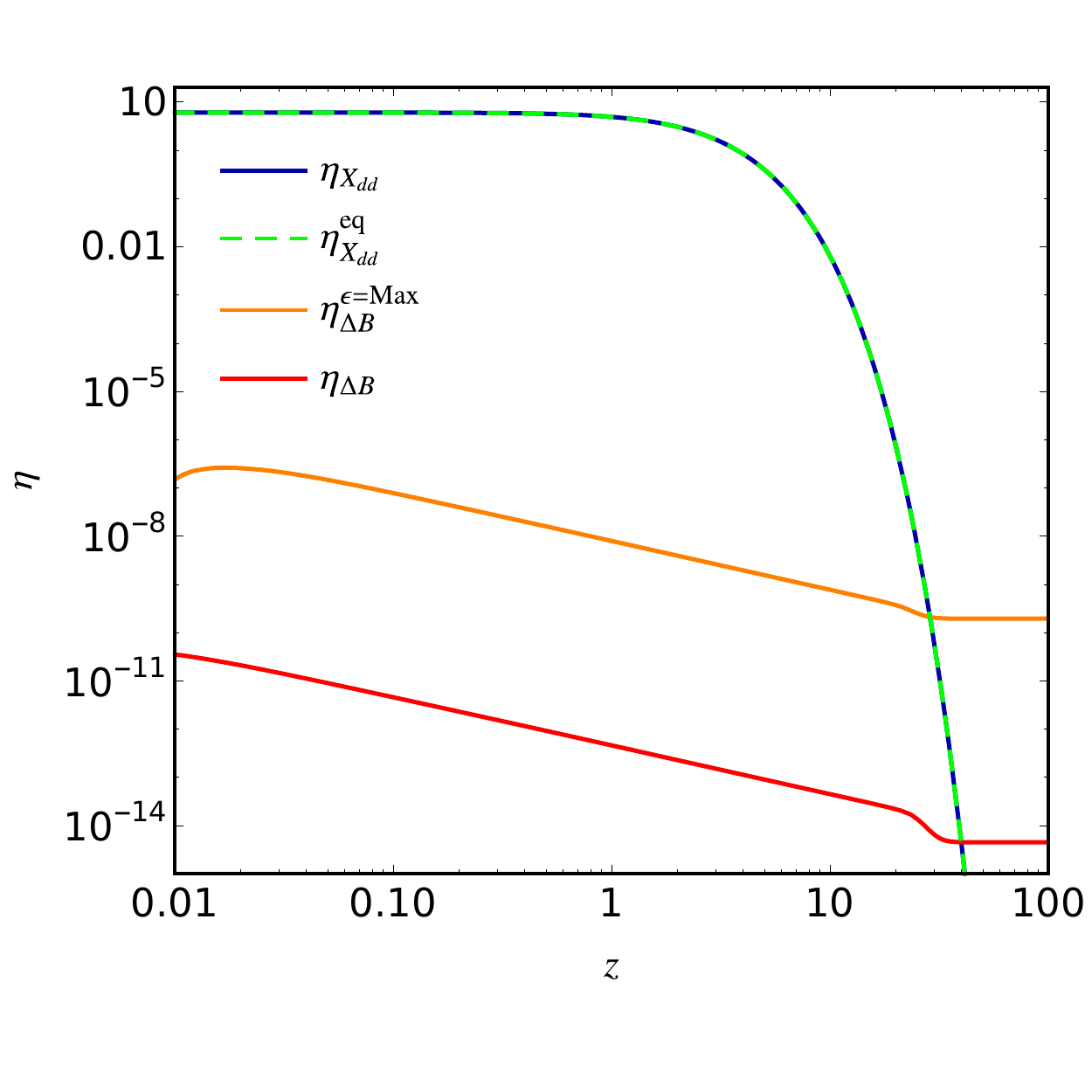}\hspace{-1mm}
		\includegraphics[height=7.8cm]{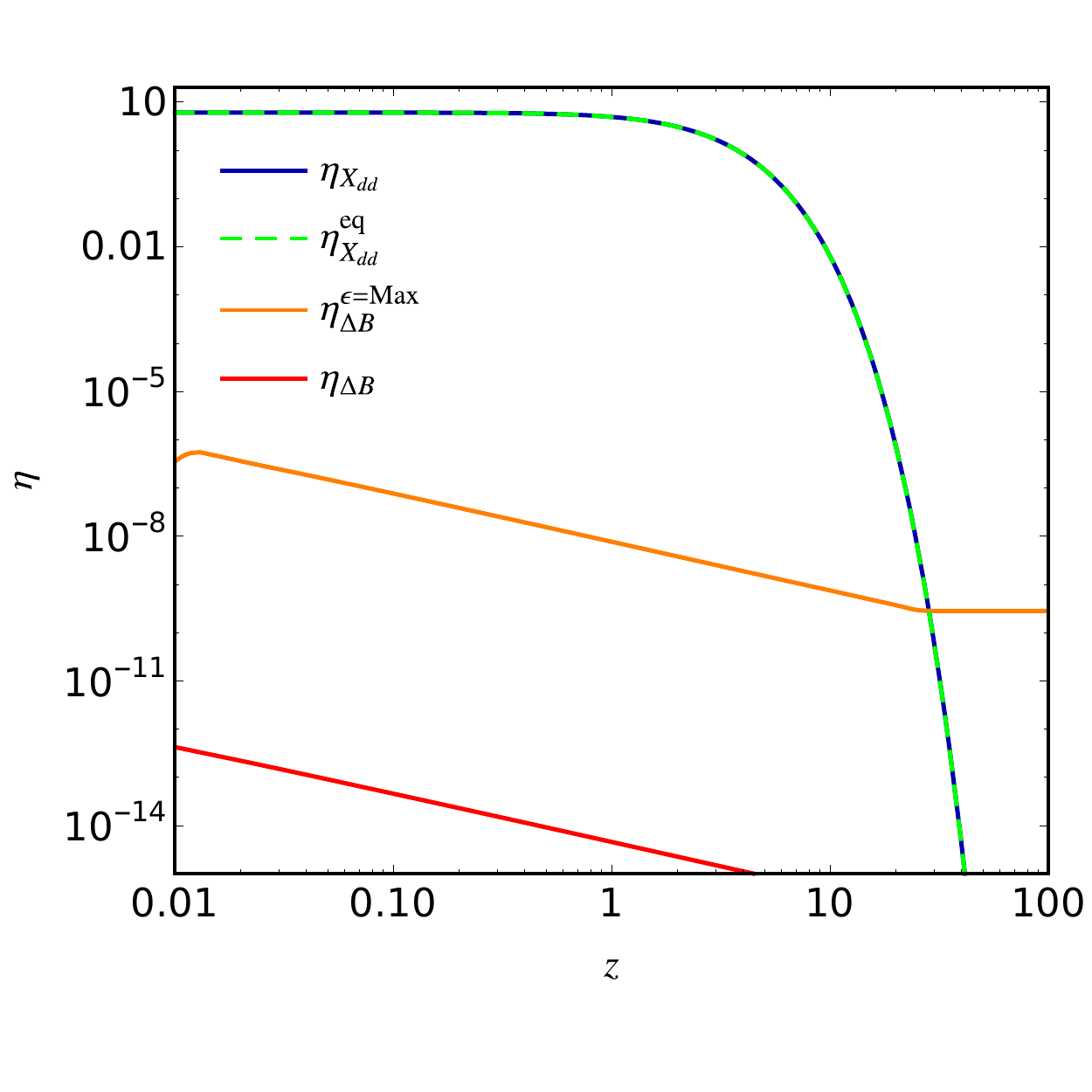}}
	\caption{{\bf Low-scale scenario: } Evolution of baryon asymmetry $\eta_{\Delta B}$ (red lines) for the maximal value of $\epsilon$ with respect to $z=m_{X_{dd}}/T$. We fix $\lambda v'=1.2 m_{X_{dd}}$, $m_{X_{ud}} = 10^8$ GeV, and the diquark couplings are chosen as follows: \textit{Left: }$f_{ud}=f_{dd}=0.05$. \textit{Right:} $f_{ud}= 10f_{dd}=0.05$. The orange solid line indicates the baryon asymmetry for $\epsilon=2 r$, and the red solid line  assumes $\epsilon\approx 3.6\times 10^{-3}$ (\textit{left}) and $\epsilon\approx 3.6\times 10^{-5}$ (\textit{right}) as given by Eq.~\ref{eq:eps2} with $x=0.2$. Blue solid and green dashed lines depict the number density and equilibrium number density of $X_{dd}$ (normalised to the photon number density), respectively.}
	\label{fig:1Dplots_midscale}
\end{figure}
To illustrate the typical dynamics of the baryon asymmetry generation in the low-scale baryogenesis scenario, in Fig.~\ref{fig:1Dplots_midscale}, we show the evolution of $\eta_{\Delta B}$ with $z$ for two different coupling hierarchies $f_{ud}=f_{dd}=0.05$ (left figure) and $f_{ud}=10 f_{dd}=0.05$ (right figure). The rest of the parameters are fixed to the benchmark values $m_{X_{ud}}=10^8$ GeV, $m_{X_{dd}}= 3 m_{X_{ud}}$ and $\lambda v'=1.2 m_{X_{dd}}$. The evolution of the baryon asymmetry for the maximum $\epsilon$ case (orange curve) as well as $\epsilon$ corresponding to Eq.~\eqref{eq:eps2} with $x=0.2$ (red curve) are shown in each plot. An illustration of the evolution of the decay rates and scattering rates in the low-scale baryogenesis scenario is shown in Fig.~\ref{fig:rates_lowscale}, using the same parameters as in the the left plot in Fig.~\ref{fig:1Dplots_midscale}. Similar to the high-scale case (cf. Fig.~\ref{fig:rates_highscale}), Fig.~\ref{fig:rates_lowscale} shows for the low-scale case the overall dominance of the decay rate until some high $z$ when the $X_{s(t)}$ washout takes over.

\begin{figure}[H]
	\centering
	\includegraphics[height=6.2in]{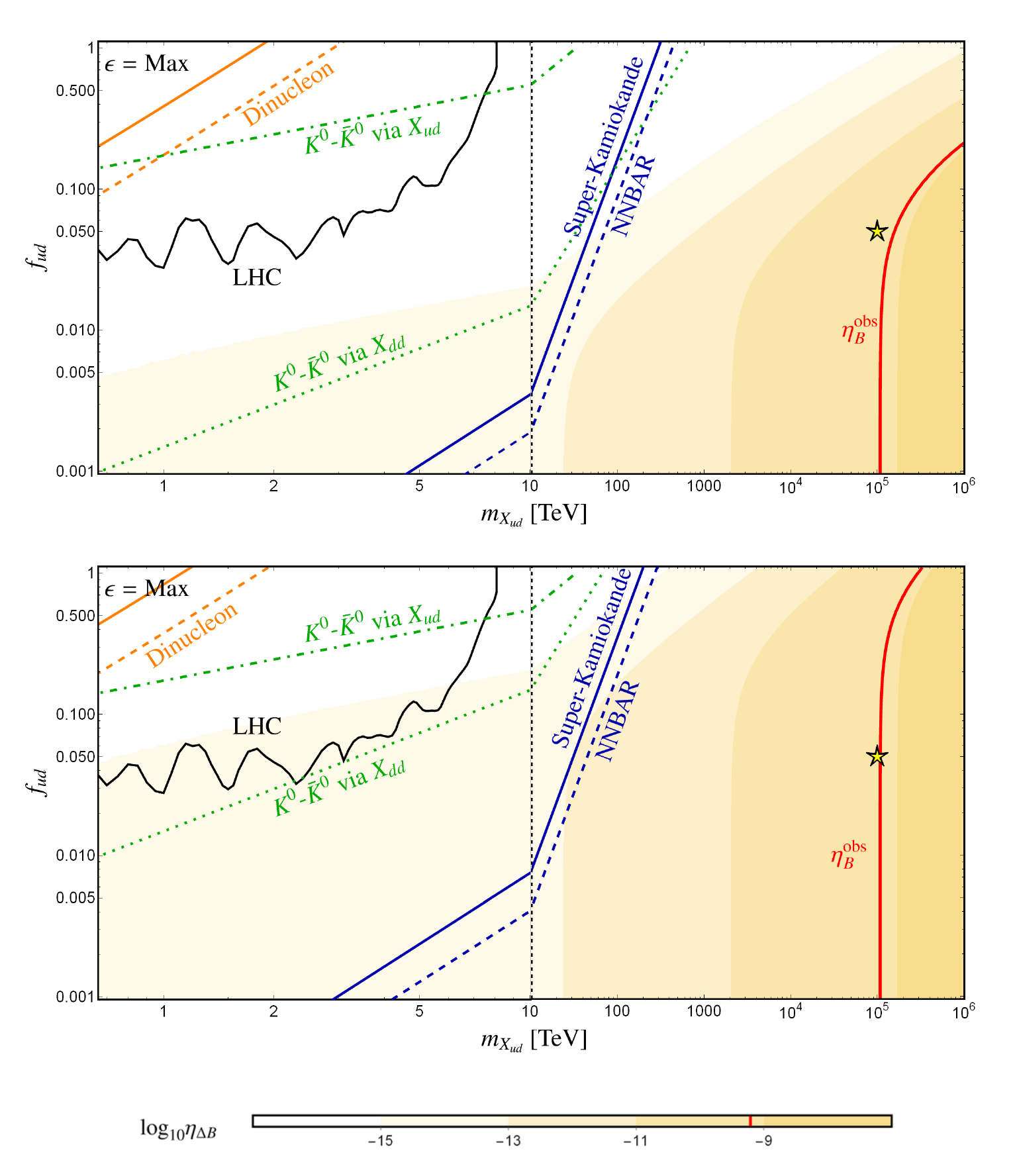}
	\caption{{\bf Low-scale scenario: } Yellow shading indicates the final baryon asymmetry $\eta_{\Delta B}$ at the time of electroweak symmetry breaking in the $m_{X_{ud}}$\---$f_{ud}$ plane. The diquark couplings are chosen as follows: \textit{Top: }$f_{ud}=f_{dd}$. \textit{Bottom:} $f_{ud}= 10f_{dd}$. In both plots, we chose $\lambda v'=1.2 m_{X_{dd}}$, $m_{X_{dd}}=3 m_{X_{ud}}$, and the maximal CP asymmetry $\epsilon=2 r$ is used. The red line indicates the observed value of the baryon asymmetry $\eta_{B}^\text{obs}$, while the orange, green, black, and blue lines correspond to experimental constraints coming from dinucleon decay, meson oscillations, the LHC, and $\nnbar$ oscillations respectively. For the dinucleon and $\nnbar$~oscillation constraints, solid lines represent current experimental bounds, while dashed lines show projected future sensitivities. For the meson oscillation constraints, dotted and dot dashed lines correspond to tree-level oscillation via a $X_{dd}$ mediator, and box-diagram oscillation involving an $X_{ud}$ diquark, respectively (see Sec.~\ref{sec:constraints} for details). For the LHC limit CMS data with $\sqrt{s}=13$ TeV and 36 fb$^{-1}$ integrated luminosity was used \cite{Sirunyan:2018xlo}.
	}
	\label{fig:midscale}
\end{figure}

    In Fig.~\ref{fig:midscale}, we present the parameter space for low-scale baryogenesis scenario in the $m_{X_{ud}}$\---$f_{ud}$ plane for the maximum value of $\epsilon$, being  $\epsilon=2r$. The final baryon asymmetry is shown in yellow contours, where the low\--$m_{X_{ud}}$ end of the plots is stretched out in order to show the complementary sensitivity of the dinucleon decay, neutral kaon oscillation, and collider searches with respect to $\nnbar$ oscillations. We show two different coupling hierarchies: $f_{ud}=f_{dd}$ (Fig.~\ref{fig:midscale} \textit{top}) and $f_{ud}= 10f_{dd}$ (Fig.~\ref{fig:midscale} \textit{bottom}). The tri-linear coupling is chosen as $\lambda v'=1.2 m_{X_{ud}}$, and the mass ratio between $X_{dd}$ and $X_{ud}$ is taken to be $m_{X_{dd}}/m_{X_{ud}}=3$. Our findings suggest that in the low-scale baryogenesis scenario, even for the case of maximal CP violation, successful baryogenesis can only occur when the mass scales $m_{X_{dd}}\sim m_{X_{ud}}$ are sufficiently heavier (at least $\mathcal{O}(10^{8})$ GeV) than the electroweak scale such that the washout processes remain largely subdominant as compared to the baryon asymmetry generation through $X_{dd}$ decays. Since in this scenario $m_{X_{dd}}$ is not very heavy as compared to $m_{X_{ud}}$, the quark and $X_{ud}$ mediated washout processes $S_{s(t)}^{(0)}$ do not receive any relative suppression with respect to  $X_{dd}$ mediated $X_{s(t)}$ processes, in contrast to the high-scale scenario where the heaviness of $X_{dd}$ leads to Boltzmann suppression of the washout processes $S_{s(t)}^{(0)}$ as compared to $X_{s(t)}$. For large $f_{ud}$ and $f_{dd}$ couplings and small $X_{dd} (X_{ud})$ masses, the washout processes become too strong to generate any sizeable asymmetry. In particular, for a fixed hierarchy of $X_{ud}$ and $X_{dd}$ masses (e.g. our benchmark choice $m_{X_{dd}} = 3 m_{X_{ud}}$)  the final baryon asymmetry gets reduced for smaller masses $m_{X_{ud}}$, as can be noticed in Fig.~\ref{fig:midscale}. This can be understood to be due to $X_{dd}$ being less massive (being related to $X_{ud}$ masses by a fixed hierarchy), thereby falling out-of-equilibrium at a later time when the washout rates are less suppressed with respect to the Hubble rate. Furthermore, a larger coupling $f_{ud}$ should lead to an increased washout via $X_s$ and $X_t$, and therefore a lower final baryon asymmetry; however, since the $X_s$ and $X_t$ washout processes also depend on the coupling $f_{dd}$, such an effect is noticeably less strong in the bottom plot in Fig.~\ref{fig:midscale}, in which $f_{dd}$ is an order of magnitude smaller as compared to the top plot. On the other hand, a higher value of $\lambda v'$ would (as in the high-scale scenario) lead to an increase in both, the rate of the decay for $X_{dd}$ as well as in the scattering rate for washout processes $S_{s(t)}$ and $X_{s(t)}$. While this leads to more asymmetry being generated at an earlier time, it finally would lead to a larger washout for large $z$ once the baryon asymmetry generation freezes out, leading to a smaller final baryon asymmetry.

From Fig.~\ref{fig:midscale}, it is very clear that if $m_{X_{dd}}\sim \mathcal{O}(\text{few}) m_{X_{ud}}$ lies below $\mathcal{O}(10^{8})$ GeV then the washout processes are too strong to generate any sizeable asymmetry, even if the CP violation is maximal. Therefore, an observation of $X_{dd}$ and $X_{ud}$ at the LHC together with a signal for $\nnbar$~oscillations would imply that the low-scale baryogenesis scenario is completely ruled out as a mechanism behind the observed baryon asymmetry. On the other hand, for $m_{X_{dd}}\sim \mathcal{O}(\text{few}) m_{X_{ud}}>\mathcal{O}(10^{8})$ GeV the low-scale baryogenesis mechanism remains a viable option, however, the small couplings required together with the mass scale make this part of the parameter space inaccessible to all current and near-future experimental efforts. In this region, all washout processes involving scatterings are small because of the Boltzmann suppression due to the heaviness of  $X_{dd}$ (and $X_{ud}$) and the smallness of the couplings. For instance, the dominant washout processes $X_s$ and $X_t$ are small due to the Boltzmann suppression coming from the two $X_{ud}$ in the outer legs. Therefore, an initially generated asymmetry due to the CP violating decay of $X_{dd}$ can survive. In the contrary, if the hierarchy between $m_{X_{dd}}$ and $m_{X_{ud}}$ is larger ($ m_{X_{dd}}/m_{X_{ud}}\gtrsim 4$), then the Boltzmann (or phase-space) suppression of the washout processes (e.g. $X_s$ and $X_t$), due to $X_{ud}$ in the external legs, get lifted making the washout very strong, and consequently the baryon asymmetry gets rapidly wiped out\footnote{Naively the Boltzmann suppression is independent of the mass hierarchy. However, in the low-scale scenario, the successful generation of a baryon asymmetry relies on the minimisation of the $X_{dd}$-mediated washout processes $X_s$ and $X_t$, which generally come into effect after the most active decay of $X_{dd}$, but before the temperature has dropped sufficiently low for the Boltzmann suppression from $X_{ud}$ to play a major role. The window of washout due to $X_s$ and $X_t$  is therefore proportional to the mass hierarchy between $X_{dd}$ and $X_{ud}$. This window can be seen as a small kink appearing in the red line in the bottom left of Fig.~\ref{fig:1Dplots_midscale}, where for a short period of time, the washout is prominent (this effect is also present, although less visibly so, in the orange line of the same figure, corresponding to maximal CP violation).}.

In case the CP-violating parameter $\epsilon$ would be given by Eq.~\eqref{eq:eps2} rather than being fixed to the maximum value as in Fig.~\ref{fig:midscale}, we find that the entire parameter space of the low-scale baryogenesis scenario cannot reach the observed baryon asymmetry of the Universe. The asymmetry is particularly small for lower values of the coupling $f_{dd}$, as a consequence of less asymmetry being initially generated due to $\epsilon$ being smaller. Apart from this effect, $\epsilon$ given by Eq.~\eqref{eq:eps2} shows a similar behavior as in Fig.~\ref{fig:midscale} for the low-scale scenario with respect to a decreased asymmetry for lower masses $m_{X_{ud}}$. This is caused by the washout rates being larger than the Hubble rate for low masses of ${X_{dd}}$.

To summarise, if $m_{X_{dd}}\sim \mathcal{O}(\text{few}) m_{X_{ud}}$ is less than $\mathcal{O}(10^{8})$ GeV, then the baryon number washout processes are too strong to generate any sizeable asymmetry even if the CP violation is enhanced to its theoretically allowed maximal value. Consequently, the possibility of a low-scale baryogenesis to explain the correct baryon asymmetry can be ruled out in this scenario if both the diquark states $X_{dd}$ and $X_{ud}$ are discovered at the LHC or future colliders together with a positive signal for $\nnbar$ oscillations. On the other hand, successful baryogenesis can be achieved for some parts of the parameter space of the low-scale baryogenesis scenario, however, then the corresponding masses of both diquarks are generally too high and the couplings are too small to probe such a scenario at the LHC, dinucleon decays or $\nnbar$~oscillation searches.

\section{Conclusions}\label{sec:conclusion}
Within the SM, baryon number is an accidental global symmetry and only violated at finite temperature via non-perturbative sphaleron interactions. On the other hand, new physics is required in order to explain the observed baryon asymmetry via a mechanism that complies with the three Sakharov conditions by violating $B-L$, C and CP while leading to a departure from thermal equilibrium. Hence, it is natural to search for new $B$-violating interactions that might have far-reaching implications on the mechanism behind the baryon asymmetry.

While the dim-6 operators with $|\Delta B|=1$ and $|\Delta (B-L)|=0$ are tightly constrained by limits from proton decay, dim-9, $|\Delta B|=|\Delta (B-L)|=2$ interactions are less stringently constrained. However, in the future, new facilities such as the Deep Underground Neutrino Experiment and the European Spallation Source, will probe those interactions via $\nnbar$~oscillation experiments to an unprecedented sensitivity and make this possibility timely to be reconsidered.
As a discovery of $\nnbar$~oscillations would directly imply baryon-number violation, it might have a tight link to  the mechanism of baryogenesis. Moreover, its oscillation time $\tau_{\nnbar}$ would give us a hint on the scale of new physics $\Lambda_{\mathrm{NP}}$ of the effective interaction.

In this work, we have studied the implications of such a discovery for baryogenesis in a comprehensive manner. First, we performed a model-independent EFT analysis of the new dim-9, $|\Delta B|=2$ operator $uudddd$. Assuming no inherent CP violation source connected to this new operator, its interaction would lead to washout processes that could wipe out a potential pre-existing baryon asymmetry. Under the assumption of a discovery with a given oscillation time  $\tau_{\nnbar}$, we estimated the corresponding NP scale $\Lambda_{\mathrm{NP}}$ and calculated the corresponding washout strength $\Gamma_W/H$. In order to relate the low scale, where $\nnbar$~oscillations are expected to occur, with the high scale, where the NP is active and the washout takes place, we considered the running of the Wilson coefficients from the scale of NP down to the scale $\mu_0=2~\mathrm{GeV}$, where the nuclear matrix elements are provided by lattice calculations. We demonstrate that the observation of $\nnbar$~oscillations at future experiments such as DUNE or NNBAR, would imply a strong washout down to around $100~\mathrm{TeV}$. With the LHC already probing scales up to $1-10~\mathrm{TeV}$, we would predict, under the made assumptions, new physics between $10-100~\mathrm{TeV}$, possibly detectable at a future $100~\mathrm{TeV}$ collider.

In order to accommodate the possibility of different hierarchies of NP within the effective operator $uudddd$ and an additional new source of CP violation, we explored a simplified set-up that realises one of the two possible UV-complete topoligies, featuring two new types of scalar diquarks $X_{ud}$ and $X_{dd}$ with couplings $f^{ud}$ and $f^{dd}$ to SM quarks, respectively. After $B-L$ breaking, both carry a baryon number $-2/3$ such that the out-of-equilibrium decay of the heavier one, $X_{dd} \rightarrow X_{ud}X_{ud}$, violates baryon number and generates (together with a new CP phase) a baryon asymmetry. Due to CPT invariance, also the decay $X_{dd} \rightarrow d^cd^c$ contributes indirectly to this asymmetry. In order to describe the final amount of the baryon asymmetry, we presented a comprehensive framework of Boltzmann equations.

Moreover, we discussed in detail all experimental constraints which this set-up could be subject to. We performed a dedicated study of current LHC, but also future HE-LHC and FCC-hh limits on the scalar sextet diquarks. While current LHC limits exclude these new particles for $f^{ud, dd} \approx \mathcal{O}(1)$ up to around $10~\mathrm{TeV}$, smaller couplings ($f^{ud, dd} \approx \mathcal{O}(0.01)$) allow for masses heavier than $6-7~\mathrm{TeV}$. Depending on the mass scale of the diquarks, meson oscillations can strongly constrain the respective couplings. As $X_{ud}$ can contribute to meson oscillations only via loop diagrams, limits on the coupling $f^{ud}$ are generally weak. In contrast, due to the direct tree-level contribution of $X_{dd}$, the coupling $f^{dd} \gtrsim \mathcal{O}(10^{-3})$ is strongly constrained for $m_{X_{dd}} \approx \mathcal{\mathrm{TeV}}$. In particular for third generation couplings and comparable masses of $X_{ud}$ and $X_{dd}$, the dinucleon decay can set more stringent limits than $\nnbar$~oscillations, as they contribute at the two-loop instead of the three-loop level. This leads with current limits from Frejus to $f_{33}^{ud} < 1$ for $m_{X_{ud}}  \approx m_{X_{dd}}/3 \approx 2~\mathrm{TeV}$ and is expected to improve to $f_{33}^{ud} < 0.3$ for $m_{X_{ud}}  \approx m_{X_{dd}}/3 \approx 2~\mathrm{TeV}$ or $f_{33}^{ud} < 1$ for $m_{X_{ud}}  \approx m_{X_{dd}}/3 \approx 4~\mathrm{TeV}$ with Hyper-Kamiokande. Moreover, we discussed considerations from colour preserving vacua and commented on further limits from electroweak precision observables and electric dipole moments from which we do not expect any major constraints due to chirality and loop suppression.

Finally, we compared the prospects to generate the observed baryon asymmetry with the allowed parameter space from all considered experimental constraints. In our first scenario, we assumed a heavy  $m_{X_{dd}} \approx 10^{13}~\mathrm{GeV}$ and a lighter $m_{X_{ud}} \approx \mathcal{O}(\mathrm{TeV})$. Depending on the actual size of the CP violation, the observed baryon asymmetry can be produced with respective couplings, e.g. $f^{ud} = f^{dd} \approx 0.3$ for $\epsilon=0.01$ or $f^{ud} = f^{dd} \approx 0.4$ for maximal $\epsilon$. In this scenario, an interesting interplay between the LHC and future $\nnbar$ experiments can take place: In order to account for the observed baryon asymmetry, a discovery of a diquark at the LHC $tb$ final state would still allow for a successful high-scale baryogenesis scenario. If the diquark is also discovered at the LHC in the lighter generation quark dijet final states, a signal at $\nnbar$~oscillation experiments would give valuable insight into the flavour dynamics of a possible high-scale baryogenesis scenario, and would motivate a comprehensive study of flavour effects. However, if a corresponding signal at both the LHC and $\nnbar$~oscillation experiments is discovered and naively applying the maximal washout corresponding to the first generation $\nnbar$~oscillations to our single flavoured analysis suggests that the high-scale scenario will still remain a viable option.

In our second scenario, we analysed a less hierarchical set-up with $m_{X_{ud}} = m_{X_{dd}}/3$. In this case, the observed baryon asymmetry can only be generated with $m_{X_{ud}} = m_{X_{dd}}/3 \approx 10^5~\mathrm{TeV}$, which is out of reach of all experiments. This confirms our na\"ive EFT estimate of a strong washout $\Gamma_W/H$ even in the presence of a CP violating source at low scales. However, we have shown that the future NNBAR experiment will set stringent limits on the parameter space ranging from $f^{ud} = f^{dd} = 1$ for $m_{X_{ud}} = m_{X_{dd}}/3 \approx 700~\mathrm{TeV}$ to 
$f^{ud} = f^{dd} = 0.001$ for $m_{X_{ud}} = m_{X_{dd}}/3 \approx 10~\mathrm{TeV}$. This demonstrates the excellent future prospects of this experiment, making it possible to directly observe $|\Delta B| = 2$ interactions up to an unprecedented scale.

In summary, the discovery of $\nnbar$~oscillations can have far-reaching consequences on our understanding of physics beyond the Standard Model. Not only would it point to new $B$-violating interactions, it could also have significant implications on the mechanism behind the baryon asymmetry. The studied less hierarchical set-up confirms the na\"ive EFT estimate of a too strong washout in order to generate the observed baryon asymmetry in case of an observation. Moreover, it even demonstrates that a possible CP violation of the new low-scale interactions could not balance out this strong washout. Hence, in order to generate the observed baryon asymmetry, we would expect some other new physics between the LHC exclusion and the scale of strong washout in case the $|\Delta B| = 2$ interaction does not involve any new CP-violating interaction. In contrast, the analysis of our high-scale scenario demonstrates that for a large hierarchy of new degrees of freedom that feature additionally a new CP phase, absorbed in the effective $|\Delta B| = 2$ interaction $uudddd$, we are actually able to generate the observed baryon asymmetry in case of some new physics being at a low scale. This would imply that an observation of $\nnbar$~oscillations might point towards new physics experimentally reachable at the LHC or future colliders.

%------------------------------------------------------------------------------
%
	\section*{Acknowledgements}
The authors acknowledge support from the DFG Emmy Noether Grant No. HA 8555/1-1. K.F. additionally acknowledges support from the DFG Collaborative Research Centre “Neutrinos and DarkMatter in Astro- and Particle Physics” (SFB 1258).

%%%%%%%%%%%%%%%%%%%%%%%%%%%%%%%%%%%%%%%%%%%%%%%%%%%%%%%%%%%%%%%%%%%%%%%%%%%%%%%%%%%%%%%%%%%%%%%%%%%%%%%%%%%%%%%%%%%%%%%%%%%%%%%%%%%%%%%%%%%%%%%%%%%%%%%%%%%%%%%%%%%%%%%%%%%%%%%%%%%%%%%%%%%%%%%%%%%%%%%%%%%%%%%%%%%%%%

\begin{appendix}
\section{A possible UV completion motivated from GUTs}{\label{subs:UV}}
Diquarks can naturally arise in many UV completion of the SM gauge group, e.g. in Grand Unified Theory (GUT) embeddings such as the Pati-Salam group $G_{\text{PS}}\equiv SU(2)_L \times SU(2)_R \times SU(4)_c$, $SO(10)$, or $E_6$. Here, we will mainly focus on one of the extensively studied example of $SO(10)$.

The effective Yukawa couplings and quartic scalar interactions given in Eq.~\eqref{lag:top2} can naturally arise in $SO(10)$ as discussed in Ref.~\cite{Babu:2012vc}. There are a number of possibilities for the breaking route of $SO(10)$ to the SM gauge group. Some of them rely on the intermediate gauge symmetry $G_{\text{PS}}$, which provides the possibility of unifying quarks with leptons with the extended colour gauge group $SU(4)_c$ accommodating leptons as a fourth colour. Also using the left-right symmetric gauge subgroup $G_{LR}\equiv SU(3)_c \times SU(2)_L \times SU(2)_R \times U(1)_{B-L}$ as an intermediate symmetry in $SO(10)$ GUT is fairly common. In order to discuss the realisations of the effective diquark couplings in the context of $SO(10)$, we write down the decomposition of some relevant $SO(10)$ multiplets under $G_{\text{PS}}$:
\begin{eqnarray}
10 &=& (2,2,1) + (1,1,6) \nonumber \\
16&=&(2,1,4)+(1,2,\overline{4}) \nonumber \\
45&=&(3,1,1)+(1,3,1)+(1,1,15)+(2,2,6) \nonumber \\
54&=&(1,1,1)+(3,3,1)+(1,1,20)+(2,2,6) \nonumber \\
120 &=& (2,2,1) + (1,1,10)+(1,1,\overline{10})+(3,1,6)+(1,3,6)+(2,2,15)\nonumber \\
126 &=& (1,1,6) + (3,1,10) + (1,3,\overline{10}) + (2,2,15)\nonumber \\
210 &=& (1,1,1) + (1,1,15) + (3,1,15) +  (1,3,15) + (2,2,6) + (2,2,10)+ (2,2,\overline{10}) ~.
\end{eqnarray}
\renewcommand{\arraystretch}{1.25}
\begin{table*}[t]
\centering
\begin{tabular}{ccc}
\hline \hline
$G_{\text{PS}}$
& $G_{\text{LR}}$
& $G_{\text{SM}}$
\\
\hline
 $\left({ 2,1,4} \right)$
& $\left({ 3,2,1},\frac{1}{3} \right)$
& $\left({ 3,2},\frac{1}{6}  \right)$
\\
\null
& $\left({ 1,2,1},-1 \right)$
& $\left({ 1,2},-\frac{1}{2}  \right)$
\\
 $\left({ 2,1,\overline{4}} \right)$
& $\left({ \overline{3},1,2},-\frac{1}{3} \right)$
& $\left({ \overline{3},1},\frac{1}{3} \right)$ $\oplus$ $\left({ \overline{3},1},-\frac{2}{3} \right)$
\\
\null
&  $\left({ 1,1,2},1 \right)$
&  $\left({ 1,1},1 \right)$ $\oplus$ $\left({ 1,1},0 \right)$
\\
\hline \hline
\end{tabular}
\caption{Decomposition for the representation $16$ of $SO(10)$ under $G_{\text{PS}}\equiv SU(2)_L \times SU(2)_R \times SU(4)_c$, $G_{LR}\equiv SU(3)_c \times SU(2)_L \times SU(2)_R \times U(1)_{B-L}$
and the SM gauge group $G_{\text{SM}}\equiv SU(3)_c \times SU(2)_L  \times U(1)_{Y}$ .}
\label{tab:16d}
\end{table*}
\renewcommand{\arraystretch}{1.25}
\begin{table*}[ht]
\centering
\begin{tabular}{ccc}
\hline \hline
$G_{\text{PS}}$
& $G_{\text{LR}}$
& $G_{\text{SM}}$
\\
\hline
 $\left({ 1,3,1} \right)$
& $\left({ 1,1,3},0 \right)$
& $\left({ 1,1},+1 \right)$ $\oplus$ $\left({ 1,1},0 \right)$ $\oplus$ $\left({ 1,1},-1 \right)$
\\
$\bf{\left({ 3,1,1} \right)}$
& $\bf{\left({ 1,3,1},0 \right)}$
& $\bf{\left({ 1,3},0 \right)}$
\\
$\left({2,2,6} \right)$
&  $\left({ 3,2,2},-\frac{2}{3} \right)$
&  $\left({ 3,2},+\frac{1}{6} \right)$ $\oplus$ $\left({ 3,2},-\frac{5}{6} \right)$
\\
\null
&  $\left({ \overline{3},2,2},+\frac{2}{3} \right)$
&  $\left({ \overline{3},2},+\frac{5}{6} \right)$ $\oplus$ $\left({ \overline{3},2},-\frac{1}{6} \right)$
\\
$\left({1,1,15} \right)$
& $\left({ 1,1,1},0 \right)$
& $\left({ 1,1},0 \right)$
\\
\null
&  $\left({ 3,1,1},+\frac{4}{3} \right)$
&  $\left({ 3,1},+\frac{2}{3} \right)$
\\
\null
&  $\left({ \overline{3},1,1},-\frac{4}{3} \right)$
&  $\left({ \overline{3},1},-\frac{2}{3} \right)$
\\
\null
&  $\left({ 8,1,1},0 \right)$
&  $\left({ 8,1},0 \right)$
\\
\hline \hline
\end{tabular}
\caption{Decomposition for the representation $45$ of $SO(10)$ under $G_{\text{PS}}$, $G_{\text{LR}}$,
and $G_{\text{SM}}$. We highlight the representations relevant to the discussion in the main text in bold.  }
\label{tab:45d}
\end{table*}
All SM fermion fields (for a given generation or family) can be accommodated within the multiplet $16$ of $SO(10)$, as can be seen from Tab.~\ref{tab:16d}. In addition it contains a SM singlet, which can be associated with a right-handed neutrino.
Note that the Higgs multiplets which can couple to the fermion bilinears of interest $16_i 16_j$ (containing all the SM fermions) are $10_H$, $126_H$ and $120_H$, with the couplings of the $10_H$ and $126_H$ being symmetric in flavour indices $(i, j)$ and those of the $120_H$ being antisymmetric. In general, the SM Higgs doublet $(1,2,1/2)$ is a possible linear combination of $h$: $(1,2,1/2)$ and $h^\dagger$: $(1,2,-1/2)$ which belongs to the $10$ and $\overline{126}$ multiplets of $SO(10)$. In Tables \ref{tab:45d}, \ref{tab:54d} and \ref{tab:126d}, we present the decomposition of the three representations $45$, $54$ and $126$ of $SO(10)$ via the breaking chain $G_{\text{PS}} \rightarrow G_{\text{LR}}\rightarrow G_{\text{SM}}$, which will be essential for our discussion. In what follows we will briefly highlight how the criterion of gauge coupling unification can naturally motivate one of the choices for diquark mass scales that we subsequently use to study the scenario of high-scale baryogenesis.

Before moving onto the discussion of gauge coupling unification let us briefly summarise the possibility of intermediate symmetries which can be realised by choosing appropriate scalar multiplets (which will also contain the scalar diquark fields) relevant for realising a given symmetry breaking chain. The possibility of realising an intermediate gauge symmetry group is dictated by the choice of Higgs multiplets of $SO(10)$ used for symmetry breaking. In particular, a real $45$, a real $54$, or a real $210$ can be used along with a complex $126$ to achieve the breaking to the SM gauge group. Using the $210$ leads to the intermediate symmetry $G_{\text{PS}}$ but with a broken discrete parity symmetry (subsequently, leading to an asymmetric gauge couplings for $SU(2)_{L(R)}$: $g_L\neq g_R$ ), often referred to as $D$-parity, while $54$ can break $SO(10)$ down to $G_{\text{PS}}$ symmetry preserving $D$-parity. On the other hand, the intermediate symmetry corresponds to a left-right symmetric gauge group $G_{\text{LR}}$ with a broken $D$-parity \cite{Chang:1983fu}.
\renewcommand{\arraystretch}{1.25}
\begin{table*}[ht]
		\centering
\begin{tabular}{ccc}
\hline \hline
$G_{\text{PS}}$
& $G_{\text{LR}}$
& $G_{\text{SM}}$
\\
\hline
 $\left({ 1,1,1} \right)$
& $\left({ 1,1,1},0 \right)$
& $\left({ 1,1},0 \right)$
\\
$\bf{\left({ 3,3,1} \right)}$
& $\bf{\left({ 1,3,3},0 \right)}$
& $\left({ 1,3},-1 \right)$ $\oplus$ $\bf{\left({ 1,3},0 \right)}$ $\oplus$ $\left({ 1,3},+1 \right)$
\\
$\left({2,2,6} \right)$
&  $\left({ 3,2,2},-\frac{2}{3} \right)$
&  $\left({ 3,2},+\frac{1}{6} \right)$ $\oplus$ $\left({ 3,2},-\frac{5}{6} \right)$
\\
\null
&  $\left({ \overline{3},2,2},+\frac{2}{3} \right)$
&  $\left({ \overline{3},2},+\frac{5}{6} \right)$ $\oplus$ $\left({ \overline{3},2},-\frac{1}{6} \right)$
\\
$\left({1,1,20'} \right)$
& $\bf{\left({ 6,1,1},+\frac{4}{3}  \right)}$
& $\bf{\left({ 6,1},+\frac{2}{3} \right)}$
\\
\null
&  $\bf{\left({  \overline{6},1,1},-\frac{4}{3} \right)}$
&  $\bf{\left({ \overline{6},1},-\frac{2}{3} \right)}$
\\
\null
&  $\left({ 8,1,1},0 \right)$
&  $\left({ 8,1},0 \right)$
\\
\hline \hline
\end{tabular}
\caption{Decomposition for the representation $54$ of $SO(10)$ under $G_{\text{PS}}$, $G_{\text{LR}}$,
and $G_{\text{SM}}$. We highlight the representations relevant to the discussion in the main text in bold.}
\label{tab:54d}
\end{table*}
\renewcommand{\arraystretch}{1.25}
\begin{table*}[ht]
		\centering
\begin{tabular}{ccc}
\hline \hline
$G_{\text{PS}}$
& $G_{\text{LR}}$
& $G_{\text{SM}}$
\\
\hline
$\left({ 1,1,6} \right)$
& $\left({ \overline{3},1,1},+\frac{2}{3} \right)$
& $\left({ \overline{3},1},+\frac{1}{3} \right)$
\\
\null
& $\left({ 3,1,1},-\frac{2}{3} \right)$
& $\left({ 3,1},-\frac{1}{3} \right)$
\\
$\left({ 3,1,10} \right)$
& $\left({ 1,3,1},-2 \right)$
& $\left({ 1,3},-1 \right)$
\\
\null
& $\left({ 3,3,1},-\tfrac{2}{3} \right)$
& $\left({ 3,3},-\tfrac{1}{3} \right)$
\\
\null
& $\left({ 6,3,1},+\tfrac{2}{3} \right)$
& $\left({ 6,3},+\tfrac{1}{3} \right)$
\\
$\bf{\left({ 1,3,\overline{10}} \right)}$
& $\left({ 1,1,3},+2 \right)$
& $\bf{\left({ 1,1},0 \right)}$ $\oplus$ $\left({ 1,1},+1 \right)$ $\oplus$ $\left({ 1,1},+2 \right)$
\\
\null
& $\left({ \overline{3},1,3},+\tfrac{2}{3} \right)$
& $\left({ \overline{3},1},-\tfrac{2}{3} \right)$ $\oplus$ $\left({ \overline{3},1},+\tfrac{1}{3} \right)$ $\oplus$ $\left({ \overline{3},1},+\tfrac{4}{3} \right)$
\\
\null
& $\bf{\left({ \overline{6},1,3},-\tfrac{2}{3} \right)}$
& $\bf{\left({ \overline{6},1},-\tfrac{4}{3} \right)}$ $\oplus$ $\bf{\left({ \overline{6},1},-\tfrac{1}{3} \right)}$ $\oplus$ $\bf{\left({ \overline{6},1},+\tfrac{2}{3} \right)}$
\\
$\left({ 2,2,15} \right)$
& $\left({ 1,2,2},0 \right)$
& $\left({ 1,2},-\tfrac{1}{2} \right)$ $\oplus$ $\left({ 1,2},+\tfrac{1}{2} \right)$
\\
\null
& $\left({ \overline{3},2,2},-\tfrac{4}{3} \right)$
& $\left({ \overline{3},2},-\tfrac{7}{6} \right)$ $\oplus$ $\left({ \overline{3},2},-\tfrac{1}{6} \right)$
\\
\null
& $\left({ 3,2,2},+\tfrac{4}{3} \right)$
& $\left({ 3,2},+\tfrac{7}{6} \right)$ $\oplus$ $\left({ 3,2},+\tfrac{1}{6} \right)$
\\
\null
& $\left({ 8,2,2},0 \right)$
& $\left({ 8,2},-\tfrac{1}{2} \right)$ $\oplus$ $\left({ 8,2},+\tfrac{1}{2} \right)$
\\
\hline \hline
\end{tabular}
\caption{Decomposition for the  representation $126$ of $SO(10)$ under $G_{\text{PS}}$, $G_{\text{LR}}$
and $G_{\text{SM}}$. We highlight the representations relevant to the discussion in the main text in bold.}
\label{tab:126d}
\end{table*}

 Effective $(B-L)$-violating interactions can naturally occur in GUT theories like $SO(10)$, $G_{\text{PS}}$ or $G_{LR}$, where $(B-L)$ is part of the local gauge symmetry and broken by the vacuum expectation value of a scalar field carrying non-vanishing $(B-L)$ charge. In $SO(10)$ and $G_{\text{PS}}$  the effective $(B-L)$-violating couplings can arise, when the $(B-L)$-symmetry is broken by giving a vacuum expectation value to the complete SM singlet field. For instance, it can be contained in the $\overline{126}_H$ multiplet of $SO(10)$ transforming as (1,1,0) under the SM gauge group $G_{\text{SM}}$ while transforming as $(1,3,\overline{10})$ under $G_{\text{PS}}$, see bold-print in Table~\ref{tab:126d}. Note that this field, denoted by $\xi$ in our Lagrangian in Eq.~\eqref{lag:top2} carries $(B-L)=-2$. In $SO(10)$ GUT $\xi$ can potentially also generate large Majorana masses for the right-handed neutrinos through the couplings of the form $\nu^{c}\nu^{c}\xi$. Note that we assume the scale of $B-L$ breaking and consequently any intermediate symmetry like $G_{\text{PS}}$ or $G_{LR}$ to lie around the unification scale to simplify the RG running. In case of a low-scale (few TeV) breaking of such intermediate symmetries the running will be affected and additional care must be taken to include the effects of new light degrees of freedom (e.g. a right handed $W_R$ gauge boson that can lead to large washout effects in baryogenesis \cite{Frere:2008ct,Deppisch:2013jxa,Dev:2014iva,Deppisch:2015yqa,Dhuria:2015wwa,Dhuria:2015cfa,Dev:2015vra}).

 The effective Yukawa couplings given in our Lagrangian in Eq.~\eqref{lag:top2} can be obtained from the Yukawa coupling $16_i 16_j \overline{126}_H$ in the $SO(10)$ GUT theory, where the $\overline{126}_H$ contains the scalar diquark fields of interest
 $X_{uu}:\left({ \overline{6},1},-\tfrac{4}{3} \right)$, $X_{ud}:\left({ \overline{6},1},-\tfrac{1}{3} \right)$ and  $X_{dd}:\left({ \overline{6},1},+\tfrac{2}{3} \right)$. These fields belong to the $\left({ \overline{6},1,3},-\tfrac{2}{3} \right)$  of $G_{LR}$ and $\left({ 1,3,\overline{10}} \right)$ multiplet of $G_{\text{PS}}$. Hence, the quartic scalar coupling in Eq.~\eqref{lag:top2}, can be generated through $\lambda \overline{126}_H^4$. 
 
 Note that the multiplet 54 of $SO(10)$ also contains the field $X_{dd}:\left({ \overline{6},1},+\tfrac{2}{3} \right)$, see Table~\ref{tab:54d}. Therefore, it can mix with the aforementioned other copy of the field with the same quantum number, $\mu \overline{126}^2 54$,  at a scale below the $SO(10)$ breaking and can hence play an essential role in providing  a second copy of $X_{dd}$ that, together with a CP-violating coupling, can fulfil the three Sakharov conditions needed for successful baryogenesis (see Sec.~\ref{sec:boltzmann} and the discussion around Eq.~\eqref{lag:Xddp}).

The possibility of achieving a gauge coupling unification with in a GUT framework can be verified by using the RGEs governing the evolution for running coupling constants $g_{i}$. At one-loop level the RGEs for the gauge couplings can be expressed as
\begin{equation}{\label{4.1}}
\mu\,\frac{\partial g_{i}}{\partial \mu}=\frac{b_i}{16 \pi^2} g^{3}_{i},
\end{equation}
which can be expressed as
\begin{equation}{\label{4.2}}
\frac{1}{\alpha_{i}(\mu_{2})}=\frac{1}{\alpha_{i}(\mu_{1})}-\frac{b_{i}}{2\pi} \ln \left( \frac{\mu_2}{\mu_1}\right),
\end{equation}
where $\alpha_{i}=g_{i}^{2}/4\pi$ corresponds to the coupling of the 
$i$--th gauge group, $\mu_1, \mu_2$ are the energy scales with
$\mu_2 > \mu_1$. The relevant beta-coefficients $b_i$ at one-loop order can be obtained using \cite{Jones:1981we}
\begin{eqnarray}{\label{4.3}}
	&&b_i= - \frac{11}{3} \mathcal{C}_{2}(G)
				 + \frac{2}{3} \,\sum_{R_f} T(R_f) \prod_{j \neq i} d_j(R_f)
  + \frac{1}{3} \sum_{R_s} T(R_s) \prod_{j \neq i} d_j(R_s)\, ,
\label{oneloop_bi}
\end{eqnarray}
where $\mathcal{C}_2(G)$ corresponds to the quadratic Casimir operator for the
gauge bosons in their adjoint representation
\begin{equation}{\label{4.4}}
	\mathcal{C}_2(G) \equiv \left\{
	\begin{matrix}
    N & \text{if } SU(N)\, , \\
    0 & \text{if } \phantom{S}U(1)\, .
	\end{matrix}\right.
\end{equation}
$T(R_f)$ and $T(R_s)$ correspond to the Dynkin indices of the
irreducible representation $R_{f,s}$ for a given fermion and scalar field, respectively, e.g.
\begin{equation}{\label{4.5}}
	T(R_{f,s}) \equiv \left\{
	\begin{matrix}
    \frac{1}{2} & \quad \quad \,\, \text{if $R_{f,s}$ is fundamental}, \\
    N   & \text{if $R_{f,s}$ is adjoint}, \\
    0   & \text{if $R_{f,s}$ is singlet},
	\end{matrix}\right.
\end{equation}
and $d(R_{f,s})$ is the dimension of the representation $R_{f,s}$
under all gauge groups except the $i$-th~gauge group under
consideration. An additional factor of $1/2$ is multiplied to $T(R_{s})$ in the case $R_s$ is a real Higgs representation.

\begin{figure}[hbt!]
\begin{center}
    \includegraphics[width = 0.45 \textwidth]{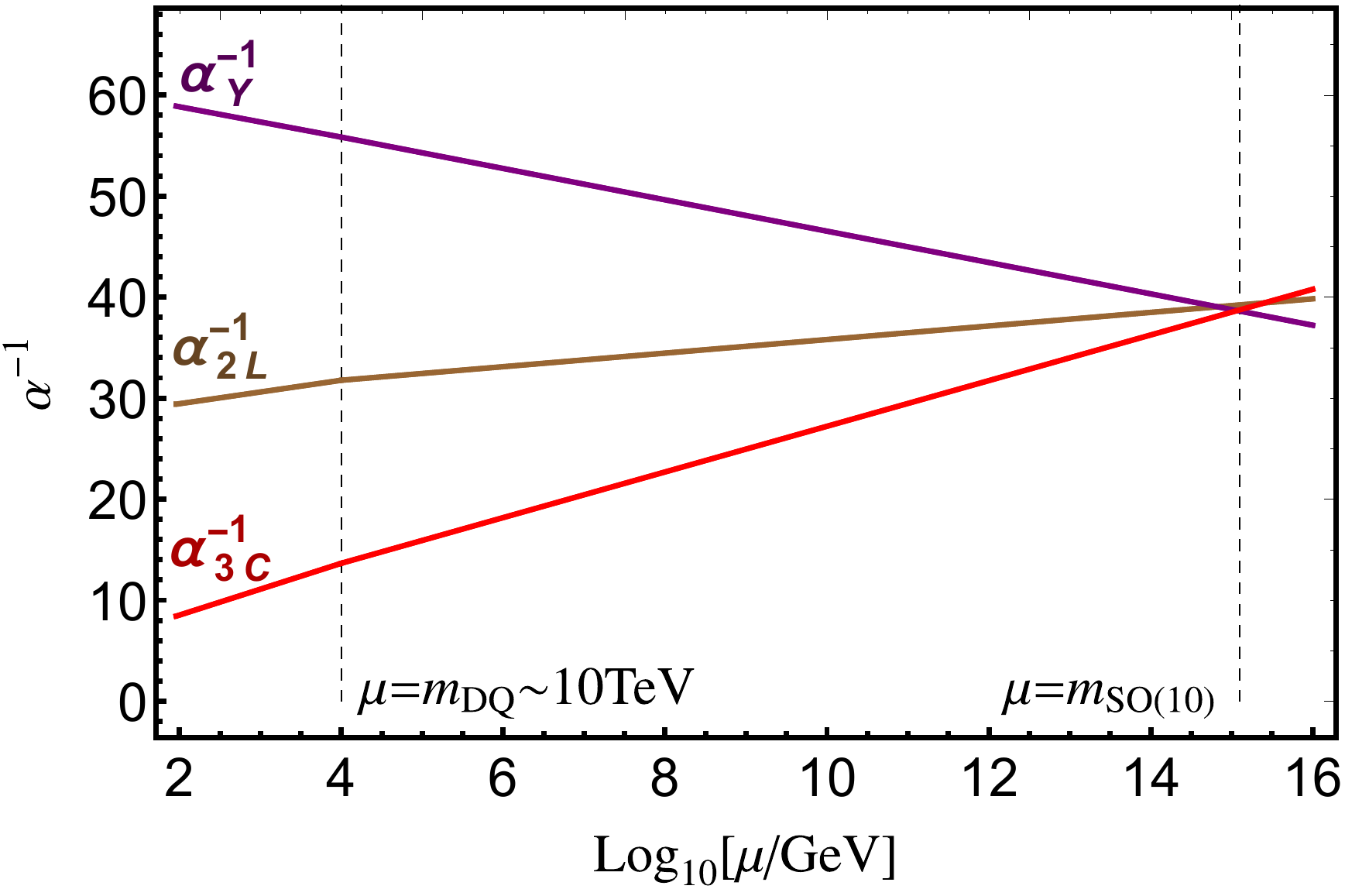}
     \includegraphics[width = 0.45 \textwidth]{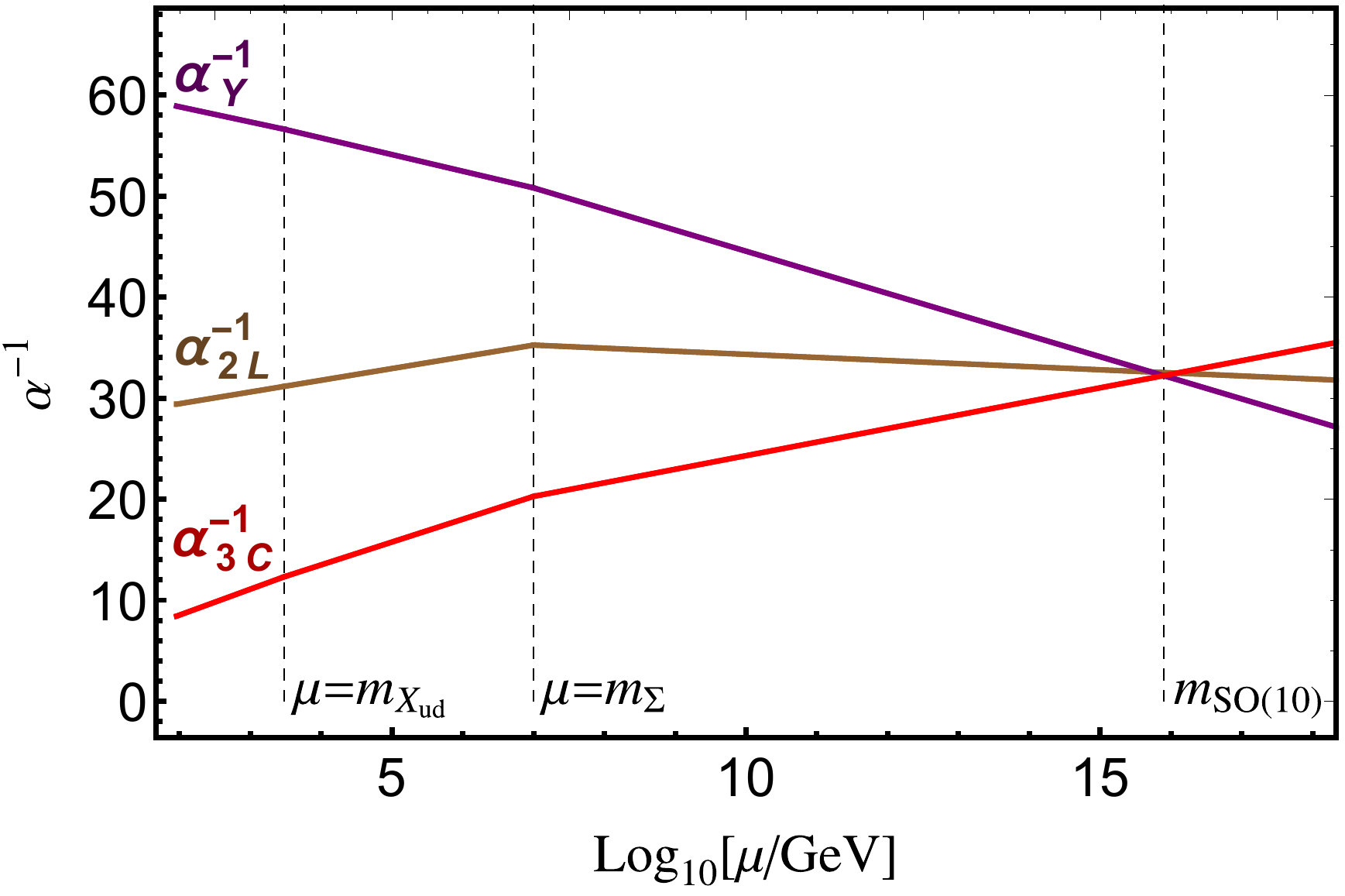}
\caption{Evolution of SM gauge couplings with a colour sextet $X_{ud}$ and two copies of $\Delta:\left({ 1,3},0\right)$ both around the TeV scale {\it (left)}, or a colour sextet $X_{ud}$  at the TeV scale and a $SU(2)_L$ triplet field $\Sigma:\left({ 6,3},\frac{1}{3}\right)$ at an intermediate scale {\it (right)} demonstrating a possible gauge coupling unification.
}
\label{fig:GUT2}
\end{center}
\end{figure}

Let us now consider a $SO(10)$ GUT scenario where $X_{ud}$ is the only light field ($m_{X_{ud}} \sim \mathcal{O}$(TeV) scale) affecting the running of gauge coupling constants at one-loop level (in addition to the SM field content), while the remaining diquarks ($X_{dd}$ and $X_{uu}$) and other new fields lie around the unification scale. This scenario is particularly interesting because it allows for a high-scale baryogenesis scenario to be realised through the decay $X_{dd}\rightarrow X_{ud}^* X_{ud}^*$ as discussed in Sec.~\ref{sec:boltzmann}. Additionally, a relatively light ${X_{ud}}$ can lead to a sizeable contribution to the $\nnbar$~oscillation rate and can be directly searched for at the LHC and future colliders. To simplify the discussion, let us first assume that (if present) the intermediate gauge symmetry $G_{\text{PS}}$ also lies very close to the $SO(10)$ unification scale. In such a scenario where $X_{ud}$ is the only light field with a mass around $\mathcal{O}$(TeV), the gauge couplings do not unify. However, if one also includes two copies of $\Delta:\left({ 1,3},0\right)$ around a $\mathcal{O}$(TeV) scale a successful $SO(10)$ unification can be achieved as first pointed out in~\cite{Babu:2012vc}. Note that the field $\Delta:\left({ 1,3},0 \right)$ can be obtained from a $45$ or $54$ representation of $SO(10)$, see Tab.~\ref{tab:45d} and \ref{tab:54d}. In Fig.~\ref{fig:GUT2} (left panel), we show the evolution of the gauge couplings for $m_{X_{ud}}\sim m_{\Delta}=3$ TeV. With increasing $m_{X_{ud}}\sim m_{\Delta}$ (beyond a few TeV) a unification triangle starts to develop (not visible in the figure due to the scaling). However, subject to the threshold corrections, unification can still be considered as a marginally successful. This leads to $m_{X_{ud}} \lesssim 10$ TeV as a necessary condition for a successful $SO(10)$ gauge coupling unification.

Another alternative scenario that provides successful unification is $X_{ud}$ as the only light field with mass $\mathcal{O}$(TeV) and an additional field $\Sigma:\left({ 6,3},\frac{1}{3}\right)$  at an intermediate scale $m_{\Sigma}\sim 10^4$ TeV. In Fig.~\ref{fig:GUT2} (right panel), we show the evolution of the gauge couplings for $m_{X_{ud}}=3$ TeV and $m_{\Sigma}= 10^4$ TeV. 

%%%%%%%%%%%%%%%%%%%%%%%%%%%%%%%%%%%%%%%%%%%%%%%%%%%%%%%%%%%%%%%%%%%%%%%%%%%%%%%%%%%%%%%%%%%%%%%%%%%%%%%%%%%%%%%%%%%%%%%%%%%%%%%%%%%%%%%%%%%%%%%%%%%%%%%%%%%%%%%%%%%%%%%%%%%%%%%%%%%%%%%%%%%%%%%%%%%%%%%%%%%%%%%%%%%%%%%
\section{Chemical potential relations}{\label{app:cp}}

In the early Universe, the chemical potential of a particle species $a$ can be related to the ratio of number density $\eta_a$ to equilibrium number density $\eta^{\eq}_a$, such that \cite{Harvey:1990qw,Deppisch:2017ecm}
\begin{equation}
\frac{\eta_a}{\eta^{\eq}_a} \approx e^{\mu_a/T} \approx 1 + \frac{\mu_a}{T},
\label{eq:muapprox}
\end{equation}
where $\mu_a$ is the chemical potential of the species $a$ and $T$ is the temperature of the Universe. Furthermore, in writing Eq.~\eqref{eq:muapprox}, we have assumed that the number density of the particle species is close to equilibrium such that $|\eta_a-\eta_{\bar a}| \ll \eta_a^\eq$, where $\bar a$ is the antiparticle of $a$, and where the chemical potential of a particle species $a$ is related to the chemical potential of its antiparticle via the relation $\mu_a = -\mu_{\bar a}$. If an interaction is in equilibrium, a relation can be found between the chemical potentials of the particle species involved. For example, the reaction $W^- \leftrightarrow \bar u_L + d_L$ being in equilibrium leads to the relation $\mu_W + \mu_{u_L} = \mu_{d_L}$. At temperatures above the scale of electroweak symmetry breaking $T \gtrsim \Lambda_{\text{EW}}$, the total electric charge $Q$ and the $z$-component of isospin $T_3$ must both be zero. The latter constraint leads to $\mu_W = 0$ \cite{Harvey:1990qw}, such that (suppressing flavour indices)
\begin{equation}
\label{eq:chempot_ewud}
\mu_{u_L} = \mu_{d_L} \equiv \mu_{Q}.
\end{equation}
Similarly, from the reactions $W^- + \nu_L \leftrightarrow e_L$ and $W^- \leftrightarrow h^- + h^0$, we find the relations
\begin{equation}
\mu_{\nu_L}=\mu_{e_L}\equiv\mu_L,\quad \mu_{h^0}=-\mu_{h^-}\equiv\mu_H.
\end{equation}
From the Yukawa interactions $h^0 + \bar{e^c} \leftrightarrow e_L$, $h^0 + u_L \leftrightarrow \bar{u^c}$, and $h^0 + \bar{d^c} \leftrightarrow d_L$, we obtain
\begin{equation}
	\mu_{H} = \mu_{L} - \mu_{\bar{e^c}}, \quad	\mu_{H} = -\mu_{Q} + \mu_{\bar{u^c}}, \quad	\mu_{H} = \mu_{Q} - \mu_{\bar{d^c}},
\end{equation}
and the electroweak sphaleron interaction in equilibrium leads to the relation
\begin{equation}
\label{eq:chempot_sphaleron}
	3\sum_i\mu_{Q_i} = -\sum_i\mu_{L_i}\,,
\end{equation}
where $i$ denote the flavour indices. Assuming flavour universality, henceforth we replace the sum over flavours by the number of SM quark generations in thermal equilibrium $N$. Lastly, since we are interested in the evolution of the number density of the diquark $X_{ud}^*$ in Sec.~\ref{sec:boltzmann}, we will consider the reaction $X_{ud}^* \leftrightarrow \bar{u^c} + \bar{d^c}$ to be in equilibrium, thereby obtaining the relation
\begin{equation}
\label{eq:chempot_xud}
	\mu_{X_{ud}^*} = \mu_{\bar{u^c}} + \mu_{\bar{d^c}}.
\end{equation}
In terms of chemical potentials, the total electric charge $Q$ of the Universe can then be expressed as
\begin{equation}
\label{eq:totcharge}
	Q =2N\left(\mu_{Q}+\mu_{\bar{u^c}}\right)-N\left(\mu_{Q}+\mu_{\bar{d^c}}\right)-N\left(\mu_{L}+\mu_{\bar{e^c}}\right)+2 \mu_{H}+2\frac{C_{X^*_{ud}}}{3}\mu_{X_{ud}^*},
\end{equation}
where $C_{X^*_{ud}}$ is the colour multiplicity of $X_{ud}^*$, $N$ is the number of fermion generations in thermal equilibrium\footnote{We keep the number of generations of the SM fermions as a free parameter since, depending on the scale of baryogenesis, different numbers of the SM fermion generations can be in thermal equilibrium at different temperatures~\cite{Davidson:2008bu}.}. 
Since the total electric charge $Q$ and the $z$-component of isospin $T_3$ of the Universe must be zero above the scale of electroweak symmetry breaking, the individual chemical potential relations can be combined with equation \eqref{eq:totcharge} to relate the chemical potential of the quark singlets $\bar{u^c}$ and $\bar{d^c}$ to the diquark $X_{ud}^*$. 
Using the notation
\begin{equation}
\mu_a = x_a\mu_{X_{ud}^*},
\end{equation}
for any particle species $a$, we obtain the relations
\begin{equation}{\label{app:eq:x}}
x_{\bar{u^c}} = \frac{3-6N-2C_{X_{ud}^*}}{12N+6},\quad x_{\bar{d^c}} = \frac{3+18N+2C_{X_{ud}^*}}{12N+6}.
\end{equation}
To define the net baryon number density of the Universe, we include the particle species $X_{ud}$, $\bar{d^c}$ and $\bar{u^c}$ as well as their corresponding antiparticles, and assume that the other SM fields do not contribute to a baryon number asymmetry generation or washout\footnote{We note that even though $u_L$ and $d_L$ do not participate directly in any baryon number violating interactions for our scenario, they can indirectly affect the baryon asymmetry through spectator processes, which are taken into account by a final sphaleron conversion factor \cite{Buchmuller:2001sr,Nardi:2005hs,Davidson:2008bu}.}.
Such a prescription leads to the net baryon number difference of the Universe, cf. Eq.~\eqref{eq:def_B_X}, given by
\begin{equation}
\label{eq:nbdiff}
	n_b-\bar n_b \equiv \frac{T^2}{6}\left(N \mu_{\bar{d^c}}+N \mu_{\bar{u^c}}+2N \mu_{Q}+\frac{4}{3}C_{X_{ud^*}}\mu_{X^*_{ud}}\right)=\frac{T^3}{6}\frac{6N +4 C_{X_{ud}^*}}{3}\frac{\mu_{X_{ud}^*}}{T}.
\end{equation}
Under similar assumptions, the equilibrium baryon number density then can be defined in terms of the number densities of the relevant species as
\begin{equation}
\label{eq:nbeq}
	n_b^\eq \equiv \frac{2}{3}C_{X_{ud}^*}n_{X_{ud}^*}^\eq+N n_{\bar{d^c}}^\eq+N n_{\bar{u^c}}^\eq+N n_{u_L}^\eq+N n_{d_L}^\eq = \frac{\zeta(3)T^3}{\pi^2}\frac{8C_{X_{ud}^*}+36N}{12}.
\end{equation}
To derive the two above equations, we have used
\begin{equation}
\label{eq:neweq}
n_{i}^{\rm eq}=
\frac{g_i T^3}{ \pi^2}\times \left\{ \begin{array}{cc}
\zeta(3) + \frac{\mu_i}{T} \zeta(2) + ... & ({\rm bosons})\\
\frac{3}{4} \zeta(3) + \frac{\mu_i}{T} \frac{\zeta(2)}{2} + ... & ({\rm
	fermions})
\end{array}\right.
\end{equation}
with $\zeta(s)$ denoting the Riemann zeta function. Combining equations \eqref{eq:nbdiff} and \eqref{eq:nbeq}, we finally obtain a relation between baryon-to-photon density and the chemical potential of $X_{ud}^*$
\begin{equation}
	\frac{n_b-\bar n_b}{n_b^\eq}=\frac{\eta_b-\bar \eta_b}{\eta_b^\eq} =  C_B\frac{\mu_{X_{ud}^*}}{T},\quad C_B \equiv  \frac{\pi^2}{3\, \zeta(3)}\frac{6N +4C_{X^*_{ud}}}{18N+ 4 C_{X_{ud}^*}}.
\end{equation}
For the relevant cases in our analysis, we have $C_{X_{ud}^*}=6$, and depending on the number of generation in thermal equilibrium ($N$), $C_B$ takes the values
\begin{equation}
\label{eq:neweq2}
C_B=
\frac{\pi^2}{3\, \zeta(3)}\times \left\{ \begin{array}{cc}
\frac{5}{7} & (N=1)\\
\frac{7}{13}  & (N=3)\, .
\end{array}\right. 
\end{equation}
%
%%%%%%%%%%%%%%%%%%%%%%%%%%%%%%%%%%%%%%%%%%%%%%%%%%%%%%%%%%%%%%%%%%%%%%%%%%%%%%%%%%%%%%%%%%%%%%%%%%%%%%%%%%%%%%%%%%%%%%%%%%%%%%%%%%%%%%%%%%%%%%%%%%%%%%%%%%%%%%%%%%%%%%%%%%%%%%%%%%%%%%%%%%%%%%%%%%%%%%%%%%%%%%%%%%%%%%%
\section{Details of the Boltzmann equation for the evolution of number density }{\label{app:boltzmann}}
The equilibrium number density of a particle species $X$ is given by
\begin{equation}
\eta^{\text{eq}}_{X}(z) = \frac{g_X}{g_\gamma}z^2K_2\lr{z},
\end{equation}
where $K_2(z)$ is the modified Bessel function of the second kind and $z=m_X/T$. For a two-body decay and inverse decay channel $ X\leftrightarrow a_i + b_i$, the reaction density $\gamma_{D_i}$ is given by
\begin{equation}
\gamma_{D_i}^X = \eta^{\text{eq}}_{X}n_{\gamma}^{\text{eq}}\frac{K_1\lr{z}}{K_2\lr{z}}\Gamma_i,
\end{equation}
where the decay width is given by
\begin{equation}
\label{eq:decaywidth}
\Gamma_i = \frac{1}{1+\delta_{ab}}\frac{m_{X}^2-m_{a_i}^2-m_{b_i}^2}{16\pi m_{X}^3}|\overline{\mathcal{M}}_i|^2,
\end{equation}
with $\mathcal{M}_i$ denoting the matrix element corresponding to the process and the symmetry factor in front with $\delta_{ab}$ takes care of identical initial (final states), and the full reaction rate for (inverse) decays is the sum of all allowed channels,
\begin{equation}
\gamma_{D}^X=\sum_i\gamma_{D_i}^X.
\end{equation}
The reaction rate $\gamma_{1,2\leftrightarrow 3,4}$ due to the $s$- and $t$-channel scattering  $1,2\leftrightarrow 3,4$ is given by
\begin{equation}
\gamma_{1,2\leftrightarrow 3,4} = \frac{m_{X}}{64\pi^4z}\int_{s_{\text{min}}}^{\infty}ds \sqrt{s}\hat{\sigma}_{1,2\leftrightarrow 3,4}\lr{s}K_1\lr{z\frac{\sqrt{s}}{m_{X}}},
\end{equation}
with $s_{\text{min}} = \max((m_1+m_2)^2,(m_3+m_4)^2)$, and $\hat{\sigma}\lr{s}$ is given by \cite{Luty:1992un}
\begin{equation}
\label{eq:sigmahat}
\hat{\sigma}\lr{s} = \frac{1}{8\pi s}\int_{t_{\text{+}}}^{t_{\text{-}}}dt|\overline{\mathcal{M}}|^2,
\end{equation}
where the limits on $t$ can be obtained as
\begin{equation}
t^\pm = \frac{(m_1^2-m_2^2-m_3^2+m_4^2)^2}{4s}-\lr{\sqrt{\frac{(s+m_1^2-m_2^2)^2}{4s}-m_1^2}\pm\sqrt{\frac{(s+m_3^2-m_4^2)^2}{4s}-m_3^2}}^2,
\end{equation}
and where $s$ and $t$ denote the usual Mandelstam variables.

%%%%%%%%%%%%%%%%%%%%%%%%%%%%%%%%%%%%%%%%%%%%%%%%%%%%%%%%%%%%%%%%%%%%%%%%%%%%%%%%%%%%%%%%%%%%%%%%%%%%%%%%%%%%%%%%%%%%%%%%%%%%%%%%%%%%%%%%%%%%%%%%
\section{Matrix elements}{\label{app:A}}
In this appendix we provide the expressions for the matrix elements that are relevant for the decay and scattering processes discussed in relation to the Boltzmann equation in Sec.~\ref{sec:boltzmann}. Here we will follow the same notation as Figs.~\ref{fig:gammaD1} and \ref{fig:gammaD2}, where the Feynman diagrams of the the different relevant processes are shown. For the decays, the matrix elements are given by
\begin{align}
&|\overline{\mathcal{M}}_{D_d^0}|^2 = 4f_{dd}^{ij2}\lr{m_{X_{dd}}^2-(m_{d_i}+m_{d_j})^2}\\
&|\overline{\mathcal{M}}_{D_d}|^2 = \lambda^2 v'^2,
\end{align}
where $f_{dd}^{ij}$ and $\lambda v'$ are Yukawa and trilinear scalar coupling constants respectively, and $m_i$ is the mass of particle $i$.
For scattering processes mediated by $X_{ud}^{(*)}$ the matrix elements are given by
\begin{align}
|\overline{\mathcal{M}}_{S_S}|^2 &= \frac{2(\lambda v'f_{ud}^{ij})^2\lr{s-\lr{m_{u_i}+m_{d_j}}^2}}{\lr{s-m_{X_{ud}}^2}^2+\Gamma_{X_{ud}}^2m_{X_{ud}}^2},\\
|\overline{\mathcal{M}}_{S_T}|^2 &= \frac{2(\lambda v'f_{ud}^{ij})^2\lr{t-\lr{m_{u_i}+m_{d_j}}^2}}{\lr{t-m_{X_{ud}}^2}^2},
\end{align}
where $t$ and $s$ are the usual Mandelstam variables, and $\Gamma_i$ is the decay width of particle $i$. Similarly, for the diagrams meditated by $X_{dd}^{(*)}$ the matrix elements are given by
\begin{align}
|\overline{\mathcal{M}}_{X_S}|^2 &= \frac{2(\lambda v'f_{dd}^{ij})^2\lr{s-\lr{m_{d_i}+m_{d_j}}^2}}{\lr{s-m_{X_{dd}}^2}^2+\Gamma_{X_{dd}}^2m_{X_{dd}}^2},\\
|\overline{\mathcal{M}}_{X_T}|^2 &= \frac{2(\lambda v'f_{dd}^{ij})^2\lr{t-\lr{m_{d_i}+m_{d_j}}^2}}{\lr{t-m_{X_{dd}}^2}^2}.
\end{align}
Finally, the matrix elements corresponding to scatterings with a quark mediator are given by
\begin{align}
|\overline{\mathcal{M}}_{S_S^0}|^2 = \nonumber \frac{(f_{dd}^{ik}f_{ud}^{kj})^2}{\left(s-m_{d_k}^2\right)^2}&\Bigg\{m_{X_{dd}}^2 \left(m_{X_{ud}}^2-m_{u_j} (m_{u_j}+m_{d_k})-s \right)+\\ \nonumber
&m_{d_i}^2 \left(m_{u_j} (m_{u_j}+m_{d_k})-m_{X_{ud}}^2\right)+\\ \nonumber
&m_{d_i} m_{d_k} \left(-m_{X_{ud}}^2+m_{u_j}^2+2 m_{u_j}m_{d_k}+s \right)+\\ 
&s \left(-m_{X_{ud}}^2+m_{u^j} m_{d^k}+s+t \right)\Bigg\},\\ 
|\overline{\mathcal{M}}_{S_T^0}|^2 = \frac{(f_{dd}^{ik}f_{ud}^{kj})^2}{\left(t-m_{d_k}^2\right)^2} \nonumber &\Bigg\{m_{X_{dd}}^2 \left(m_{d_i} (m_{d_k}-m_{d_i})+m_{X_{ud}}^2\right)+\\ \nonumber
&t \left(m_{d_i}^2-m_{d_k} (m_{d_i}+m_{u_j})+m_{u_j}^2-s \right)+\\ \nonumber
&m_{u_j} \bigg[m_{d_k}\left(m_{X_{ud}}^2-m_{d_i} (m_{d_i}+m_{u_j})\right)+\\ 
&m_{u_j} (m_{d_i}-m_{X_{ud}})(m_{d_i}+m_{X_{ud}})+2 m_{d_i} m_{d_k}^2\bigg]\Bigg\}.
\end{align}

\end{appendix}

\bibliographystyle{JHEP}
\bibliography{References}

\end{document}